%% file: sentinel_arxiv.tex
\pdfoutput=1
\documentclass[pdflatex,sn-mathphys-num]{sn-jnl}
\usepackage{graphicx}\usepackage{multirow}\usepackage{amsmath,amssymb,amsfonts}\usepackage{mathrsfs}
\usepackage[title]{appendix}\usepackage{xcolor}\usepackage{textcomp}\usepackage{booktabs}\usepackage{caption}
\usepackage{hyperref}\usepackage{xurl}\AtBeginDocument{\let\burl\url}
\usepackage{geometry}
\usepackage{mathpazo}\selectfont
\definecolor{inkdeep}{RGB}{18,52,86}\definecolor{inkteal}{RGB}{13,102,110}\definecolor{inkbrick}{RGB}{150,54,42}\definecolor{inkmute}{RGB}{110,118,128}
\makeatletter
\def\sectionfont{\reset@font\fontfamily{\rmdefault}\fontsize{14bp}{16bp}\bfseries\selectfont\raggedright\boldmath\color{inkdeep}}
\def\subsectionfont{\reset@font\fontfamily{\rmdefault}\fontsize{12bp}{14bp}\bfseries\selectfont\raggedright\boldmath\color{inkteal}}
\makeatother
\newcommand{\rhtext}{Community Odour Reports as a Forecastable Sensor of Fugitive Emissions}
\makeatletter
\def\ps@headings{\let\@oddfoot\@empty\let\@evenfoot\@empty
  \def\@oddhead{\parbox[b]{\textwidth}{\footnotesize\color{inkmute}\rhtext\hfill\raisebox{0.1ex}{\color{inkteal}\thepage}\\[2.5pt]{\color{inkteal}\rule{\textwidth}{0.4pt}}}}%
  \let\@evenhead\@oddhead\let\@mkboth\markboth}
\makeatother
\usepackage{microtype}
\hypersetup{colorlinks=true,linkcolor=inkdeep,citecolor=inkteal,urlcolor=inkbrick,breaklinks=true}
\newif\ifshowmarkers \showmarkersfalse
\makeatletter
\newcommand{\oocnum}[1]{\@latex@warning{OOC-NUMBER: \detokenize{#1}}\ifshowmarkers\textcolor{orange}{#1}\else#1\fi}
\newcommand{\oocref}[1]{\@latex@warning{OOC-REFERENCE: \detokenize{#1}}\ifshowmarkers\textcolor{orange}{[ref: #1]}\else[ref]\fi}
\newcommand{\open}[1]{\@latex@warning{OPEN: \detokenize{#1}}\ifshowmarkers\textcolor{magenta}{[OPEN: #1]}\fi}
\makeatother

\newcommand{\snote}[1]{\refstepcounter{section}\section*{Supplementary Note \thesection. #1}\addcontentsline{toc}{section}{Supplementary Note \thesection. #1}}
\makeatletter
\newcommand{\GAwidth}{\linewidth}
\let\GA@printabstract\printabstract
\def\printabstract{\GA@printabstract\par\vspace{4pt}\centerline{\includegraphics[width=\GAwidth]{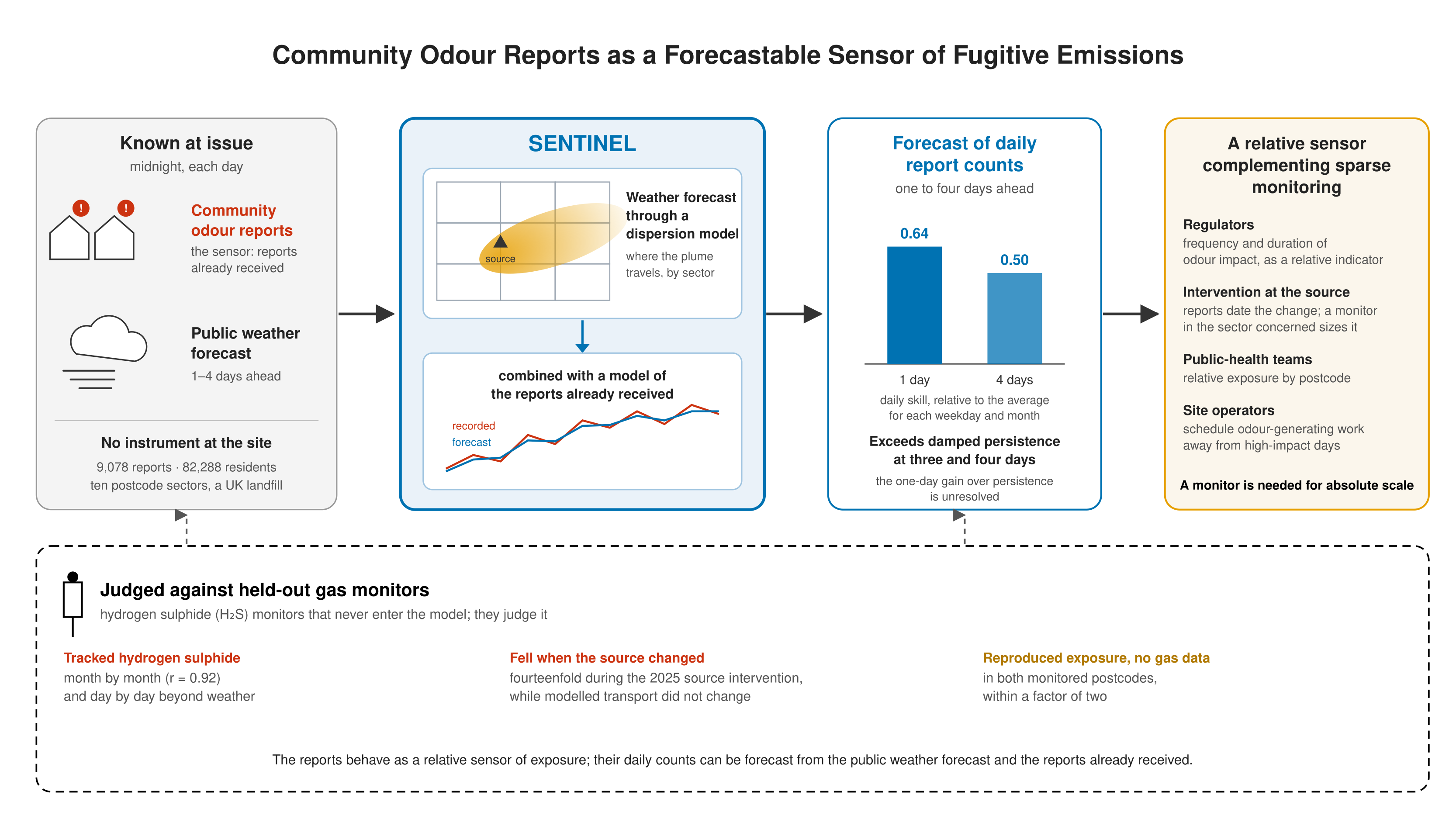}}\par\vspace{0pt}}
\def\addressfont{\reset@font\fontsize{9.5bp}{11.5bp}\selectfont\titraggedcenter}%
\def\abstractfont{\reset@font\fontsize{9bp}{11bp}\selectfont\leftskip=0pt\rightskip=0pt\parfillskip=0pt plus 1fil}%
\renewcommand\abstracthead{\@startsection{section}{1}{\z@}{-6pt \@plus0ex \@minus0ex}{2pt}{\abstractheadfont}}
\renewcommand{\@maketitle}{\newpage\null\vskip-12pt%
    \hsize\textwidth\parindent0pt%
    \ifx\@title\empty\else\removelastskip\vskip4pt\nointerlineskip{\Titlefont\@title\par}\fi%
    \ifnum\aucount>0
        \global\punctcount\aucount\vskip8pt\artauthors\par%
        {\vskip3pt\addressfont\auaddress\par\removelastskip\vskip6pt%
        \ifnum\emailcnt>0\relax\ifx\corrauthemail\@empty\else{\ifnum\aucount>1*\fi}Corresponding author(s). E-mail(s): \corrauthemail\par\fi%
          \ifx\authemail\@empty\else Contributing authors:\ \authemail\fi\fi%
        \removelastskip\vskip2pt}%
    \fi%
    {\printabstract\par}{\printkeywords\par}\removelastskip\vskip8pt}
\makeatother
\newcommand{\sifiglabelfont}{\bfseries\color{inkteal}}
\providecommand{\sifiglabelfont}{\bfseries}
\newcommand{\sifigpart}[2]{\clearpage\begin{center}#1\\[4pt]
  {\footnotesize\textbf{\figurename~\ref{#2}, continued on the next page.}}\end{center}}
\newcommand{\sifig}[4]{\clearpage\begingroup\refstepcounter{figure}\label{#3}%
  \begin{center}#1\end{center}\vspace{-0.4em}
  \footnotesize\noindent{\sifiglabelfont\figurename~\thefigure.}~\textbf{#2.} #4\par\endgroup\addvspace{\bigskipamount}}
\newcounter{smeth}
\newcommand{\smethod}[1]{\refstepcounter{smeth}\section*{Supplementary Methods \thesmeth. #1}\addcontentsline{toc}{section}{Supplementary Methods \thesmeth. #1}}
\begin{document}

\title{\centering\fontsize{15}{17.7}\selectfont Community Odour Reports as a Forecastable Sensor of Fugitive~Emissions}

\author*[1]{\fnm{Timothy C.} \sur{Pearce}}\email{t.c.pearce@le.ac.uk}
\author[2]{\fnm{David J. T.} \sur{Smith}}\email{toby.smith@ukhsa.gov.uk}
\author[2]{\fnm{Alec} \sur{Dobney}}\email{alec.dobney@ukhsa.gov.uk}
\author[2]{\fnm{Alessia} \sur{Freddo}}\email{alessia.freddo@ukhsa.gov.uk}

\affil*[1]{\orgdiv{Biomedical Engineering Research Group, School of Engineering}, \orgname{University of Leicester}, \orgaddress{\city{Leicester}, \country{United Kingdom}}}
\affil[2]{\orgdiv{Environmental Hazards and Emergency Department}, \orgname{UKHSA}, \orgaddress{\city{Nottingham}, \country{United Kingdom}}}

\abstract{%
Ambient air quality monitors (AAQMs) measure exposure to emissions from waste facilities at few points; for most residents the only record is their own odour reports, used by regulators as evidence of impact but not typically as measurements of exposure\cite{bokowa2021,nicell2009,brancher2017}. Calibrated public reports are accepted measurements in seismology\cite{wald2012,atkinson2007,worden2012}, but whether community odour reports can serve as a sensor of exposure, and be forecast, is untested against independent monitors. Here we show they can, at a UK landfill, with SENTINEL, which combines the public weather forecast, run through a dispersion model, with the reports already received and no instrument at the site. Judged against the independent monitors, monthly report frequency was correlated with hydrogen sulphide ($r=0.92$) and daily reports remained associated with it beyond weather; they fell fourteenfold after a change at the source while modelled transport did not; and, with no gas data as input, gave exposure in both monitored postcodes within a factor of three of that measured. SENTINEL forecast daily report counts with skill 0.64 one day and 0.50 four days ahead, exceeding damped persistence four days ahead; the one-day gain over persistence was unresolved. Community odour reports can act as a relative sensor complementing sparse monitoring; a monitor is needed for absolute scale.%
}

\maketitle

\section*{Introduction}
At landfills the principal odorant is usually hydrogen sulphide (H$_2$S)\cite{kim2005landfill,atsdr2001landfillgas}, detectable at concentrations far below those of acute toxicological concern\cite{govuk_nd_h2s,nrc2010acute}; the health effects of chronic low-level exposure remain unresolved\cite{guidotti2010,nuvolone2019}, but odour itself can produce symptoms through annoyance and stress\cite{schiffman2005,hirasawa2019subjective}. Waste facilities are disproportionately sited near deprived communities\cite{martuzzi2010,mohai2009,carson2024malodors}, yet epidemiological studies of their effects usually rely on distance from the site as a surrogate for exposure\cite{elliott2001}. Regulators assess odour impact through its frequency, intensity, duration, offensiveness and location (FIDOL)\cite{EAH4_2011,IAQM2018,nicell2009}. Ambient air quality monitors (AAQMs) measure exposure at only a few points, and monitoring networks often under-represent the populations they serve\cite{LITPedde2024Representativeness}; for most residents the only record of exposure is their own odour reports, which regulators use as evidence of odour impact but not typically as measurements of exposure\cite{bokowa2021,brancher2017}. Around the landfill studied here, 77.4~\% of residents live in postcode sectors without a monitor.

A community's odour reports could serve as a sensor of its exposure, the role that volunteered public observations already play in other fields\cite{goodchild2007}; in seismology, public reports of felt shaking, calibrated against instruments, are an accepted measure of ground-motion intensity\cite{wald2012,atkinson2007,worden2012}. Two bodies of work bear on this. First, reported odour increases with measured H$_2$S in diary studies of recruited residents\cite{wing2008,heaney2011}, resident odour records agree with an instrumental odour index\cite{lee2013}, and odour annoyance is related to modelled exposure\cite{blanesvidal2012}; closest to this study, community reports fell with measured H$_2$S during a documented change at a source\cite{rico2025}. Second, odour impact has been forecast with dispersion models driven by numerical weather prediction and checked against incidents logged by recruited observers\cite{Cartelle2016,Jimenez2016} or plumes traced by field inspectors\cite{capelli2018}, and with statistical models of weather and air-quality data\cite{Ustun2026,Gavros2024,Hsu2020}, including up to three days ahead with H$_2$S monitors at the site\cite{Mulrow2020}; citizen odour reports have also been combined with dispersion modelling to test emission inventories\cite{Veratti2026}. No study has yet tested, with AAQMs excluded from the model, whether the reports a regulator already receives can serve as a sensor of exposure and be forecast.

Whether a community's odour reports can serve as a sensor of its exposure, and be forecast days ahead from the public weather forecast, with no instrument at the site and the AAQMs used only to evaluate the result, is untested. Two difficulties make this a demanding test. Reports and gas concentrations share meteorological drivers, so a correlation between them does not show that reports are related to exposure beyond what the weather explains. Daily report counts are also strongly autocorrelated, so a forecast that largely repeats the previous day can appear skilful; persistence is therefore a demanding benchmark, as analyses of influenza estimates from internet searches have shown\cite{ginsberg2009,lazer2014}, and a useful forecast must beat references built from the report record alone: damped persistence and a blend of climatology and persistence.

SENTINEL runs the public weather forecast through an atmospheric dispersion model for the ten postcode sectors around the landfill and combines the result with a state-space model of the reports already received; no instrument at the site enters it. To convert reports into exposure, we describe the community's response with a psychometric reporting model, a probit dose--response of the kind used in psychophysics\cite{wichmann2001} and in studies of odour annoyance\cite{nicell2009,blanesvidal2012}: the probability that residents of a sector report in a six-hour window rises with the H$_2$S concentration around a threshold, with a spread that reflects differences in sensitivity and in readiness to report. A network of AAQMs measuring H$_2$S, never used as model input, provides the measured exposure with which we compare the reports. Earlier work at this site took the opposite approach, nowcasting the H$_2$S measured by the monitors from on-site meteorology\cite{Pearce2026}. Our analyses indicate that, judged against AAQMs that never entered the model, the reports were related to the landfill's hydrogen sulphide beyond what the weather explains, fell after a change at the source while modelled transport did not, gave the pattern of exposure in the two monitored postcodes at a level within a factor of three with no gas data as input, and could be forecast one to four days ahead from the public weather forecast and the reports already received, adding information beyond the report history and exceeding damped persistence four days ahead.

\section*{Results}

\subsection*{Odour reports and fugitive emissions}\label{sec:r1}
After removing duplicates, we analysed 9,078 odour reports made between March 2023 and January 2026 by residents of the ten postcode sectors around the landfill (82,288 residents); 9,069 of them were made by July 2025. An earlier record, dated by when reports were filed rather than when the odour occurred, was analysed separately (Methods; Supplementary Note~\ref{snote:legacy}). Throughout, exposure means exposure to the landfill's odorous emissions, indexed by H$_2$S, and the AAQMs are never used as model inputs; we use them only to evaluate the results.

\begin{figure}[!tbp]\centering
\makebox[\textwidth][c]{\includegraphics[width=0.8\paperwidth,height=0.6\textheight,keepaspectratio]{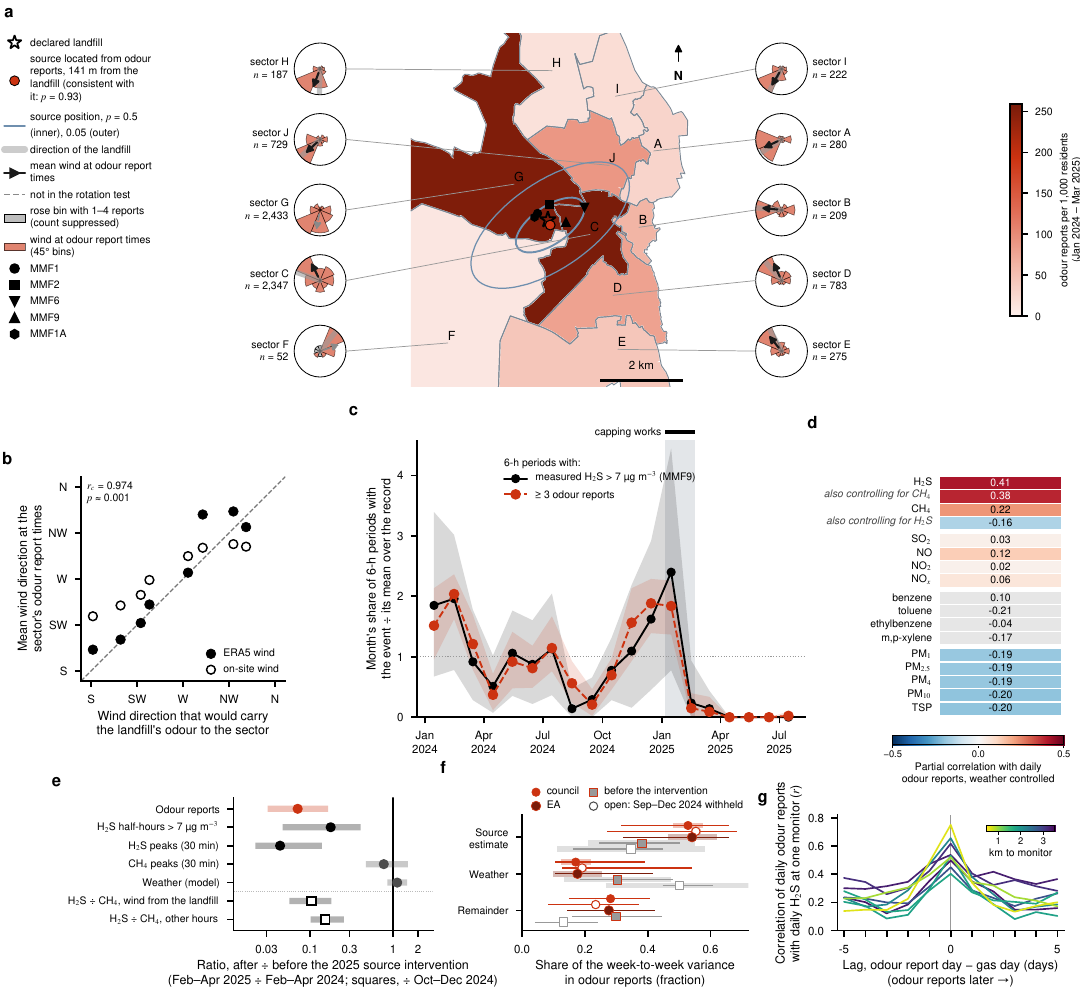}}
\caption[Temporal patterns in odour reports, monitored gas concentrations and the 2025 source intervention.]{\textbf{Temporal patterns in odour reports, monitored gas concentrations and the 2025 source intervention.}  \input{legend_Fig1}}\label{fig:1}
\end{figure}

Between January 2024 and March 2025, most of the 7,517 reports came from the two sectors next to the landfill: 2,433 from sector~G and 2,347 from sector~C (Fig.~\ref{fig:1}a). Locating the source from the reports alone placed it 141\,m from the declared landfill, consistent with the landfill being the source ($p=0.93$). Wind directions at the time of reporting gave the same result (Fig.~\ref{fig:1}b; Supplementary Fig.~\ref{fig:s1}a). In the 8 sectors with at least 150 reports, the mean wind direction during reports was correlated with the direction from the landfill to each sector (circular correlation, a correlation between two directions, $r=0.974$; one-sided permutation test against 10,000 records with the wind shifted in time, $P$~\ensuremath{\approx}~0.001; Supplementary Note~\ref{snote:attribution}).

Over months, the share of six-hour periods with at least three reports was correlated with the share of periods in which hydrogen sulphide (H$_2$S) at monitor MMF9 exceeded 7\,\textmu g\,m$^{-3}$ ($r=0.92$; 99\,\% interval 0.82–0.99; 15~months; Fig.~\ref{fig:1}c). This was the strongest of the 84 definitions of an odour episode that we tested on this window (median 0.52), and its interval does not allow for that selection; the definition fixed before the analysis gave $r=0.85$, and leaving out any one month gave 0.91–0.94. Of the pollutants measured at the monitor, H$_2$S had the largest partial correlation with daily reports after allowing for weather, season, day of the week and recent reporting (partial $r=0.41$; 0.37 on unstable and 0.40 on stable nights; Fig.~\ref{fig:1}d). The reports were associated with H$_2$S and with co-emitted methane, and not detectably with sulphur dioxide, nitrogen oxides or particles. They therefore indicate exposure to the landfill's odorous emissions, not total chemical exposure. With the psychometric reporting model, we converted the reports into a relative estimate of the state of the source (Methods, equation~\eqref{eq:state}); below, we compare this estimate with the exposure measured at the monitors. These relations held with each of the four four-month periods of the scored window (January--April 2024, May--August 2024, September--December 2024 and January--April 2025) withheld in turn (monthly frequency $r$ at least 0.91; Supplementary Table~\ref{tab:periods}).

From January 2024 to March 2025, a source estimate derived from the monitors accounted for the largest share of the week-to-week variance in reports (0.53~[0.32,~0.66]; 65~weeks), mostly through the step change of the 2025 source intervention (Fig.~\ref{fig:1}f). Because this estimate is built from measured gas, it includes the transport at the monitors (Methods). In each of the nine sectors with enough reports, the correlation of daily reports with H$_2$S at the nearest monitor was highest when both were taken on the same day (Fig.~\ref{fig:1}g).

Only two of the ten sectors have a monitor. In the other eight, the correlations between monitored gas and reporting over time are dominated by an event shared by all sectors in January 2025 and do not show that each sector's reports varied with its own monitor (Supplementary Note~\ref{snote:attribution}), and exposure can be estimated only with the dispersion model. A change at the source provides a further test of the reports as a sensor.

\subsection*{Response to a change in the source}\label{sec:rint}
The 2025 source intervention provided a natural experiment, in which a change at the source can be separated from the weather\cite{Grange2019}. In early 2025 the last open areas of the landfill were capped (the capping works, 6 January to 21 February 2025, from the cover records; Fig.~\ref{fig:1}c, shaded). In February--April 2025, compared with the same months of 2024, reports fell fourteen-fold (95\,\% interval 6.0–32.4), half-hours with H$_2$S above 7\,\textmu g\,m$^{-3}$ fell six-fold and 30-minute H$_2$S peaks (the 99th percentile of 30-minute means) fell to 0.04~[0.02,~0.14] of their earlier level, while 30-minute methane peaks (0.77~[0.47,~1.51]) and modelled atmospheric transport (ratio after to before 1.12~[0.86,~1.48]) showed no detectable change (Fig.~\ref{fig:1}e; Supplementary Fig.~\ref{fig:s1}b--e). Between October--December 2024 and February--April 2025, H$_2$S fell relative to co-emitted methane both with wind from the landfill (0.10~[0.06,~0.18]) and in other hours (0.15~[0.10,~0.25]), with a step in the daily ratio on 22~February~2025; the point estimate fell in each wind quadrant (after-to-before ratios 0.07~[0.05,~0.17] to 0.23~[0.09,~1.35], the southern interval including no change), consistent with a change at the source across most of the site. Capping, gas extraction and flaring cannot be separated with the records available, and the order of the gas and report steps is not resolved.

The report-based exposure estimate fell further than measured exposure (estimate to measurement, after to before, 0.42~[0.25,~0.72] in sector~C and 0.38~[0.22,~0.71] in sector~G; Extended Data Fig.~\ref{fig:ed4}c), and residents also reported less per unit of measured H$_2$S (0.69~[0.44,~1.12], an interval that includes no change, and 0.14~[0.03,~0.65]; about~two~thirds of sector~C's excess fall at the point estimate). In sector~C the change in reports per unit of H$_2$S was within the month-to-month variation of the record (0.80); in sector~G it was lower than at any other month boundary (0.20; Extended Data Fig.~\ref{fig:ed4}d). Correcting one sector's estimate by the change measured at the monitor in the other sector did not recover its measured change to within 0.8--1.25 (0.79~[0.57,~1.08] and 1.89~[1.14,~3.20], although the first interval includes 1; Extended Data Fig.~\ref{fig:ed2}e); these month-bootstrap intervals are too narrow (Methods). The timing of the change could therefore be found from the reports, but measuring its size needed a monitor in the sector concerned. Forecast report counts fell with the recorded counts through the report record rather than the weather forecast (Extended Data Fig.~\ref{fig:ed4}b), and trained without January--April 2025 the forecast scored best on that period (daily skill 0.83 one day ahead).

\subsection*{Odour-impact factors from the reports}\label{sec:r2}
\begin{figure}[!tbp]\centering
\includegraphics[width=\textwidth,height=0.5\textheight,keepaspectratio]{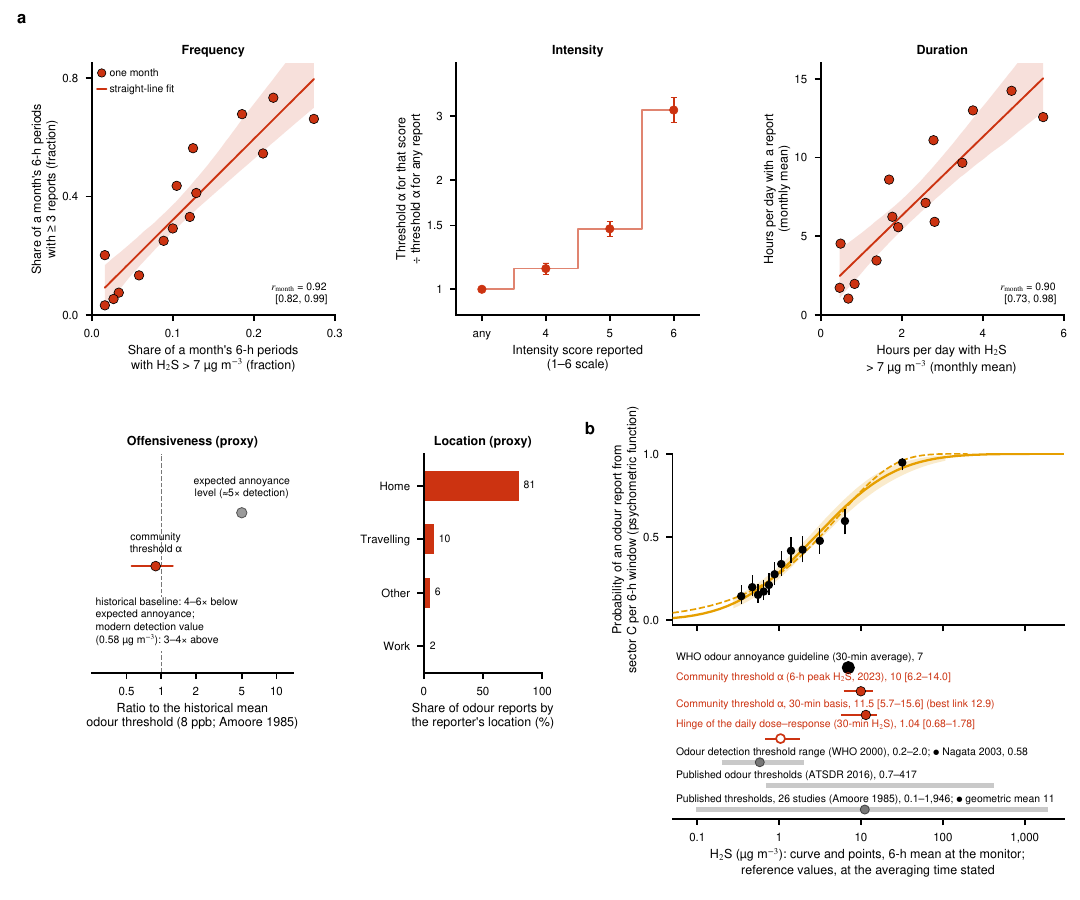}
\caption[Characteristics of reported odour impacts and reporting patterns.]{\textbf{Characteristics of reported odour impacts and reporting patterns.}  \input{legend_Fig2}}\label{fig:2}
\end{figure}

We derived each of the five odour-impact factors used by UK regulators (FIDOL)\cite{EAH4_2011,IAQM2018} from the reports and compared it with H$_2$S at monitor MMF9, which the forecast does not use (Fig.~\ref{fig:2}a; Supplementary Fig.~\ref{fig:s2}). Three factors could be measured directly from the reports and two only approximated. Monthly frequency is the episode measure of Fig.~\ref{fig:1}c ($r=0.92$). Duration, the number of hours a day with a report, correlated with the number of hours above that level ($r=0.90$; 99\,\% interval 0.73–0.98). Reported intensity was only weakly related to H$_2$S (Spearman 0.28~[0.22,~0.35]). Because the landfill emits almost continuously, these three factors are correlated with one another and do not provide independent confirmation; similarly, in field studies odour frequency predicts annoyance and intensity adds little\cite{sucker2007b}. We used the monitors individually as references, because each sits in its own microclimate and their average is a poor measure of exposure.

Offensiveness was approximated by the community's reporting threshold relative to published odour thresholds, whose direction depends on which detection threshold is taken as the baseline, and location by where residents were when they reported; most of the 8,275 reports made from all postcodes between January 2024 and April 2025 were made from home (Supplementary Note~\ref{snote:fidol}).

We estimated the community's reporting threshold with the psychometric reporting model (Fig.~\ref{fig:2}b; Methods). The probability of at least one report from sector~C in a six-hour window rose with the six-hour mean H$_2$S at MMF9 along a cumulative normal curve, with its midpoint at 2.7\,\textmu g\,m$^{-3}$ (5.5\,\textmu g\,m$^{-3}$ with a complementary log--log curve). Measured on the 30-minute averaging time of the World Health Organization (WHO) odour annoyance guideline\cite{who2000airquality}, the probability of a report from sector~C reached half its fitted upper limit when the peak 30-minute H$_2$S was 11.5~[5.7,~15.6]\,\textmu g\,m$^{-3}$ (averaged over link functions; the interval includes the guideline value), about 1.5--2 times the guideline value of 7\,\textmu g\,m$^{-3}$ and of the same order; the interval for the best-fitting link (12.9~[10.1,~16.4]\,\textmu g\,m$^{-3}$) excludes 7. The threshold is a property of this sector and monitor, combining residents' sensitivity, their readiness to report and the number of residents, and it refers to the landfill-gas mixture indexed by H$_2$S: the absence of any association with benzene, toluene and xylenes (Fig.~\ref{fig:1}d) is consistent with the reported odour being its sulphurous component. The WHO guideline was set ``to avoid substantial complaints about odour annoyance'' and describes annoyance rather than detection\cite{who2000airquality}. The model predicted daily reports better than the daily share of time above 7\,\textmu g\,m$^{-3}$ (difference in mean squared error \ensuremath{-}0.22~[\ensuremath{-}0.33,~\ensuremath{-}0.12]); comparisons with published detection thresholds are in Supplementary Note~\ref{snote:fidol}.

The reports are therefore related to exposure to the landfill's odorous emissions, mainly H$_2$S. They are related to methane only through H$_2$S and not detectably to the other pollutants measured, and they do not measure a health-relevant dose. They also differ from warning tiers based on gas concentrations: one day ahead, forecast tiers agreed with gas tiers with $\kappa=0.42$, close to 0.41 for persistence. After allowing for season, weekday, weather and recent reporting, daily reports were still correlated with H$_2$S (Supplementary Note~\ref{snote:beyondweather}).

\subsection*{Exposure estimated from reports alone}\label{sec:r3}
\begin{figure}[!tbp]\centering
\makebox[\textwidth][c]{\includegraphics[width=0.8\paperwidth,height=0.6\textheight,keepaspectratio]{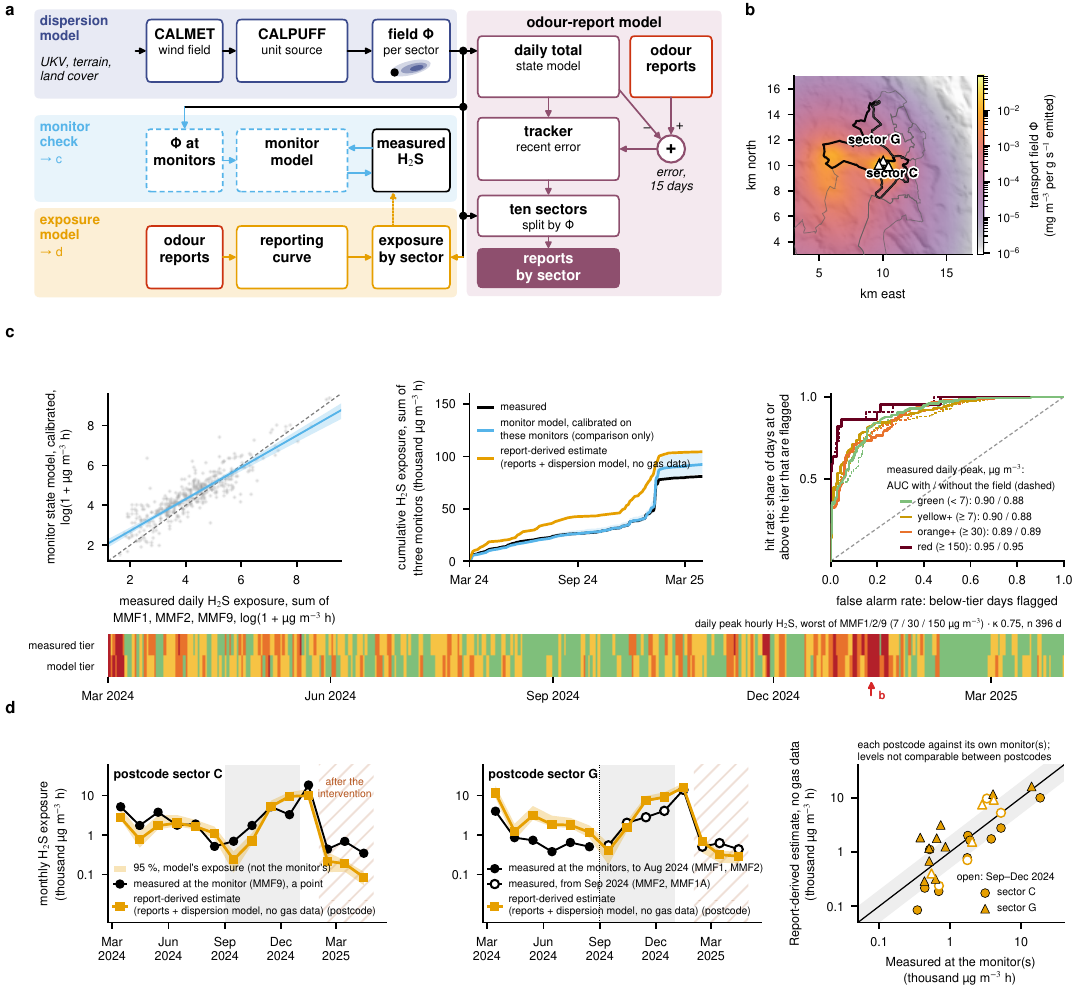}}
\caption[Spatial estimates of exposure derived from odour reports and atmospheric dispersion modelling.]{\textbf{Spatial estimates of exposure derived from odour reports and atmospheric dispersion modelling.} \input{legend_Fig3}}\label{fig:3}
\end{figure}

The modelling chain uses only public inputs (Fig.~\ref{fig:3}a; Methods). The CALMET meteorological model and the CALPUFF puff dispersion model\cite{Scire2000calmet,Scire2000calpuff}, run with the Met Office's public UK forecast (UKV), give a dispersion field, $\Phi$, for each postcode sector (Fig.~\ref{fig:3}b; Supplementary Movie~1); a state-space model divides each day's reports among the sectors according to $\Phi$; and an exposure model converts reports into concentrations through the psychometric reporting model. No instrument at the site enters the chain, and the AAQMs are used only to evaluate the results.

We first checked the dispersion field at the three monitors near the source (Fig.~\ref{fig:3}c). A model of each monitor, based on its own past H$_2$S and on $\Phi$, agreed with measured daily exposure ($r=0.87$; slope 0.80~[0.72,~0.86]; cumulative exposure 1.13\ensuremath{\times}~[1.02,~1.28] the measured value) and the warning tiers ($\kappa=0.75$; area under the receiver operating characteristic curve 0.89 to 0.95). Most of this agreement comes from each monitor's own past readings, whose six-hour persistence was more accurate still (Supplementary Note~\ref{snote:monitormodel}). This checks the dispersion field near the source, but not transport to the sectors where most reports are made.

Over 29~months, the number of reports per 1,000 residents was correlated with the H$_2$S exposure measured in both monitored postcodes (Fig.~\ref{fig:3}d; Extended Data Fig.~\ref{fig:ed2}a,b). Monthly Spearman correlations were 0.87~[0.67,~0.94] in sector~C and 0.77~[0.34,~0.94] in sector~G, and 0.89 and 0.78 after removing the seasonal cycle (weekly values in Supplementary Note~\ref{snote:exposure}). On monitors that did not change, the Spearman correlations between week-on-week changes in reports per 1,000 residents and in measured exposure were 0.73~[0.59,~0.85] in sector~C and 0.62~[0.39,~0.77] in sector~G (MMF2), and at least 0.48~[0.28,~0.70] with any four-month period withheld. These two postcodes have both the monitors and the most reports, so the relationship is untested where reports are sparse. In each of the seven other sectors with enough reports, reports were correlated with H$_2$S at the nearest monitor on the same day, but these correlations are dominated by an event shared by all sectors, and the exposure level in these sectors cannot be checked.

We then converted the reports into concentrations through the psychometric reporting model, placing the WHO guideline at its midpoint and using the dispersion model but no gas data. The estimated 14-month exposure was 0.64–0.87\ensuremath{\times} the measured level in sector~C and 1.41–1.87\ensuremath{\times} in sector~G on its unchanged monitor MMF2 (range across published spreads; 1.70\ensuremath{\times} at the published spread), with ratios across the averaging-time conversions we tested (0.38--1.33 and 0.69--2.40); with 14 months the sampling uncertainty of each postcode's level could not be bounded usefully, while summed over the three monitors it was 1.29~[0.93,~2.03] (Extended Data Fig.~\ref{fig:ed2}c). The calibration of the estimated exposure level was chosen by comparison with these monitors, which were sited where impacts were expected\cite{diggle2010}, so it is not an independent test of level; the timing comparison is. The mean estimate was close to the measured level, but weekly estimates typically differed from the measurements by a factor of 3.1 in sector~C and 2.6 in sector~G. The estimates also varied with residents' readiness to report (Supplementary Note~\ref{snote:exposure}). A monitor is needed to put the estimates on an absolute scale, and transferring its calibration to another sector or to a later period did not improve them.

Other calibrations are compared in Supplementary Note~\ref{snote:exposure}.

The dispersion model ranked the two postcodes in the opposite order to their measured exposure (Extended Data Fig.~\ref{fig:ed2}d). Distance from the source ranked the ten sectors about as well as the reports did (+0.903), as expected because the long-term pattern of the dispersion field depends mainly on distance and bearing from the source, and week-to-week shifts in where reports come from, beyond this geography, remain unexplained (Supplementary Note~\ref{snote:exposure}). Across the ten sectors, reports per 1,000 residents rose with modelled transport from the site (Spearman 0.81) and showed no detectable gradient with deprivation or age (with ten sectors, only strong gradients could be detected). Residents' readiness to report also changed over time: the Environment Agency's share of reports varied more than counting noise alone would produce (ratio of the 90th to the 10th percentile of its odds 2.06~[1.76,~2.66], against 1.38 expected from noise), and from counts alone such changes cannot be separated from changes in incidence\cite{liu2023}. The reports therefore indicate when exposure is high within a postcode, while comparing exposure between postcodes, or estimating a dose, requires a monitor. We next tested whether the same modelling chain can reproduce, and then forecast, the daily number of reports.

\subsection*{Same-day estimates with observed weather}\label{sec:r4}
\begin{figure}[!tbp]\centering
\makebox[\textwidth][c]{\includegraphics[width=0.8\paperwidth,height=0.6\textheight,keepaspectratio]{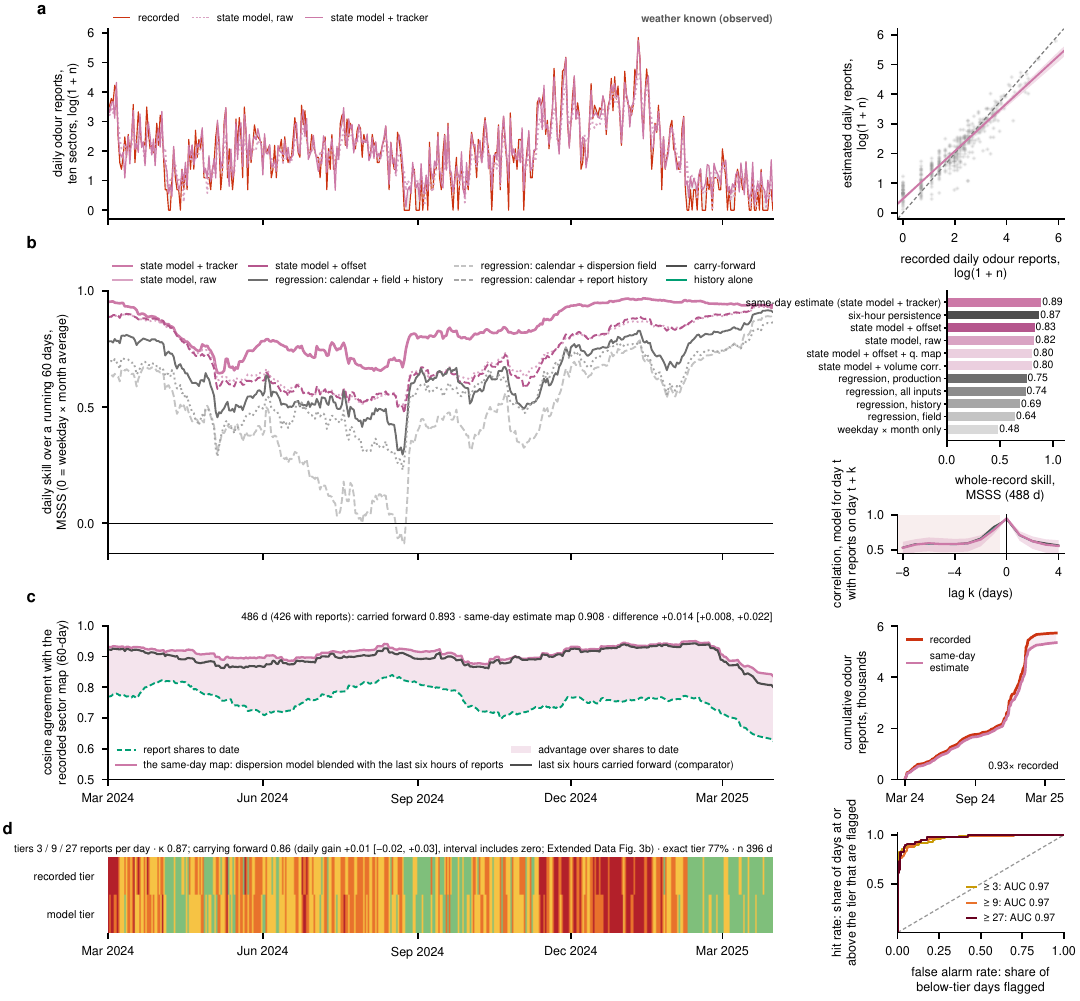}}
\caption[Same-day estimates of daily odour reports with observed weather.]{\textbf{Same-day estimates of daily odour reports with observed weather.} \input{legend_Fig4}}\label{fig:4}
\end{figure}

To see how well the modelling chain can do at best, we ran it with the observed weather in place of the forecast, which gives a same-day estimate of each day's reports: a reference run with the observed weather rather than a forecast (Methods). Its daily totals agreed closely with recorded totals ($r=0.94$; slope 0.80~[0.76,~0.84]; 396~days; Fig.~\ref{fig:4}a) and with the daily report level ($\kappa=0.87$; Fig.~\ref{fig:4}d), with overall daily accuracy similar to that of persistence (0.89 against 0.87; Fig.~\ref{fig:4}b). Its sector map, which blends the dispersion model's shares with the last six hours of reports, agreed more closely with the recorded division of reports among sectors than the last six hours of reports did (gain in daily cosine agreement +0.0145~[+0.0081,~+0.0223]; Fig.~\ref{fig:4}c), mostly because of the recent reports (Extended Data Fig.~\ref{fig:ed3}).

\subsection*{Forecasts one to four days ahead}\label{sec:r5}
\begin{figure}[!tbp]\centering
\makebox[\textwidth][c]{\includegraphics[width=0.8\paperwidth,height=0.6\textheight,keepaspectratio]{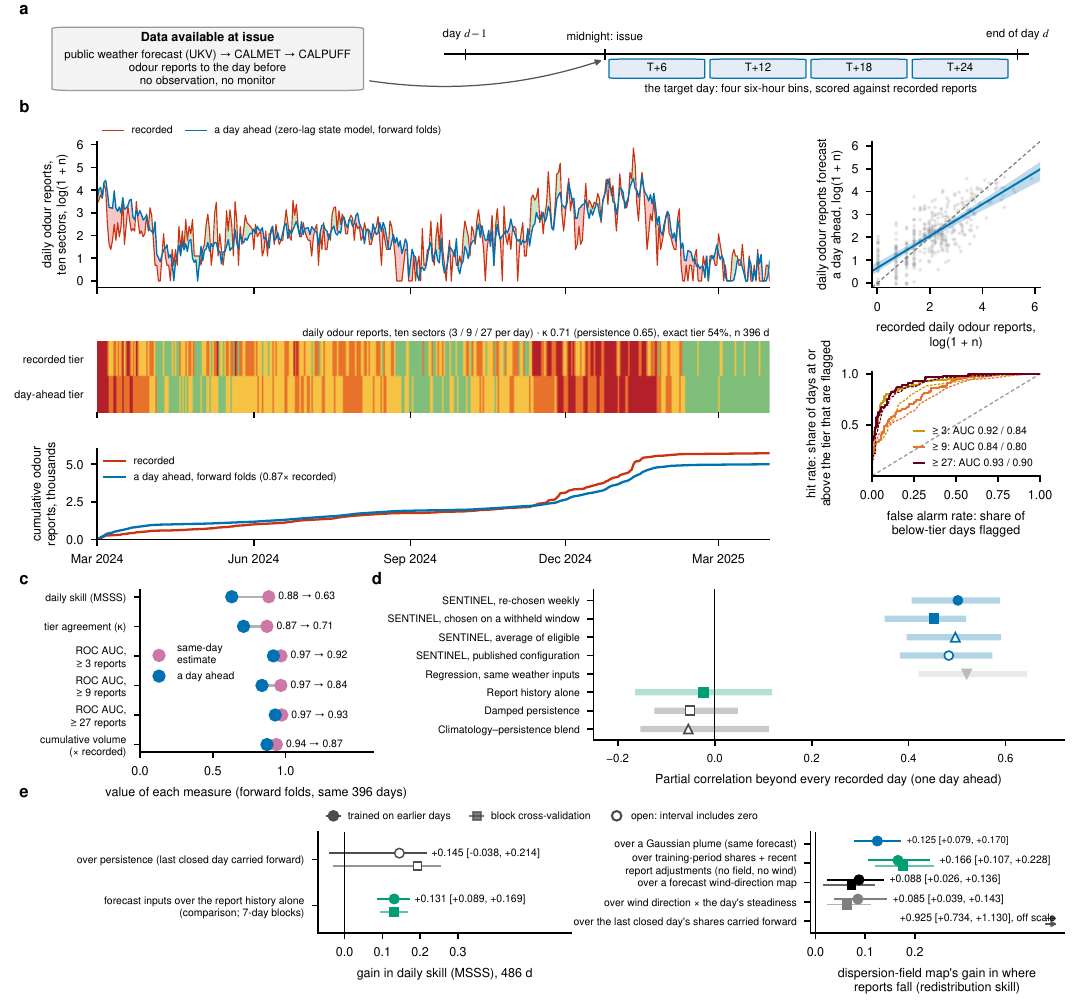}}
\caption[One-day-ahead forecasts of daily odour reports and their spatial distribution.]{\textbf{One-day-ahead forecasts of daily odour reports and their spatial distribution.}  \input{legend_Fig5}}\label{fig:5}
\end{figure}

SENTINEL issues its forecast at midnight from two inputs: the public weather forecast, processed through CALMET and CALPUFF, and the reports recorded up to the previous day (Fig.~\ref{fig:5}a; Methods). Each target day was forecast with a model trained only on earlier days. Results under block cross-validation, in which each block of days is forecast by a model trained on all other blocks, including later days, are given in parentheses. Forecasts are issued at midnight; one day ahead is the 24\,h after issue, and four days ahead the 24\,h starting 72\,h after issue. Our main measure of accuracy is daily skill: the reduction in mean squared error of $\log(1+\text{daily reports})$ relative to the average for the same weekday and month, where 0 means no improvement on that average and 1 a perfect forecast. The references are persistence (the last recorded day)\cite{murphy1988,jolliffe2007}, damped persistence (the usual count for that weekday and month plus part of the last recorded day's departure from it) and a weighted blend of the usual count and the last recorded day, both refitted each day on recorded days (Methods). SENTINEL's configuration was chosen from 226 candidates using the first half of the scored days (Methods; Supplementary Note~\ref{snote:selection}). The scored days had more reports than the dataset as a whole (Extended Data Table~2).

\begin{figure}[!tbp]\centering
\includegraphics[width=\textwidth,height=0.5\textheight,keepaspectratio]{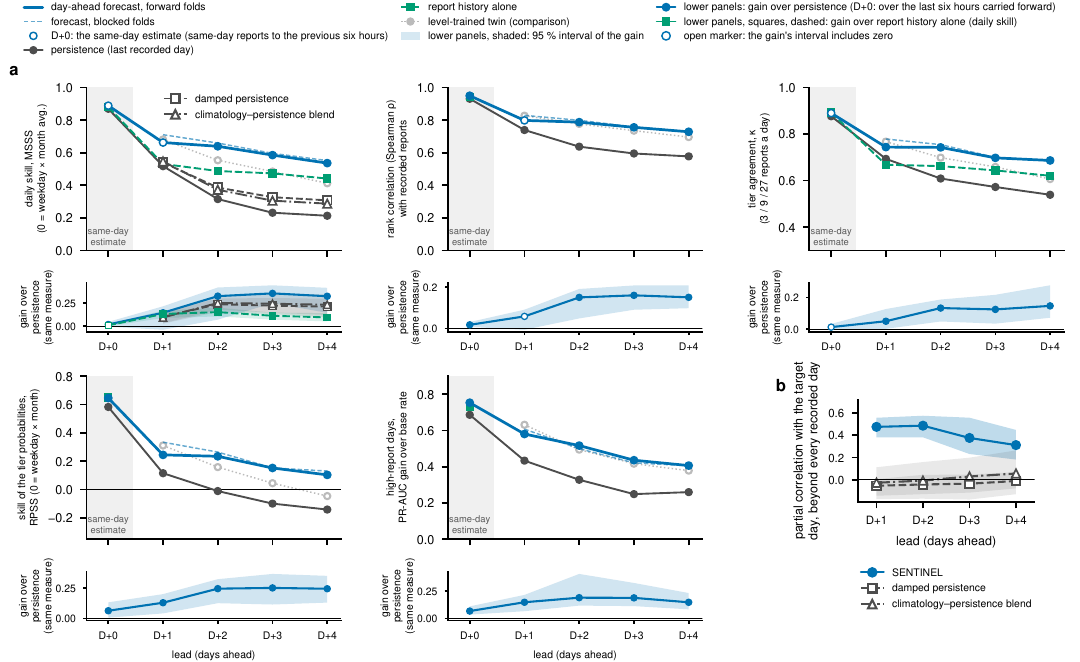}
\caption[Forecast performance across lead times against stronger references.]{\textbf{Forecast performance across lead times against stronger references.}  \input{legend_Fig6}}\label{fig:6}
\end{figure}

With operational data delays (weather from the previous forecast cycle and reports closed 24\,h before issue), daily skill was 0.642 one day ahead and 0.500 four days ahead (block cross-validation, 0.661 and 0.528). Without these delays it was 0.662~[0.569,~0.730] and 0.536~[0.403,~0.636] (Fig.~\ref{fig:6}a; Extended Data Table~1). Over the days shown in Fig.~\ref{fig:5}b, forecast and recorded daily totals were correlated ($r=0.78$; slope 0.70~[0.59,~0.76]), and the forecast agreed better than persistence with the recorded daily report tier ($\kappa=0.71$ against 0.65). Using forecast rather than observed weather lowered skill only slightly (0.676 with the observed weather); most of the drop from the same-day estimate comes from the missing reports of the target day (Fig.~\ref{fig:5}c).

Daily report counts change slowly, so we asked whether the forecast adds information about the target day beyond the report history recorded at issue: we regressed each day's reports on the 14 most recent recorded days, the mean of earlier days, weekday and season, and added the forecast (Methods). One day ahead, the public weather forecast, run through the dispersion model, added information about the next day's reports that the report history did not: SENTINEL's partial correlation with the target day was 0.50~[0.42,~0.59] after the forecast of the same model with the report history alone had also been included. Forecasts from the report record alone added none (0.04), and a regression on report history, calendar and the same weather and dispersion inputs added as much (0.53), so the information is in those inputs rather than in SENTINEL's model form. With the configuration re-chosen each week from earlier days only, the partial correlation was 0.50~[0.41,~0.59] (Fig.~\ref{fig:5}d), consistent across all 7,232 combinations of design choices (Extended Data Fig.~\ref{fig:ed5}f--j).

Daily skill declined slowly with lead time (Fig.~\ref{fig:6}a; Supplementary Fig.~\ref{fig:s5}). Four days ahead it exceeded damped persistence by 0.21~[0.10,~0.30] and the climatology--persistence blend by 0.23~[0.12,~0.34]. Three and four days ahead, all 20 paired gains over persistence (at two lead times under two training schemes) on five verification measures (daily skill, rank correlation, agreement in daily report level, the skill of the forecast probabilities for each report level and the detection of days with 27 or more reports) were positive; for forecasts trained on earlier days, all~10 exceeded zero at the 99.9\,\% level, at which a simulation with a known added effect found the joint bounds held in about~0.89 of simulated records (Methods). One day ahead, the gain in daily skill over persistence was positive but its interval included zero (+0.145, interval \ensuremath{-}0.038–0.214), and the gain over the report history alone was +0.131~[+0.057,~+0.186].

SENTINEL also forecasts how each day's reports divide among the ten postcode sectors. One day ahead, this division was more accurate than one based on the forecast wind direction (gain in redistribution skill +0.088~[+0.026,~+0.136]), than a Gaussian plume driven by the same forecast (+0.125~[+0.079,~+0.170]) and than the average shares of the training period (+0.166~[+0.107,~+0.228]; Fig.~\ref{fig:5}e; Extended Data Fig.~\ref{fig:ed6}a); two days ahead its gain over the shares of the last 28 recorded days was +0.135~[+0.044,~+0.240]. These intervals had lower coverage than their nominal level (measured coverage 0.63--0.83; Methods).

When both used the same report history, SENTINEL had higher daily skill than a regression on report history and calendar, refitted on the same schedule, with an interval excluding zero, in 7~of~8 combinations of lead time and training scheme (Supplementary Note~\ref{snote:comparators}).

Two or more days ahead, the forecast's partial correlation with the target day beyond the recorded report history declined from 0.49~[0.38,~0.58] two days ahead to 0.31 four days ahead and stayed above zero, while for damped persistence and the blend it was close to zero (\ensuremath{-}0.05~to~+0.06; Fig.~\ref{fig:6}b). Four days ahead the forecast's correlation with the target day was nevertheless lower than with the most correlated recorded day (difference \ensuremath{-}0.061~[\ensuremath{-}0.107,~\ensuremath{-}0.030]; Extended Data Fig.~\ref{fig:ed5}e), so beyond one day part of its skill comes from report levels already known at issue.

These gains have limits. Economic value, which measures the benefit of acting on a forecast\cite{zhu2002,Garner2012}, was similar for SENTINEL and persistence at an equal number of alerts, and SENTINEL's alerts for days of high gas concentration two to four days ahead were false less often (false-alarm ratio 0.79–0.86 against 0.92–0.98); at an equal number of alerts this is the same result as a higher hit rate, which was not consistent across training schemes or periods (Supplementary Note~\ref{snote:alerts}; Supplementary Table~\ref{tab:uses}). Trained with each four-month period withheld in turn, the forecast scored on that period as it did under cross-validation (daily skill 0.709 one day ahead), and one day ahead it had a partial correlation above zero beyond the recorded days in every period; it was weakest in May--August 2024 (0.35), when reports were fewest, and strongest after the source intervention (0.83, in part because the weekday-and-month reference did not include the fall in reports; Supplementary Table~\ref{tab:periods}). Three and four days ahead the gains over damped persistence and the blend had intervals that excluded zero in some periods but not in all.

Compared with the monitors, monthly frequency and duration forecast one day ahead were correlated with monitored H$_2$S, as were those of persistence (Extended Data Fig.~\ref{fig:ed4}a); the gain in catching days of high gas concentration at an equal number of alerts was not consistent across training schemes and periods and is reported with its caveat in the Supplementary (Extended Data Fig.~\ref{fig:ed6}b; Supplementary Note~\ref{snote:alerts}); and adding the dispersion model improved forecasts of the H$_2$S measured at the monitors (Extended Data Fig.~\ref{fig:ed6}c). Adding the local measurements we tested did not detectably improve the forecast; equivalence was shown only one and two days ahead, at about 87\,\% one-sided confidence (Extended Data Fig.~\ref{fig:ed1}a--c; Supplementary Notes~\ref{snote:monthly}--\ref{snote:instruments}). AAQMs remain necessary to validate the forecasts, to put exposure on an absolute scale, to define episodes of high gas concentration and to establish whether a change in reporting reflects a change in emissions.

\section*{Discussion}
Judged against AAQMs that never entered the model, the odour reports made by residents around one of the most closely observed landfills in the United Kingdom behaved as a sensor of their exposure to its odorous emissions. First, the reports were related to the landfill's H$_2$S: the wind at report times pointed back to the site, the monthly frequency and duration of reports were correlated with measured H$_2$S, and day by day the reports were associated with H$_2$S beyond weather. Second, after the 2025 source intervention the reports fell fourteen-fold and H$_2$S exceedances six-fold while modelled transport did not change detectably, and H$_2$S fell relative to methane with wind from the landfill and in other hours; the change could be dated from the reports, and a monitor in the sector concerned was needed to size it. Third, with no gas data as input, exposure estimated from the reports had the pattern of measured exposure in both monitored postcodes, at a level within a factor of three of that measured; the level is not a blind test, because its calibration was chosen by comparison with the monitors. Fourth, SENTINEL forecast daily report counts one to four days ahead from the public weather forecast and the reports already received (skill 0.642 and 0.500 with operational data delays), exceeding damped persistence four days ahead; the public weather forecast, run through the dispersion model, added information about the next day's reports that the report history did not (partial correlation 0.50~[0.42,~0.59]); the one-day gain over persistence was unresolved.

Calling the reports a sensor is a claim about seven properties, and the monitors let five of them be checked. Response: reports were correlated with H$_2$S month by month (frequency and duration, $r$ 0.92 and 0.90; 0.85 for frequency with the episode definition fixed in advance), week by week in both monitored sectors on monitors that did not change (Spearman 0.73~[0.59,~0.85] and 0.62~[0.39,~0.77]; at least 0.48~[0.28,~0.70] with any four-month period withheld), and day by day beyond weather, on stable and unstable nights alike (partial $r$ 0.40 and 0.37). Selectivity: reports were associated with H$_2$S and the methane transported with it, and not detectably with the other pollutants measured. Threshold: the probability of a report from the nearest sector reached half its upper limit at a peak 30-minute concentration of about 11.5~[5.7,~15.6]\,\textmu g\,m$^{-3}$, of the order of the WHO odour-annoyance guideline. Spatial response: reporting per head rose with modelled transport from the landfill across the ten sectors (Spearman 0.81) and showed no gradient with deprivation or age, although with ten sectors only strong gradients could be detected. Drift: reporting per unit of gas fell after the source intervention, within its ordinary variation in one monitored sector (0.80) and beyond it in the other (0.20), so the conversion from reports to exposure must be reset against a monitor from time to time. Two properties are missing. Absolute scale: the calibration of the exposure conversion was chosen by comparison with a monitor, not tested blind, and its level ranges within a factor of three (0.38--2.40 across sectors and conversions). Generality: one site. This is the position calibrated public reports hold in seismology, which give shaking intensity directly and ground motion only through conversions calibrated against instruments\cite{wald2012,atkinson2007,worden2012}.

Reports have been related to measured exposure before: reported odour rose with measured H$_2$S in diary studies of recruited residents\cite{wing2008,heaney2011}, resident odour records agreed with an instrumental odour index\cite{lee2013}, odour annoyance was related to modelled exposure\cite{blanesvidal2012}, and, closest to this study, community reports were compared with measured H$_2$S across a change at a source\cite{rico2025}, although the reports were not forecast or used to estimate exposure without the monitors. Odour impact has also been forecast: with public weather forecasts and dispersion modelling at a fixed emission rate, checked against incidents logged by a recruited panel of residents (41.2\,\% of 49 incidents forecast within two hours)\cite{Cartelle2016,Jimenez2016}; as community reports of respiratory irritation from harmful algal blooms, one day ahead\cite{ross2022beachlevel}; and up to three days ahead from on-site H$_2$S measurements\cite{Mulrow2020}; citizen odour reports have also been combined with dispersion modelling to check emission inventories\cite{Veratti2026}. Here the reports are both the sensor and the quantity forecast, the inputs are public, and the monitors were used only to evaluate the results. Earlier work at this site took the opposite approach, using on-site weather to nowcast the gas measured by the monitors and comparing the result with odour reports\cite{Pearce2026}.

The dispersion model adds little to the daily total, most of whose skill comes from the report history, but it improves how reports are divided among sectors and the model of the gas at the monitors: here the physical model contributes more to validity than to total skill\cite{Kirchner2006}. Driven by the same forecast, CALPUFF divided reports among sectors better than a lead-matched Gaussian plume (+0.125~[+0.079,~+0.170]) and improved the monitor model over a Gaussian plume (Fig.~\ref{fig:3}c). One possible explanation for our results is that residents are more likely to report odour when they are within the plume, so that reporting varies with the concentration of H$_2$S, an odorous tracer, and with the direction and timing of transport from the landfill.

The results suggest uses, each with a condition (Supplementary Table~\ref{tab:uses}). Regulators could use the reports they already receive as a relative indicator of the frequency and duration of odour impact; absolute concentrations need a monitor. The timing of a change at the source can be found from the reports, but measuring its size needs a monitor in the sector concerned. Environmental-health teams could use weekly or monthly relative exposure by postcode in exploratory studies of exposure and health; these estimates are more reliable for timing within a postcode than for comparisons between postcodes, and in individual weeks they can differ substantially from measured exposure. Site operators could schedule odour-generating work, such as waste handling and the capping works, away from days of high forecast impact. For alerts, the economic value of the forecast was similar to that of persistence at an equal number of alerts; air-quality alerts can improve health outcomes when they prompt action\cite{aldegunde2024pollutionalerts}, a benefit not shown here. Reports were associated with methane, an odourless co-emitted gas, only through H$_2$S (Fig.~\ref{fig:1}d), so they index the odorous fraction of the plume; whether they also indicate exposure to odourless constituents could be tested only by measuring plume composition where people live\cite{atsdr2001landfillgas,kim2005landfill}. H$_2$S concentrations here were far below toxicological thresholds\cite{govuk_nd_h2s,nrc2010acute}.

SENTINEL forecasts odour reports rather than gas concentrations. At this site, community odour reports acted as a relative sensor that complemented sparse monitoring: at least one monitor is needed to put exposure on an absolute scale, to validate the forecasts and to size the effect of an intervention in the sector where it is measured, since a monitor in another sector did not reliably recover a sector's measured change. Because SENTINEL needs only the public weather forecast and the report record that regulators already keep, it could extend a sparse network to the sectors without a monitor, although exposure there has not been verified, and inform where the next monitor goes.

The study has several limitations. The method has yet to be tested at another site; within this record, the correlations, skill and gains in the summary paragraph were re-estimated with each four-month period withheld (the monthly frequency relation stayed at or above 0.91), and the forecast was re-trained with each period withheld (one-day skill on the withheld period from 0.35 in May--August 2024 to 0.83 after the source intervention) (Supplementary Table~\ref{tab:periods}). All interval methods fell short of their nominal coverage on these data (Methods). Reporting per unit of gas fell after the source intervention, within its ordinary variation in sector~C and beyond it in sector~G, so the number of reports per unit of gas changes over time and the conversion must be reset against a monitor. The weather forecast represents calm, stable nights poorly (Supplementary Note~\ref{snote:chain}), and on nights with H$_2$S episodes the gain over persistence was not distinguishable from zero. The dispersion model assumes a constant emission rate, so changes in emissions reach the forecast only through the reports. Our data come from the council's reporting channel only. The Environment Agency received 2.8–4.9 times as many contacts, including repeat contacts from the same address, although the weekly counts of the two channels rank weeks almost identically (Supplementary Note~\ref{snote:records}). The exposure conversion's absolute level depends on the guideline used to set it; with the midpoint set instead at a laboratory annoyance level, the estimates are several times too low (Supplementary Note~\ref{snote:exposure}).

Two further tests would strengthen these findings: applying the method at another site, and to reports received after this record closes; with almost no reports after April 2025, the checks on withheld days within the record cannot stand in for a later period. Linking postcode-level exposure estimates to syndromic surveillance or primary-care records would then allow their association with health outcomes to be studied, in place of the distance from the site that epidemiological studies of waste facilities usually use\cite{elliott2001}. We release the code for SENTINEL (Code availability) so that the method can be tested at other sites. Community odour reports can act as a relative sensor of exposure to fugitive emissions, complementing sparse monitoring, and their daily counts can be forecast from the public weather forecast and the reports already received; a monitor is needed for absolute scale.

\section*{Methods}
\subsection*{Study area, records and data handling}\label{sec:m1}
The study area is the ten postcode sectors around the landfill, labelled A--J, with 82,288 residents (2021 Census); sectors~C and~G contain the ambient air quality monitors (AAQMs) and adjoin the site (Supplementary Note~\ref{snote:records}). The odour reports are those that residents made to the local council, which records for each report the date and time at which the odour was noticed, the postcode and an intensity rating from 0 to 6. Residents can also report to the Environment Agency, which logs reports separately and received 2.8–4.9 times as many contacts in each year, including repeat contacts from the same address; those reports are not part of our data, and we use the Environment Agency's published weekly counts only as a check (Supplementary Note~\ref{snote:records}). We removed reports delivered more than once and collapsed repeats with the same minute, postcode sector and rating, leaving 9,893 reports from all postcodes and 9,078 from the ten sectors. The time of each report is the time of the odour; the time at which it was filed is not recorded. Closing the report history 24\,h before the forecast is issued, which allows for filing delays of up to a day, lowered one-day skill under block cross-validation from 0.710 to 0.666. Counts for each period and sector are in Supplementary Table~\ref{tab:s1}, and checks of the record against published council and regulator figures are in Supplementary Note~\ref{snote:records}. The earlier record (2021--23) was collected under an older system and is dated by filing time, with a median delay of about 15\,h after the odour. We use it only for weekly and monthly comparisons, and never as a replication of the daily results (Supplementary Note~\ref{snote:legacy}). \textbf{Ethics and information governance.} \textit{[Author input needed: the ethics approval or exemption that covered the use of the council's report records (approving body and reference), and the information-governance basis under which the records were received.]} Results are reported only as counts aggregated by postcode sector, and no individual report is published.

The AAQMs (MMF1, MMF2, MMF6, MMF9 and MMF1A; MMF6 has no data in the analysis window) belong to the regulator's network and record H$_2$S in \textmu g\,m$^{-3}$ and methane in mg\,m$^{-3}$; we averaged their 15-minute values to hourly and six-hour means. We applied the regulator's recalibration of September 2023 as published\cite{EA2024h2scal}; the instruments are described in earlier work at this site\cite{Pearce2026} and in Supplementary Note~\ref{snote:monitors}. In sector~G, the monitors used as the reference changed in September 2024, and measured exposure there differs by a factor of about two before and after the change. The monitors serve only to evaluate results: they are not used as inputs to SENTINEL or to choose its thresholds. The calibration of the exposure estimate (the WHO guideline at the midpoint and the published spread) was chosen by comparison with them, so its level is not independent of them. Weather inputs come from the Met Office UKV forecast\cite{LITVosper2013High}, used at lead times of one to four days; the headline forecasts allow for operational data delays: they use the previous forecast cycle, with the report history closed 24\,h before issue; the ERA5 global reanalysis and the on-site wind measurements were used only for checks (Supplementary Note~\ref{snote:chain}). Data handling is described in Supplementary Methods~\ref{smeth:records}.
\subsection*{The dispersion field}\label{sec:m2}
CALMET was run with forecast UKV fields as its only meteorological input, without observations, and CALPUFF released a constant unit emission ($Q=1$\,g\,s$^{-1}$) from the landfill\cite{Scire2000calmet,Scire2000calpuff}. For postcode sector~$s$, six-hour window~$t$ and lead time~$L$, the dispersion field $\Phi$ is the mean concentration per unit emission over the model receptors in the sector (equation~\eqref{eq:phi}), where $A_s$ is the set of receptors in sector~$s$ and $\bar C_r$ the six-hour mean concentration at receptor~$r$.
\begin{equation}
\label{eq:phi}
\Phi_{s}(t;L)=\frac{1}{|A_s|}\sum_{r\in A_s}\bar C_r(t;L)\Big|_{Q=1}
\end{equation}
Each window uses the dispersion field from the shortest available forecast lead time at or beyond its time since issue (Supplementary Methods~\ref{smeth:chain}). Models use $\log_{10}\Phi$ with a small floor. $\Phi$ describes transport only: it is neither an emission rate nor a measured concentration. Any factor that multiplies it in a model, such as the source strength, a common error in the modelled transport or residents' readiness to report, is absorbed by the latent state and cannot be separated. We used the sector mean as the main summary and the upper quartile and maximum over receptors as checks (Supplementary Note~\ref{snote:chain}; Supplementary Fig.~\ref{fig:s3}). The configuration of both models is given in Supplementary Methods~\ref{smeth:chain} and Supplementary Table~\ref{tab:chain}.
\subsection*{The SENTINEL forecasting model}\label{sec:m3}
SENTINEL runs the public weather forecast through the CALMET and CALPUFF dispersion models and uses the resulting dispersion field, together with the history of odour reports, in a forecasting model. Forecasts are issued at midnight; one day ahead refers to the day that begins at that midnight, and $k$ days ahead to the $k$th such day. The model has five parts; its design, the alternatives tested and the search that selected it are described in Supplementary Methods~\ref{smeth:sentinel}.
\textbf{State-space model.} For the total over the ten sectors in six-hour window~$t$, $y_t=\log(1+N_t)$ is modelled as a local level $\mu_t$ plus a first-order autoregressive term $a_t$ with coefficient $\psi$ and regressors $\mathbf x_t$ with coefficients $\boldsymbol\beta$ (equation~\eqref{eq:ssm}), with independent disturbances $\eta_t\sim N(0,\sigma^2_\eta)$ and $\zeta_t\sim N(0,\sigma^2_\zeta)$\cite{harvey1989}.
\begin{equation}
y_t=\mu_t+a_t+\mathbf x_t^{\top}\boldsymbol\beta,\qquad \mu_t=\mu_{t-1}+\eta_t,\qquad a_t=\psi\,a_{t-1}+\zeta_t \label{eq:ssm}
\end{equation}
We left out a separate observation-noise term because, with it, scaling all variances by the same factor gives the same forecasts, so the variances could not be estimated from forecast errors. The regressors $\mathbf x_t$ are indicators for the window of the day, the forecast wind speed and stability at lead time~$L$, and a standardised day-level index $\hat A_t$. This index is a ridge regression, fitted on the training days only, of $y_t$ on weekday, window of day, an annual harmonic pair, $\log_{10}\Phi_s(t;L)$ for the ten sectors, wind speed and stability.

\textbf{Forecast issue time.} The state is estimated from reports up to the midnight of issue. A window $k$ steps later is forecast by carrying this state forward $k$ steps, and the day's total is formed from its four windows (equation~\eqref{eq:fc}), where $\hat{\boldsymbol\xi}_{t-k\mid t-k}$ is the estimate at issue of the state vector $\boldsymbol\xi_t=(\mu_t,a_t)^{\top}$, $\mathbf G=\operatorname{diag}(1,\psi)$ is its transition matrix, $\mathbf z=(1,1)^{\top}$, and $\hat y_{d,b}$ is the forecast for window~$b$ of day~$d$.
\begin{equation}
\hat y_t=\mathbf z^{\top}\mathbf G^{k}\,\hat{\boldsymbol\xi}_{t-k\mid t-k}+\mathbf x_t^{\top}\hat{\boldsymbol\beta},\qquad
\hat D_d=\log\Big(1+\sum_{b=1}^{4}\max\big(0,e^{\hat y_{d,b}}-1\big)\Big) \label{eq:fc}
\end{equation}

\textbf{Model training.} We fitted the regression coefficients and $\sigma^2_\zeta$ by maximum likelihood, and then chose the remaining dynamics, $\theta=(\log\sigma^2_\eta/\sigma^2_\zeta,\ \operatorname{atanh}\psi)$, on the training days by minimising equation~\eqref{eq:loss} with $\lambda=1.5$.
\begin{equation}
\mathcal L(\theta)=\overline{(D_d-\hat D_d)^2}+\lambda\,\overline{\big[(\hat D_d-\hat D_{d-1})-(D_d-D_{d-1})\big]^2} \label{eq:loss}
\end{equation}
Here $D_d$ is the recorded $\log(1+\text{daily total})$ and $\hat D_d$ the one-day-ahead forecast from the filter run on the training data, and the bars denote means over the training days\cite{deVos2005,deVos2008}. A squared error on the day-to-day change alone has the same minimum as a squared error on the daily level, because the two differ only by a term known at issue. The second term in equation~\eqref{eq:loss} instead compares the forecast's own day-to-day step with the recorded step, and so penalises forecasts whose steps repeat the previous day's change. This moves the peak of the forecast's correlation with recorded reports onto the target day.

\textbf{Error correction.} We then shifted the forecast by an exponentially weighted mean of past daily log errors, with a half-life of $\tau=15$ days, using only days already recorded at issue (equation~\eqref{eq:offset}), where $o_d$ is the correction for target day~$d$, $L$ is the lead time in days and $\hat N_d$ is the forecast number of reports.
\begin{equation}
o_d=\operatorname{EWMA}_{\tau}\big\{D_{d'}-\hat D_{d'}:\ d'\le d-L\big\},\qquad \hat N_d=\max\big(0,\,e^{\hat D_d+o_d}-1\big) \label{eq:offset}
\end{equation}
We applied no volume calibration.

\textbf{Sector forecasting.} Each window's total is divided among the sectors by a multinomial logit (equation~\eqref{eq:map}).
\begin{equation}
p_{s,t}=\frac{\exp\big(b_s+\gamma\sum_j W_{sj}(\nu)\,\tilde\phi_{j,t}+o_{s,d}\big)}{\sum_{s'}\exp\big(b_{s'}+\gamma\sum_j W_{s'j}(\nu)\,\tilde\phi_{j,t}+o_{s',d}\big)},\qquad W_{sj}\propto e^{\nu\cos(\vartheta_s-\vartheta_j)} \label{eq:map}
\end{equation}
Here $p_{s,t}$ is the share of window~$t$'s reports forecast for sector~$s$, $b_s$ is a sector intercept, $\tilde\phi_{j,t}$ is $\log\Phi_j$ centred across sectors and $\gamma$ its coefficient, $\vartheta_s$ is the bearing of sector~$s$ from the landfill, the kernel $W$ with concentration~$\nu$ spreads each sector's field to neighbouring bearings, and $o_{s,d}$ is a daily residual log-share estimated from days recorded before issue. The model is fitted by multinomial likelihood on the training windows, and the forecast for sector~$s$ is $\hat N_{s,t}=p_{s,t}\hat N_t$.

We chose SENTINEL from 226 configurations, combining base models, offset half-lives and volume calibrations (not every combination was run), using a fixed rule: among configurations whose forecasts were more correlated with the target day than with the previous day, the one with the highest tier agreement on the first half of the days, confirmed on the second half, on all days and with training on earlier days only (Supplementary Methods~\ref{smeth:sentinel}; Supplementary Table~\ref{tab:search}; Supplementary Note~\ref{snote:selection}); selecting on skill alone tends to select lagged forecasts\cite{Akyol2023Avoiding}. Re-applied each week from earlier days only, the rule gave the same information beyond recorded days (0.50~[0.41,~0.59]), and choosing the configuration on the scored week would have raised daily skill by +0.090~[+0.048,~+0.137] (Extended Data Fig.~\ref{fig:ed5}h). The comparison models are persistence (the last recorded day or, for the same-day estimate, the last six hours with their daily cycle); damped persistence (the weekday $\times$ month average plus the last recorded day's departure from it, shrunk by the lag autocorrelation estimated on recorded days); the climatology--persistence blend (least-squares weights on the last recorded day and the weekday $\times$ month average, refitted each day on recorded days); the same model using the report history alone, without weather or dispersion inputs; the same model trained by likelihood alone; and the same model given the observed weather, which serves as a reference (Supplementary Methods~\ref{smeth:comparators}; Supplementary Note~\ref{snote:comparators}). The same-day estimate in Fig.~\ref{fig:4} is the six-hour model run with the observed weather. Its sector map blends the dispersion model's shares with the last six hours of reports, with the blend weight trained on earlier days (Supplementary Note~\ref{snote:nowcast}). Because three of a day's four six-hour windows are already recorded, the same-day estimate's peak on the day is a consequence of its inputs and is not evidence for the method.
\subsection*{Psychometric reporting model and exposure estimate}\label{sec:m4}
\textbf{Psychometric reporting model.} The psychometric reporting model (Fig.~\ref{fig:2}b), a probit dose--response of the kind used in psychophysics\cite{wichmann2001} and in studies of odour annoyance\cite{blanesvidal2012,nicell2009}, gives the probability of at least one report from sector~C in a six-hour window as a function of the six-hour H$_2$S concentration $c$ at MMF9 (equation~\eqref{eq:curve}).
\begin{equation}
P(\text{report}\mid c)=q\,F_{\mathcal N}\!\Big(\frac{\log_{10}c-\log_{10}\alpha}{\sigma}\Big) \label{eq:curve}
\end{equation}
$F_{\mathcal N}$ is the standard normal distribution function, $\alpha$ is the community's threshold (the concentration at which the probability reaches half its upper limit~$q$) and $\sigma$ is the spread on a $\log_{10}$ scale. We fitted the model by maximum likelihood without a background reporting rate and used a complementary log--log link as a sensitivity check. On the guideline's 30-minute basis the threshold uses the six-hour peak of 30-minute means and is averaged over the probit, logit and complementary log--log links with Akaike weights (Supplementary Note~\ref{snote:fidol}).

\textbf{Source estimation from reports.} In each six-hour window~$w$, the indicator $R_{s,w}$ that sector~$s$ makes at least one report takes the value 1 with the probability in equation~\eqref{eq:state}, where $C_{s,w}$ is the dispersion field for a nominal emission, $g_s$ a background rate, $q_s$ an upper limit, $\mu$ and $\sigma$ the midpoint and spread on the $\log_{10}$ scale, and $x_w$ a latent source state that follows a first-order autoregression with coefficient $\rho$ and Gaussian innovations $\omega_w$.
\begin{equation}
\Pr(R_{s,w}=1)=g_s+(q_s-g_s)\,F_{\mathcal N}\!\Big(\frac{x_w+\log_{10}C_{s,w}-\mu}{\sigma}\Big),\qquad x_w=\rho\,x_{w-1}+\omega_w \label{eq:state}
\end{equation}
We fitted this model on the training windows with an exact grid filter, first with a constant source and then with the dynamics. It uses no gas measurements.

\textbf{Exposure estimation.} We placed the WHO odour annoyance guideline, 7\,\textmu g\,m$^{-3}$ over 30 minutes\cite{who2000airquality}, at the midpoint of the psychometric reporting model and converted each window's standardised position $z_{s,w}$ back to a concentration using the published spread $\sigma_{\rm pub}=0.55$ (equation~\eqref{eq:exposure}), where hats denote fitted values and $\Phi^{\rm mean}_{s,w}$ and $\Phi^{\rm max}_{s,w}$ are the mean and the hourly peak of the sector field in the window.
\begin{equation}
c_{s,w}=7\cdot10^{\,\sigma_{\rm pub}z_{s,w}}\,\frac{\Phi^{\rm mean}_{s,w}}{\Phi^{\rm max}_{s,w}},\qquad z_{s,w}=\frac{\hat x_w+\log_{10}C_{s,w}-\hat\mu}{\hat\sigma} \label{eq:exposure}
\end{equation}
The final factor converts the hourly peak of the sector field to its mean. Monthly exposure is $\sum_w c_{s,w}\times 6$\,h (\textmu g\,m$^{-3}$\,h), for months in which at least 60\,\% of windows are valid. The guideline is an annoyance value, based on odour-detection data and the concentrations at which annoyance is expected, rather than a detection or psychophysical threshold\cite{nicell2009}. The guideline sets the average level, and the variation over time comes from the reports. Because the calibration (the guideline at the midpoint and the published spread) was chosen by comparison with these monitors, the level is not an independent test; across averaging-time conversions it ranged from 0.38 to 2.40 times the measured level. Results with other calibrations (published detection thresholds with an annoyance multiplier) and other spreads are in Supplementary Note~\ref{snote:exposure}; the fitting is described in Supplementary Methods~\ref{smeth:psych}.
\subsection*{Monitor model and monitor-based source estimate}\label{sec:m5}
\textbf{Monitor model.} The monitor model (Fig.~\ref{fig:3}c) tests the dispersion field where gas is measured. For monitor~$j$, log six-hour H$_2$S, $y_j(t)$, is modelled as a shared source state~$u(t)$, a local state~$l_j(t)$, an intercept~$a_j$, the dispersion field at the monitor, $\Phi_j(t)$, with coefficient~$b_j$, weather terms $\mathbf X_t$ with coefficients $\boldsymbol\beta_j$, and an error $\varepsilon_j(t)$ (equation~\eqref{eq:monitor}).
\begin{equation}
\label{eq:monitor}
y_j(t)=u(t)+l_j(t)+a_j+b_j\log_{10}\Phi_j(t)+\mathbf X_t\boldsymbol\beta_j+\varepsilon_j(t)
\end{equation}
The model is fitted on each monitor's own past H$_2$S, so it tests the contribution of the dispersion field near the source. It does not test transport to the sectors where most reports are made, and it is not designed to improve nowcasts of the gas. \textbf{Monitor-based source estimate.} The source estimate from the monitors (Fig.~\ref{fig:1}f) assumes that H$_2$S and methane at each monitor and period share one transport term; a two-gas random walk across periods then gives a relative source strength for each gas. Because it is built from measured gas, it includes the transport realised at the monitors (Supplementary Methods~\ref{smeth:monitor}; Supplementary Note~\ref{snote:monitormodel}).
\subsection*{Verification measures}\label{sec:m6}
All measures use $\log(1+\text{daily reports})$ summed over the ten sectors unless stated otherwise. \emph{Daily skill} is the mean-squared-error skill score (equation~\eqref{eq:msss})\cite{murphy1988}; its reference $C_d$ is the average for the same weekday and month, computed from reports up to the previous day after a 120-day burn-in.
\begin{equation}
\mathrm{MSSS}=1-\frac{\sum_d(\hat D_d-D_d)^2}{\sum_d(C_d-D_d)^2} \label{eq:msss}
\end{equation}
We use it as the main measure because, unlike a correlation, it penalises a forecast at the wrong level. \emph{Tier agreement} is the quadratic-weighted~$\kappa$\cite{Cohen1968} on daily tiers with boundaries at 3, 9 and 27 reports. The forecast probabilities for each tier are scored by the ranked probability skill score\cite{epstein1969scoring,gneiting2007} against tier frequencies up to the previous day. Days with at least 27 reports are scored by the area under the precision--recall curve relative to its base rate, and \emph{rank correlation} is Spearman's~$\rho$ with recorded reports. The forecast division of reports among sectors is scored by the daily cosine agreement between forecast and recorded sector shares, and by redistribution skill relative to the sector shares of the training period.

\textbf{Information beyond recorded days.} For each forecast we regressed the target day's reports on the report history recorded at issue (the 14 most recent recorded days, the mean of earlier days, weekday and season) and report the partial correlation of the forecast with the target day after that regression; in simulations, a forecast that only repeats recorded days gives zero and one that knows the target day a high value. This statistic replaced the pre-specified target-day timing test after the results of forecasts trained on earlier days had been seen. Its controls are the same model with the report history alone, a regression on report history and calendar, and a regression on report history, calendar and the same weather and dispersion inputs, each refitted on SENTINEL's schedule of training and test periods; for SENTINEL's partial correlation beyond a control, we added that control's forecast to the regression.

\textbf{Target-day timing.} Target-day timing (equation~\eqref{eq:timing}), a diagnostic of the model family (Extended Data Fig.~\ref{fig:ed5}), is the forecast's correlation with reports on the target day minus its correlation with reports on the previous day, where $r(\ell)$ is the correlation of the forecast for day~$d$ with reports on day~$d+\ell$ and $L$ is the lead time in days, following the same-day criterion used in earlier work at this site\cite{Pearce2026} and tests for lagged forecasts in hydrology and space weather\cite{deVos2005,Laperre2020}.
\begin{equation}
T=r(0)-r(-1),\qquad r(\ell)=\operatorname{corr}\big(\hat D_d,\,D_{d+\ell}\big),\qquad T_{\max}=r(0)-\max_{\ell\le -L}r(\ell) \label{eq:timing}
\end{equation}
We report $T$ and $T_{\max}$ as diagnostics, not as evidence of skill. Beyond one day ahead, several days before the target day are already recorded at issue and could be reproduced by the forecast, so $T_{\max}$ compares the target day with the best of those days. Alerts are compared at an exactly equal number of alerts, and economic value is the relative value over all cost--loss ratios\cite{zhu2002}; definitions and results are in Supplementary Note~\ref{snote:alerts}, and the formulas for all measures are in Supplementary Methods~\ref{smeth:verification}.
\subsection*{Validation design, intervals and tests}\label{sec:m7}
Our main results come from models trained only on earlier days. Training periods are in time order, as in operational use, and end before the first test day (three days before it for forecasts four days ahead), so every training day was already recorded at issue. The unit of analysis is the day (486~days scored one day ahead and 478~days four days ahead), except where weeks, months or six-hour windows are stated. We also report block cross-validation, in which each block of days is forecast by a model trained on all other blocks. Headline skill allows for operational data delays: weather from the previous forecast cycle and reports closed 24\,h before issue, trained on earlier days, with block cross-validation in parentheses. The data contain no separate period kept unseen during development. We tested robustness to any single period by re-estimating the correlations, skill and comparator gains of the summary paragraph with each four-month period withheld, and by training the forecast with each period withheld (Supplementary Table~\ref{tab:periods}).

We set the method for intervals before computing any result. Comparisons between forecasts are paired on the same days\cite{hamill1999,diebold1995}. For each family of paired comparisons we chose the interval method and block length with the highest measured coverage, assessed in two ways (in simulations with a known added effect, and with a check that uses no generating model), from moving-block percentile, basic and bias-corrected intervals with blocks of 7 to 112 days, and from subsampling\cite{kunsch1989} (Supplementary Methods~\ref{smeth:intervals}; Supplementary Note~\ref{snote:intervals}). Every method fell short of nominal 95\,\% coverage on these data (measured coverage 0.63--0.91), so all intervals cover the true value less often than their nominal level implies, and we describe a result as a gain only when its interval excludes zero. The one-day gain in daily skill over persistence depends on the interval method: its interval excludes zero with 7- or 28-day block bootstraps but includes zero with the subsampling chosen by this method. Two results were computed with short blocks and were not recalculated with this method: the difference in gain over persistence between four days and one day ahead (+0.174, $p$~0.001) and the one-day margin adjusted for selection among 226 configurations (calibrated $p$~0.024). The gain over the report history alone is given with the interval from this method in the text (+0.131~[+0.057,~+0.186] one day ahead) and with its 7-day block interval in Fig.~\ref{fig:5}e. The family-wise test of the paired gains three and four days ahead (a Romano--Wolf maximum test with one shared set of 84-day block draws) had its coverage measured in simulated records with a known effect: its joint bounds held in 0.64–0.70 of records at nominal 95\,\% and in about~0.89 at nominal 99.9\,\%, so the three-to-four-day result is based on the gains that exceed zero at the 99.9\,\% level (all~10 for forecasts trained on earlier days), not on the count of $p$ values. The month-bootstrap intervals of the intervention decomposition and the one-monitor correction cover 81--86\,\% of true values in simulation, so they are too narrow.

Analyses were specified in advance in a dated analysis plan, with three departures: the partial correlation beyond the report history replaced the pre-specified target-day timing test as the main timing measure (target-day timing is in Extended Data Fig.~\ref{fig:ed5}); in one 2023 period the criterion set in advance for abandoning the threshold estimate was met, and that estimate is reported only as a sensitivity analysis; and nine further tests were specified after the first results had been reviewed. Supplementary Note~\ref{snote:intervals} lists each pre-specified prediction, whether it held and how many comparisons passed. We applied no correction for multiple testing except where a Holm adjustment over a stated family is given, and we mark results whose lower bound lies within 0.01 of zero. We write ``no credible evidence'' only where the result is accompanied by an equivalence test against a smallest meaningful effect declared in advance. We define a detectable effect as the effect detected with 80\,\% power when a known effect is added in simulation, rather than as the half-width of an interval. For the monthly test of local instruments, the detectable effect is 0.036 (up to 0.048), larger than the 0.029 half-width, and its intervals cover 86\,\% of true values (Supplementary Note~\ref{snote:instruments}). Drift in residents' readiness to report over several years remains untested.
\subsection*{Software}\label{sec:m8}
Analyses used Python~3.12 with NumPy, SciPy, pandas, statsmodels (state-space models), scikit-learn and Matplotlib; CALMET and CALPUFF (version 5.8.5) were compiled from the distributed source without modification. Versions are listed in Supplementary Methods~\ref{smeth:software}.

\section*{Data availability}
The odour reports are personal data, logged by the local council under its statutory process, and the individual records are withheld. The aggregated counts of reports per postcode sector and six-hour period used in the analyses are available from the corresponding author on reasonable request, subject to the information-governance conditions under which the records were obtained. Hydrogen sulphide and methane concentrations at the four monitors were provided by the Environment Agency and the local council. The landfill is not named, and the monitors are identified by station code only.

Met Office UKV forecasts at 1.5\,km resolution are available to registered users from the CEDA Archive (\url{https://catalogue.ceda.ac.uk}) under the Met Office's terms. OS Terrain~50 and Code-Point Open are available from Ordnance Survey OpenData and the 2021 Census population from the Office for National Statistics, under the Open Government Licence v3.0. The UKCEH Land Cover Map 2023 is available from the Environmental Information Data Centre (\url{https://catalogue.ceh.ac.uk/id/7727ce7d-531e-4d77-b756-5cc59ff016bd}) to registered users who accept the UKCEH licence, which does not permit redistribution; the land-use files derived from it are therefore not released. Supplementary Table~\ref{tab:inputs} lists every input and its source. The UK Health Security Agency's health risk assessments for the site include monthly tables for 2021--2023.

\section*{Code availability}
The code for SENTINEL and for every analysis in this paper is available at \url{https://github.com/tcpearce/SENTINEL}. It contains the state-space model, the code for the forecast and the same-day estimate, sector forecasting and exposure estimation, the interval and timing tests, and the scripts that build each figure and generate every number printed in the main text and the Supplementary Information. Results that use the odour reports can be reproduced from this code together with the aggregated counts described under Data availability. CALMET and CALPUFF (version 5.8.5) are distributed by Exponent (\url{https://www.calpuff.org}); we compiled them from the distributed source without modification, and neither the programs nor their input and output files are redistributed. A versioned snapshot with a persistent DOI will be deposited at Zenodo on acceptance.

\section*{Acknowledgements}
This study is part-funded by the National Institute for Health and Care Research (NIHR) Health Protection Research Unit in Chemical Threats and Hazards. T.C.P., D.J.T.S., A.D. and A.F. are supported by grant award NIHR207293. The views expressed are those of the author(s) and not necessarily those of the NIHR or the Department of Health and Social Care.

\section*{Author contributions}
T.C.P., A.F., D.J.T.S. and A.D. conceived the study. T.C.P. and A.F. developed the methods, carried out the investigation and made the figures. T.C.P. and A.F. wrote the original draft, and all authors reviewed and edited the manuscript.

\section*{Competing interests}
The authors declare no competing interests.


\clearpage
\section*{Extended Data}
\renewcommand{\figurename}{Extended Data Fig.}\setcounter{figure}{0}
\begin{figure}[p]\centering
\includegraphics[width=\textwidth,height=0.62\textheight,keepaspectratio]{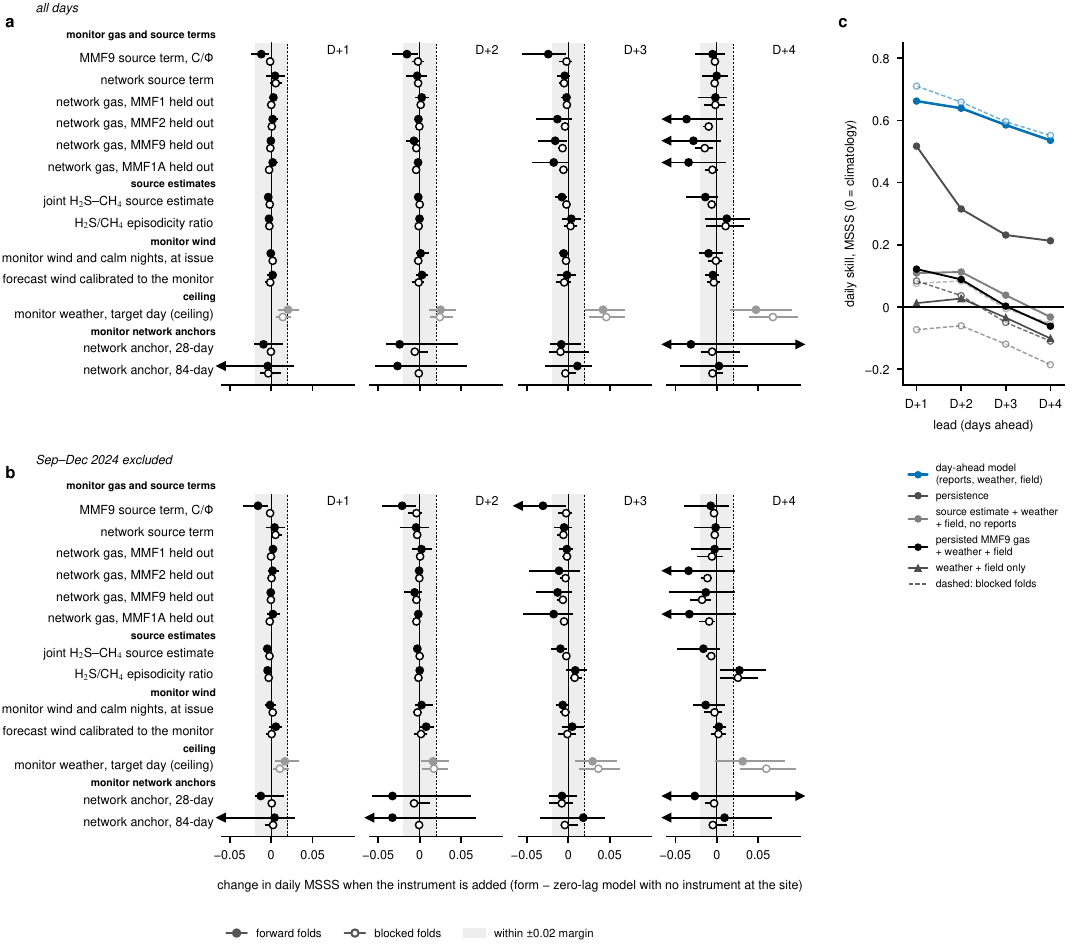}
\caption[Comparison of forecast performance with and without local measurements.]{\textbf{Comparison of forecast performance with and without local measurements.} \input{legend_ED1}}\label{fig:ed1}
\end{figure}
\begin{figure}[p]\centering
\includegraphics[width=\textwidth,height=0.62\textheight,keepaspectratio]{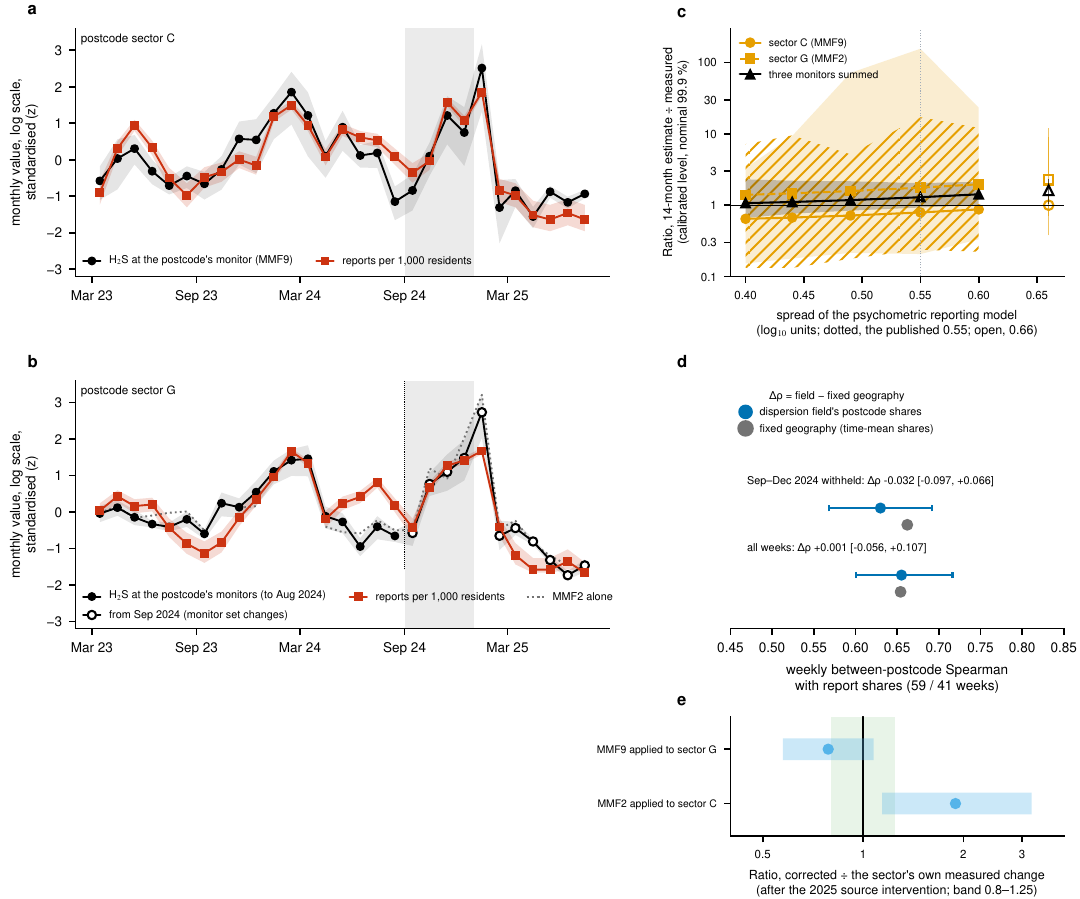}
\caption[Relationships between odour reports, monitored exposure and postcode-level variation.]{\textbf{Relationships between odour reports, monitored exposure and postcode-level variation.}  \input{legend_ED2}}\label{fig:ed2}
\end{figure}
\begin{figure}[p]\centering
\includegraphics[width=\textwidth,height=0.62\textheight,keepaspectratio]{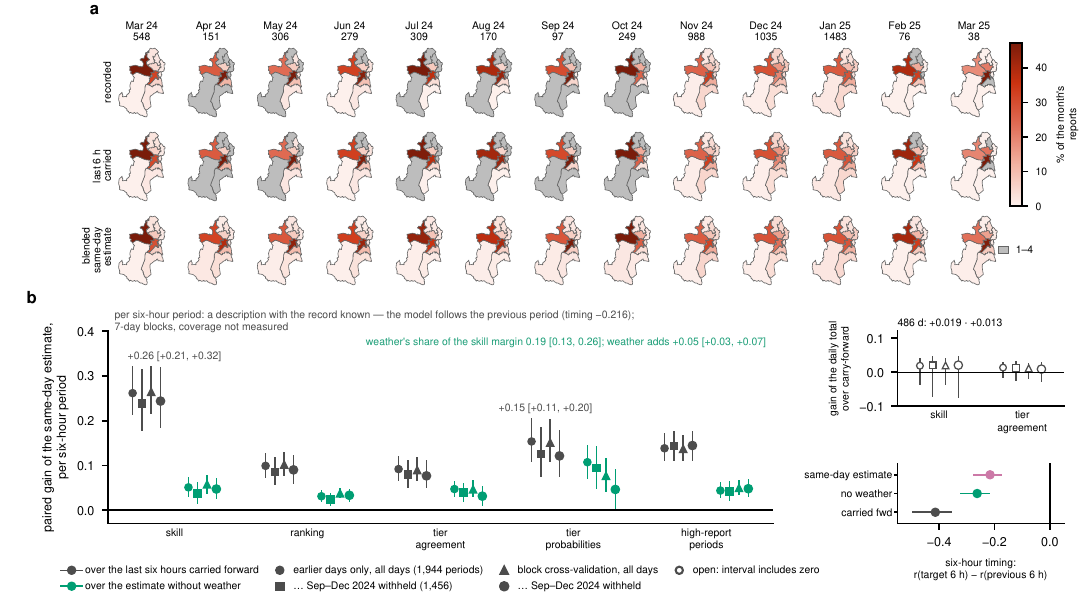}
\caption[Same-day estimates across temporal resolutions and reporting histories.]{\textbf{Same-day estimates across temporal resolutions and reporting histories.} \input{legend_ED3}}\label{fig:ed3}
\end{figure}
\begin{figure}[p]\centering
\includegraphics[width=\textwidth,height=0.62\textheight,keepaspectratio]{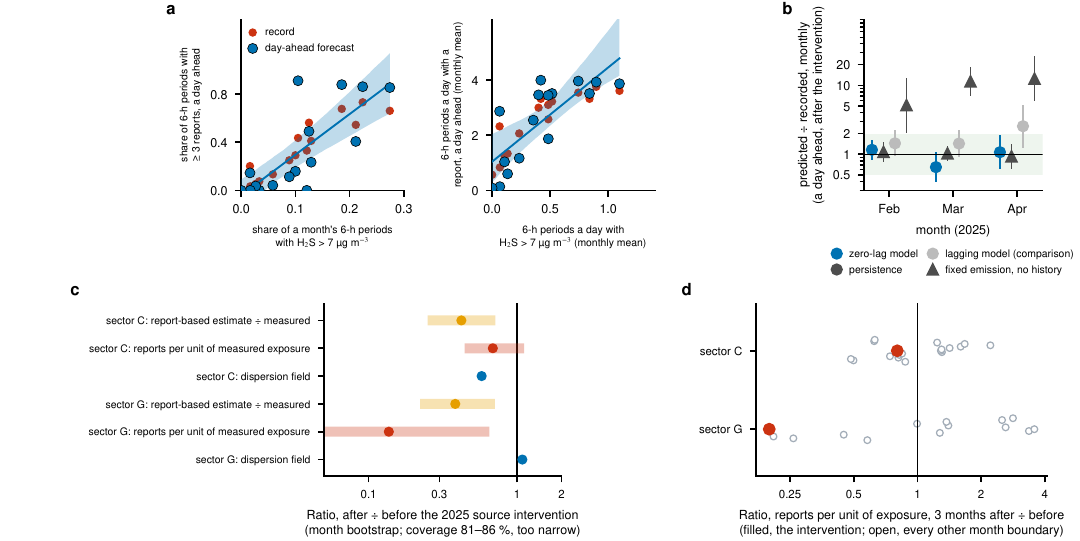}
\caption[Monthly patterns in odour reports before and after the 2025 source intervention.]{\textbf{Monthly patterns in odour reports before and after the 2025 source intervention.}  \input{legend_ED4}}\label{fig:ed4}
\end{figure}
\begin{figure}[p]\centering
\includegraphics[width=\textwidth,height=0.62\textheight,keepaspectratio]{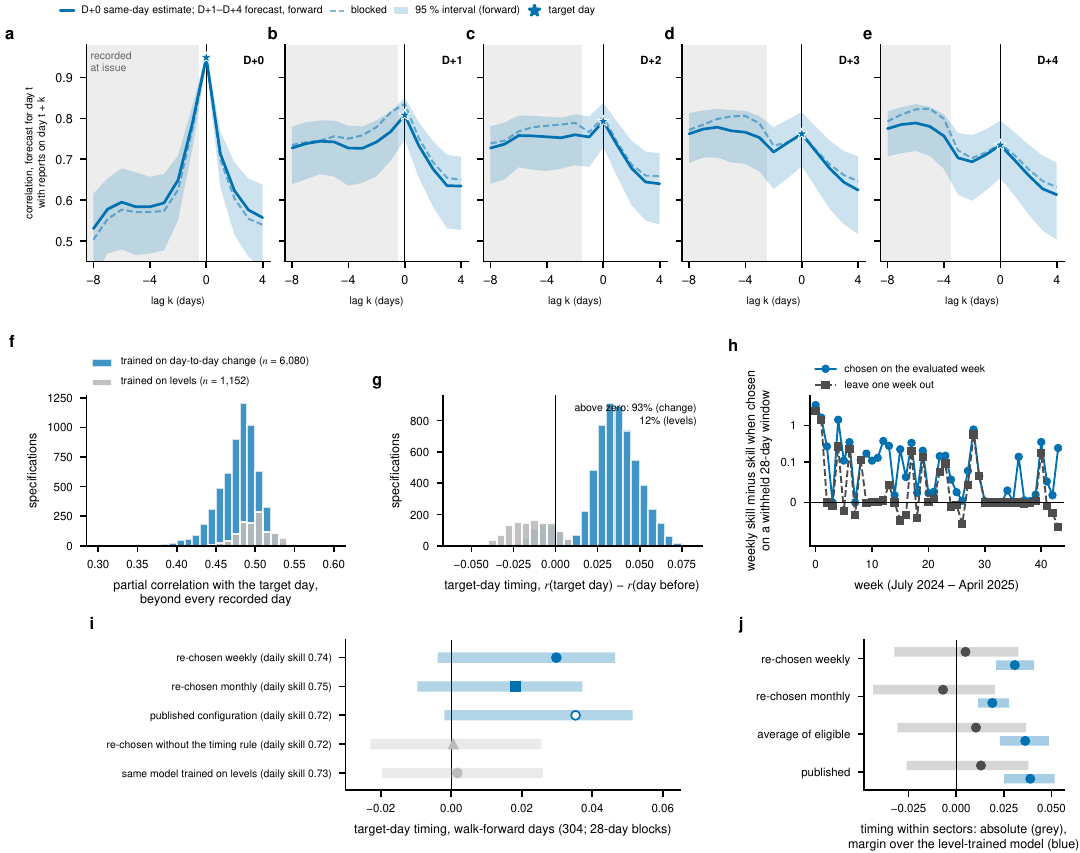}
\caption[Checks on withheld days and target-day timing of the forecast.]{\textbf{Checks on withheld days and target-day timing of the forecast.}  \input{legend_ED5}}\label{fig:ed5}
\end{figure}
\begin{figure}[p]\centering
\includegraphics[width=\textwidth,height=0.62\textheight,keepaspectratio]{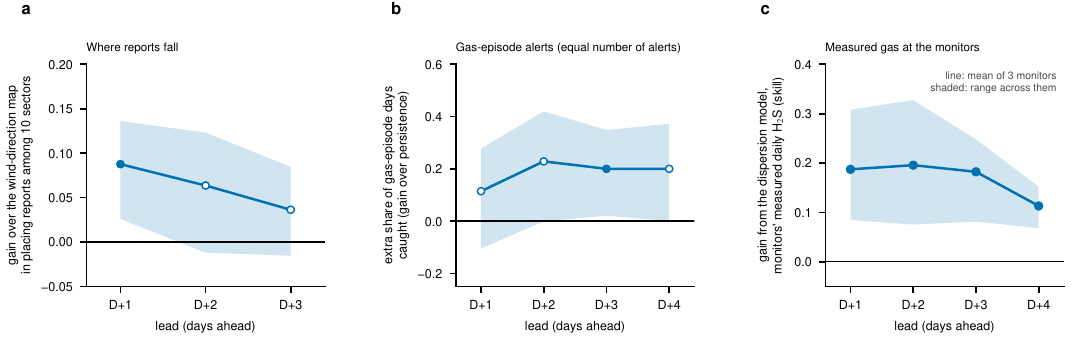}
\caption[Spatial allocation of odour reports and detection of gas episodes.]{\textbf{Spatial allocation of odour reports and detection of gas episodes.} \input{legend_ED6}}\label{fig:ed6}
\end{figure}
\clearpage
\thispagestyle{empty}\vspace*{0.24\textheight}
\begin{center}\setlength{\fboxsep}{16pt}\setlength{\fboxrule}{1pt}
\fcolorbox{inkteal}{inkdeep!4}{\begin{minipage}{0.86\textwidth}\centering
{\color{inkteal}\scshape\large Supplementary Information}\\[0.9em]
{\color{inkdeep}\Large\bfseries Community Odour Reports as a Forecastable\\ Sensor of Fugitive Emissions}\\[1.4em]
{\color{inkteal}\rule{0.34\textwidth}{0.8pt}}\\[1.2em]
\begin{minipage}{0.9\textwidth}\centering\small Supplementary Methods 1--9, Supplementary Notes 1--17, Supplementary Figures S1--S5, Supplementary Tables S1--S6, Supplementary Movie 1 and Supplementary Data 1 (all configurations of the model search) and 2 (the alternative designs and their tests).\\[0.7em] References are listed once, at the end of the main article.\end{minipage}
\end{minipage}}\end{center}
\clearpage
\renewcommand{\figurename}{Fig.}\renewcommand{\thefigure}{S\arabic{figure}}\renewcommand{\thetable}{S\arabic{table}}
\setcounter{figure}{0}\setcounter{table}{0}\setcounter{section}{0}\renewcommand{\thesection}{\arabic{section}}
\renewcommand{\rhtext}{Supplementary Information\,\textbullet\, Community Odour Reports as a Forecastable Sensor}
\captionsetup{font=footnotesize}
\noindent This Supplementary Information complements the main text. The Supplementary Methods extend the Methods, section by section, with the data handling, model configuration and model search behind them. Each Supplementary Note begins with the main-text result it supports and then gives the methods, checks and limits behind that result. The evidence is shown in Supplementary Figs~\ref{fig:s1}--\ref{fig:s5}, and the report counts for every analysis window are in Supplementary Table~\ref{tab:s1}. Numbers already given in the main text are repeated here only where a check needs them.

\smethod{Records and data handling}\label{smeth:records}
This section gives the data handling behind the Methods subsection on the study area and records; the report data and their checks against published figures are described in Supplementary Note~\ref{snote:records}. Supplementary Table~\ref{tab:inputs} lists every data input to SENTINEL and the data used only for evaluation, with their sources.

\begin{table}[tbp]\centering\footnotesize
\caption[Data inputs to SENTINEL and the data used to evaluate it]{\textbf{Data inputs to SENTINEL and the data used to evaluate it.} The upper part lists every input that SENTINEL uses to make a forecast, with its role and source; the lower part lists data used only to evaluate the results or to check the inputs, which never enter the forecast. OGL, Open Government Licence v3.0.}\label{tab:inputs}
\begin{tabular}{@{}p{0.19\textwidth}p{0.29\textwidth}p{0.28\textwidth}p{0.14\textwidth}@{}}
\toprule
data & use & source and access & licence \\
\midrule
\multicolumn{4}{@{}l}{\textit{inputs to SENTINEL}}\\
odour reports (date, time, postcode, intensity), March 2023 to January 2026 & forecast target; report history (state, error correction, sector shares) & odour-report log of the local council & personal data; aggregated counts on request \\
weather forecasts: Met Office UKV, 1.5\,km; five surface fields and five fields on six pressure levels; lead times 0--108\,h & meteorological input to CALMET; forecast wind speed and stability as regressors & CEDA Archive, registered users & Met Office terms \\
terrain: OS Terrain~50 & terrain of the CALMET grid & Ordnance Survey OpenData & OGL \\
land cover: UKCEH Land Cover Map 2023, 10\,m & land use and surface roughness of the CALMET grid & Environmental Information Data Centre, after licence acceptance & UKCEH licence; no redistribution \\
postcode sector boundaries & assigning model cells to sectors &  & \\
unit-postcode centroids: OS Code-Point Open (August 2026) & placing residents in model cells & Ordnance Survey OpenData & OGL \\
residents by unit postcode: 2021 Census & per-resident report rates and shares & Office for National Statistics (Nomis) & OGL \\
CALMET and CALPUFF, version 5.8.5 & dispersion models & Exponent (\url{https://www.calpuff.org}), compiled unmodified & distribution terms; not redistributed \\
\midrule
\multicolumn{4}{@{}l}{\textit{used only for evaluation and checks}}\\
H$_2$S and methane, 15-minute, at MMF1, MMF2, MMF6, MMF9 and MMF1A & evaluation; exposure on an absolute scale; gas tiers; monitor model & the regulator's monitoring network, with its published adjustment of 2023 &  \\
ERA5 reanalysis & checks of the weather input; weekly weather in the variance shares & Copernicus Climate Data Store & Copernicus licence \\
wind measured at the monitors & checks of the modelled wind & the regulator's monitoring network & as above \\
published council and Environment Agency report counts; council officers' odour assessments & checks of the report record; the Environment Agency's reports are a separate channel & public council and regulator documents & public documents \\
\bottomrule
\end{tabular}
\end{table}

\textbf{Report times.} Since March 2023, each report has recorded the time at which the resident noticed the odour. These times are entered by hand: they cluster on the quarter-hour and none includes seconds, so their resolution is about 15 minutes. In the earlier report sheet, the time is instead the time at which a report was logged (Supplementary Note~\ref{snote:legacy}). Residents enter the time in local clock time, and we converted it to Coordinated Universal Time (UTC), the time base of the weather and gas data. The council's published division of reports between day (07:00--23:00) and night matches our record when its times are read as local time, which confirms the time zone. Each report was assigned to the hour that ends at or after its time, the convention of the hourly output of the dispersion model and of the hourly means of the gas and weather data. Days run from midnight to midnight UTC, and the four six-hour windows begin at 00:00, 06:00, 12:00 and 18:00.

\textbf{Postcode sectors and residents.} The sector of a report is the outward code of its postcode followed by the first digit of the inward code. Residents come from the 2021 Census counts for each unit postcode, which nest within sectors, so no apportionment by area is needed; households rank the sectors in the same order as residents. For comparisons with the dispersion model, we placed each unit postcode at its Code-Point Open centroid, converted from the British National Grid to UTM zone 30N, and assigned it to the 100\,m model cell that contains it. A model cell belongs to a sector if its centre lies inside the sector boundary. Residents placed in this way reproduce the Census total of each sector to within 7\,\%.

\textbf{Gas data.} The regulator's H$_2$S analysers under-read from their installation until August 2023. The regulator published adjusted values for each instrument, combining slope, baseline and conversion-efficiency corrections derived from sulphur dioxide calibrations, together with 95\,\% ranges\cite{EA2024h2scal}, and we used the adjusted values before 1 September 2023. Methane measurements were unaffected; they are reported in mg\,m$^{-3}$, not in parts per million. Hourly means, labelled by the end of the hour, are the means of the valid 15-minute values in the hour. Six-hour means are aligned with the report windows; a window is used when at least one monitor has four or more valid 15-minute values in it, and a day of the monitor model needs at least two such windows. For the odour-impact factors, a day of half-hourly H$_2$S needs at least 36 of its 48 half-hours.

\smethod{Weather and the dispersion model}\label{smeth:chain}
This section gives the configuration behind the Methods subsection on the dispersion field (Supplementary Table~\ref{tab:chain}); the comparison of the modelled weather and dispersion with the monitors is in Supplementary Note~\ref{snote:chain}.

\textbf{Weather forecasts.} The meteorological input is the Met Office UKV forecast at 1.5\,km resolution\cite{LITVosper2013High}. An hour was used only if all five surface fields and all five profile fields on six pressure levels were present; incomplete hours were dropped and the series was split at each gap, rather than filled by interpolation, because an interpolated hour is not a forecast. Pressure levels below the ground surface were masked when vertical profiles were built. We obtained forecasts at fixed lead times of 0, 6, 12, 18, 24, 36, 48, 60, 72, 96 and 108\,h and ran the dispersion models separately for each lead time, on hourly series assembled from successive forecast cycles at that lead time. For a forecast issued at midnight, window~$b$ ($b=1,\dots,4$) of the day $k$ days ahead ends $L^{\rm req}_{k,b}$ hours after issue, and its weather and dispersion field come from the lead time $L_{k,b}$:
\begin{equation}\label{eq:lead}
L^{\rm req}_{k,b}=6\,[\,4(k-1)+b\,],\qquad L_{k,b}=\min\big\{\ell\in\mathcal L:\ \ell\ge L^{\rm req}_{k,b},\ \text{run available}\big\},\qquad L_{k,b}\le L^{\rm req}_{k,b}+24,
\end{equation}
where $\mathcal L$ is the set of lead times obtained. If no dispersion field is available for a window, we use the field for the same valid time from the next longer lead time available, which comes from an older forecast. As a result, no window uses a more recent forecast than would have been available at issue. One day ahead uses lead times of 6, 12, 18 and 24\,h, and two days ahead uses 36, 36, 48 and 48\,h.

\textbf{CALMET.} The model grid has $200\times200$ cells of 100\,m (20\,km by 20\,km) centred on the landfill and ten vertical layers. Terrain comes from OS Terrain~50 and land use from the UKCEH Land Cover Map 2023, translated into the model's land-use classes. CALMET was run with the forecast fields as its only meteorological input, with no surface, upper-air or precipitation stations, and used them as the initial guess for its diagnostic wind field, which adjusts the flow to the terrain at 100\,m. For each lead time, the models were run one calendar month at a time from that month's forecast cycles, as runs of up to 745\,h that start between 00:00 and 05:00 UTC, because CALMET's mixing-height scheme must start before dawn; a run is split where a forecast cycle is missing, and shorter runs fill the gaps. The first 5\,h of every run were discarded as spin-up: in 38 overlaps with identical input, hourly fields from a run started from an empty domain agreed with those of the longer run to within $10^{-6}$ from the fifth hour. Where runs overlap, the hours of the earlier run are kept.

\textbf{CALPUFF.} CALPUFF released an inert tracer at 1\,g\,s$^{-1}$ from the landfill, represented as area sources at a height of 2\,m. Plume rise was suppressed (exit velocity 0.001\,m\,s$^{-1}$ at ambient temperature), and chemistry and wet and dry deposition were switched off. These choices keep the modelled concentration proportional to the emission rate, so that one run serves every source strength and $\Phi$ is a transport factor: concentration per unit emission. The removal processes are slow compared with the transit time of 13--90 minutes over 1--6\,km. Concentrations were computed on the model grid and at the monitors, 2\,m above the terrain, which is the sampling height of the monitors. Hourly concentrations were averaged to six-hour means and over the cells of each sector (equation~\eqref{eq:phi} of the main text). Cells that no puff reaches have a concentration of exactly zero, so models use $\log_{10}\max(\Phi,10^{-12})$, with $\Phi$ in \textmu g\,m$^{-3}$ per g\,s$^{-1}$; the floor lies many orders of magnitude below the values in cells the plume reaches.

\begin{table}[tbp]\centering\footnotesize
\caption[Configuration of the dispersion models]{\textbf{Configuration of the dispersion models.} Settings of CALMET and CALPUFF used for every lead time, with the reason for each choice; control-file names are given for reproducibility. Neither model was modified; both were compiled from the distributed source (version 5.8.5).}\label{tab:chain}
\begin{tabular}{@{}p{0.27\textwidth}p{0.3\textwidth}p{0.37\textwidth}@{}}
\toprule
setting & value & reason \\
\midrule
grid & $200\times200$ cells of 100\,m, centred on the landfill; UTM zone 30N, WGS-84 & a 20\,km domain is needed for plumes to reach the outer sectors; 100\,m resolves the terrain around the site \\
vertical layers & 10; faces at 0, 20, 40, 80, 160, 300, 600, 1,000, 1,500, 2,200 and 3,000\,m & thinnest layers near the ground, where the receptors are \\
meteorological input & UKV forecast fields only (\texttt{NOOBS = 2}); no observation stations & no instrument at the site; observations do not exist at issue for the hours forecast \\
initial wind field & UKV winds (\texttt{IPROG = 14}), adjusted to terrain & uses the forecast flow and the 100\,m terrain \\
mixing height & 50--2,500\,m & lower and upper bounds on the diagnosed mixing height \\
time zone & UTC (\texttt{XBTZ = 0}) & matches reports, gas and weather data \\
source & area sources, release height 2\,m & surface emission through the cap \\
emission & 1\,g\,s$^{-1}$, inert, constant & output is a transport factor, proportional to any emission rate \\
plume rise & suppressed (exit velocity 0.001\,m\,s$^{-1}$, ambient temperature) & plume rise depends on the emission rate and would break proportionality \\
chemistry and deposition & off (\texttt{MCHEM = MWET = MDRY = 0}) & slow relative to transit; would make the field depend on the species \\
dispersion coefficients & Pasquill--Gifford curves (\texttt{MDISP = 3}) & one of the two schemes permitted by the model's regulatory guidance; used in every production run (one run with \texttt{MDISP = 2} was a sensitivity check) \\
terrain adjustment & partial plume path (\texttt{MCTADJ = 3}) & adjusts puff height as puffs cross terrain \\
receptors & model grid and the monitors, 2\,m above terrain & sampling height of the monitors \\
\bottomrule
\end{tabular}
\end{table}

\smethod{SENTINEL}\label{smeth:sentinel}
This section complements the Methods subsection on the SENTINEL forecasting model with the reasons for its design and the search that selected it; how the selection performed is in Supplementary Note~\ref{snote:selection}. The alternative designs and their tests are documented in Supplementary Data~2.

\textbf{Filter and forecast.} With the state vector $\boldsymbol\xi_t=(\mu_t,a_t)^{\top}$, the model of Methods equation~\eqref{eq:ssm} is
\begin{equation}\label{eq:ss}
\boldsymbol\xi_t=\mathbf G\,\boldsymbol\xi_{t-1}+\mathbf w_t,\qquad \mathbf w_t\sim N\big(\mathbf 0,\operatorname{diag}(\sigma^2_\eta,\sigma^2_\zeta)\big),\qquad y_t=\mathbf z^{\top}\boldsymbol\xi_t+\mathbf x_t^{\top}\boldsymbol\beta,
\end{equation}
with $\mathbf G=\operatorname{diag}(1,\psi)$ and $\mathbf z=(1,1)^{\top}$. The Kalman filter gives the predicted state $\hat{\boldsymbol\xi}_{t\mid t-1}=\mathbf G\hat{\boldsymbol\xi}_{t-1\mid t-1}$ with covariance $\mathbf P_{t\mid t-1}=\mathbf G\mathbf P_{t-1\mid t-1}\mathbf G^{\top}+\operatorname{diag}(\sigma^2_\eta,\sigma^2_\zeta)$, the innovation and its variance
\begin{equation}\label{eq:innov}
v_t=y_t-\mathbf z^{\top}\hat{\boldsymbol\xi}_{t\mid t-1}-\mathbf x_t^{\top}\boldsymbol\beta,\qquad F_t=\mathbf z^{\top}\mathbf P_{t\mid t-1}\mathbf z,
\end{equation}
and the update $\hat{\boldsymbol\xi}_{t\mid t}=\hat{\boldsymbol\xi}_{t\mid t-1}+\mathbf k_tv_t$, $\mathbf P_{t\mid t}=\mathbf P_{t\mid t-1}-\mathbf k_t\mathbf z^{\top}\mathbf P_{t\mid t-1}$, with gain $\mathbf k_t=\mathbf P_{t\mid t-1}\mathbf z/F_t$; windows without data are predicted but not updated. The log-likelihood used for $\boldsymbol\beta$ and $\sigma^2_\zeta$ is
\begin{equation}\label{eq:loglik}
\ell=-\tfrac12\sum_t\big(\log 2\pi F_t+v_t^2/F_t\big),
\end{equation}
and the forecast $k$ windows after issue is $\hat{\boldsymbol\xi}_{t+k\mid t}=\mathbf G^{k}\hat{\boldsymbol\xi}_{t\mid t}$ (Methods equation~\eqref{eq:fc}).

\textbf{Training on day-to-day changes.} Trained by likelihood, the model's forecast for day~$d$ is more closely correlated with reports on day~$d-1$ ($r=0.880$) than on day~$d$ ($r=0.845$). For a stationary record, the forecast is more closely correlated with the target day than with the previous day only if it reproduces more than half of the day-to-day change:
\begin{equation}\label{eq:half}
r(0)>r(-1)\iff q=\frac{\operatorname{cov}\big(\hat D_d-D_{d-1},\,D_d-D_{d-1}\big)}{\operatorname{var}\big(D_d-D_{d-1}\big)}>\tfrac12 .
\end{equation}
This follows because $r(0)-r(-1)$ has the sign of $\operatorname{cov}(\hat D_d,D_d-D_{d-1})$ and, for a stationary record, $\operatorname{cov}(D_{d-1},D_d-D_{d-1})=-\tfrac12\operatorname{var}(D_d-D_{d-1})$. The likelihood fit had $q=0.45$. A slower level, 28 forms of error correction, issuing the forecast up to 24\,h earlier and forecasting from the target day's weather and dispersion field alone all left the peak on the previous day. A squared error on the daily change cannot help, because for a forecast issued from the last recorded day it equals the squared error on the daily level,
\begin{equation}\label{eq:same}
\big(D_d-\hat D_d\big)^2=\big[(D_d-D_{d-1})-(\hat D_d-D_{d-1})\big]^2 .
\end{equation} Following work on timing errors in river-flow forecasting\cite{deVos2005}, we instead penalised the difference between the forecast's own day-to-day step and the recorded step (equation~\eqref{eq:loss} of the main text). With this penalty, the fitted dynamics move short-term variation into a fast-decaying term (autoregressive coefficient 0.71 without the penalty and 0.34 with $\lambda=0.5$), and the forecast's correlation peaks on the target day.

\textbf{Day index.} The regressors include a day index: a ridge regression, fitted on the training days only, of six-hour report counts on weekday, window of the day, an annual harmonic pair, the dispersion field of each sector at the window's lead time, wind speed and stability, standardised. With the standardised regressors $\mathbf c_t$ (six weekday indicators, three window indicators, $\sin$ and $\cos$ of the annual phase, $\log_{10}\Phi_s$ for the ten sectors at the window's lead time, wind speed and stability), the coefficients are
\begin{equation}\label{eq:ridge}
\hat{\mathbf b}=\arg\min_{b_0,\mathbf b}\ \sum_{t\in\text{training}}\big(y_t-b_0-\mathbf c_t^{\top}\mathbf b\big)^2+\lambda_{\rm r}\lVert\mathbf b\rVert^2,\qquad \lambda_{\rm r}=1,
\end{equation}
and the index $\hat A_t$ is the fitted value $\hat b_0+\mathbf c_t^{\top}\hat{\mathbf b}$, standardised over the training windows. On its own, the index forecast daily reports poorly (daily skill \ensuremath{-}0.10), because the dispersion field indicates where reports come from more than how many there will be; within the state-space model, it is how the target day's weather enters the forecast.

\textbf{Error correction and volume.} The error correction shifts the forecast by an exponentially weighted mean of past errors $e_d=D_d-\hat D_d$, updated recursively from the errors of days already recorded at issue,
\begin{equation}\label{eq:ewma}
o_d=o_{d-1}+\omega\,\big(e_{d-L}-o_{d-1}\big),\qquad \omega=1-2^{-1/\tau},
\end{equation}
with $o_d$ started at the first available error and held unchanged when an error is missing. With a half-life of one day, the correction was set almost entirely by the previous day's error, and the correlation peak moved back to the previous day; with half-lives of two weeks or more, the peak stayed on the target day. Volume calibrations rescale the forecast count with a Poisson regression on past verified days,
\begin{equation}\label{eq:vol}
\tilde N_d=\exp\big(a_d+b_d\hat D_d\big),\qquad (a_d,b_d)=\arg\max_{a,b}\sum_{d'\le d-L}w_{d'}\big[N_{d'}(a+b\hat D_{d'})-e^{a+b\hat D_{d'}}\big],
\end{equation}
with $w_{d'}=1$, or with weights that decay with age; they changed the forecast level and, in most configurations, moved the correlation peak back to the previous day (Supplementary Note~\ref{snote:selection}); SENTINEL uses none.

\textbf{Alternative designs.} We also compared three alternative designs with the state-space model: adding a population-weighted exposure index to the total (no gain; paired difference in daily skill \ensuremath{-}0.007~[\ensuremath{-}0.016,~+0.000]); a daily negative-binomial count model with a slowly varying propensity to report, which peaked on the target day but had much lower daily skill (0.245 against 0.678); and the same count model with a memory of past modelled exposure, which performed no differently. Most of the one-day skill therefore comes from day-to-day persistence in the reports themselves.

\textbf{The search.} The search covered 226 configurations drawn from 12 base models, combining training on day-to-day changes with weights $\lambda$ from 0.5 to 2 (with the dynamics parameterised as in Methods) or from 0 to 4 (with four free dynamic parameters) and training by likelihood alone, with and without the day index; nine half-lives of the error correction (none, or 10 to 120 days); and no volume calibration or one of two calibrations refitted on past verified days; not every combination was run. The selection rule was specified before the search ran: on the first half of the scored days under block cross-validation, among configurations whose correlation with reports peaked on the target day with $T>0$, the highest tier agreement, with ties within 0.001 broken by daily skill; persistence could not be selected. The selected configuration was then confirmed on the second half and on all days, with training on earlier days only, and against the likelihood-trained reference (Supplementary Table~\ref{tab:search}; all configurations in Supplementary Data~1). The same configuration is used unchanged two to four days ahead, and no selection was made at those lead times. When the configuration was re-selected each week from earlier days only, the forecast's partial correlation with the target day beyond the recorded report history was 0.50~[0.41,~0.59].

\begin{table}[tbp]\centering\scriptsize
\caption[The search over SENTINEL configurations]{\textbf{The search over SENTINEL configurations.} The ten leading configurations under the registered selection rule, ranked by tier agreement on the first half of the scored days under block cross-validation; the selected configuration is in bold. The last two rows give the best configuration of each base trained by likelihood alone. Training: on day-to-day changes with weight $\lambda$ (``4 par.'': all four dynamic parameters free) or by likelihood; offset: half-life of the error correction in days; volume: the volume calibration (``refit'': a level-dependent Poisson calibration refitted each day on past verified days; ``forget.'': the same with older days down-weighted). $\kappa$: quadratic-weighted tier agreement; skill: daily skill against the average for the same weekday and month; $T$: target-day timing $r(0)-r(-1)$; all dimensionless, higher is better. One day ahead, all ten sectors, 486 scored days; block cross-validation uses 8 blocks with a 2-day gap, and ``trained on earlier days'' uses monthly steps in time order. All 226 configurations are listed in Supplementary Data~1.}\label{tab:search}
\input{table_model_search}
\end{table}

\smethod{The psychometric reporting model and exposure}\label{smeth:psych}
This section gives the fitting behind the Methods subsection on the psychometric reporting model and the exposure estimate; the results and their checks are in Supplementary Notes~\ref{snote:fidol} and~\ref{snote:exposure}.

\textbf{Fitting the psychometric reporting model.} For six-hour window~$w$, let $R_w=1$ if sector~C made at least one report and $c_w$ be the H$_2$S concentration at MMF9, as the six-hour mean or the six-hour peak of the hourly means. The parameters of Methods equation~\eqref{eq:curve} maximise the Bernoulli log-likelihood
\begin{equation}\label{eq:bern}
\ell(\alpha,\sigma,q)=\sum_w\Big[R_w\log P(c_w)+(1-R_w)\log\big(1-P(c_w)\big)\Big],\qquad P(c)=q\,F_{\mathcal N}\Big(\frac{\log_{10}c-\log_{10}\alpha}{\sigma}\Big).
\end{equation}
As a sensitivity check, $F_{\mathcal N}$ was replaced by the complementary log--log function $1-\exp(-e^{u})$, whose midpoint differs from~$\alpha$. The threshold on the six-hour peak of hourly means comes from a joint fit over March--August and September--December 2023 (the threshold on the six-hour peak of 30-minute means in the main text is fitted on January 2024 -- March 2025), with one spread $\sigma$ and a separate upper limit $q$ for each period. For each candidate threshold on a grid from 2 to 100\,\textmu g\,m$^{-3}$, which includes 7 and 56\,\textmu g\,m$^{-3}$, $\sigma$ was chosen to maximise the joint likelihood (bounded on $\log\sigma$ between 0.05 and 5 decades); the estimate is the grid value with the highest profile likelihood, and its interval is the 2.5th to 97.5th percentile of that value over 1,000 resamples of seven-day blocks.

\textbf{Source estimation from reports.} The latent source state of Methods equation~\eqref{eq:state}, $x_w=\rho\,x_{w-1}+\omega_w$ with $\omega_w\sim N(0,\sigma_\omega^2)$, is the logarithm (base 10) of the source strength relative to a nominal emission rate. $C_{s,w}$ is the six-hour mean of the sector field multiplied by the nominal emission rate, floored at $10^{-3}$\,\textmu g\,m$^{-3}$. We computed the likelihood exactly by a forward recursion on a grid of 241 values of $x$ from $-3$ to $3$:
\begin{equation}\label{eq:grid}
\pi_w(x_i)=\frac{p(\mathbf R_w\mid x_i)\sum_j T_{ij}\,\pi_{w-1}(x_j)}{\sum_i p(\mathbf R_w\mid x_i)\sum_j T_{ij}\,\pi_{w-1}(x_j)},\qquad T_{ij}\propto\exp\Big[-\frac{(x_i-\rho x_j)^2}{2\sigma_\omega^2}\Big],
\end{equation}
where $p(\mathbf R_w\mid x)=\prod_s\Pr(R_{s,w}\mid x)$ over the sectors, the rows of $T$ sum to one, and $\pi_0$ is the stationary distribution $N\big(0,\sigma_\omega^2/(1-\rho^2)\big)$. The log-likelihood is the sum over training windows of the logarithm of the denominator; windows outside the training set are prediction steps only. The fit has two stages. First, with a constant source, $\mu$, $\sigma$, $g_s$ and $q_s$ were fitted by maximum likelihood, with $g_s$ and $q_s$ parameterised on the logistic scale so that $0<g_s<q_s<1$. Second, with $g_s$ and $q_s$ held, $\mu$, $\sigma$, $\rho$ and $\sigma_\omega$ were fitted on the filter likelihood. On simulated reports from a known source state, the second stage recovered the autoregressive coefficient (0.893 against 0.900) and the spread (0.601 against 0.600), whereas the constant-source stage alone overestimates the spread.

\textbf{Exposure.} The filtered state gives each window's standardised position $z_{s,w}$ and concentration $c_{s,w}$ (Methods equation~\eqref{eq:exposure}). Weekly exposure is $\sum_w c_{s,w}\times6$\,h over weeks with at least 20 of their 28 windows; monthly exposure uses months with at least 60\,\% of windows. Measured exposure is summed over the same windows, from MMF9 for sector~C and from the monitors in sector~G (Supplementary Note~\ref{snote:monitors}). The calibrations based on detection thresholds replace the WHO guideline with
\begin{equation}\label{eq:theta}
c_{s,w}=\theta\cdot10^{\,\sigma_{\rm pub}z_{s,w}},\qquad \theta=\frac{m\,c_{\rm det}}{r_{\rm pm}},
\end{equation}
where $c_{\rm det}$ is a published detection threshold, $m$ the multiplier from detection to annoyance (4--5), and $r_{\rm pm}$ the ratio of short-term peak to hourly mean concentration for an area source. Concentrations given in parts per billion are converted at 1.417\,\textmu g\,m$^{-3}$ per ppb throughout (20\,$^\circ$C and 101.3\,kPa, the reference conditions of UK air-quality standards).

\smethod{The monitor model and the monitor-based source estimate}\label{smeth:monitor}
This section gives the specification behind the Methods subsection on the monitor model and the source estimate (Figs~\ref{fig:1}f and~\ref{fig:3}c).

\textbf{Monitor model.} In Methods equation~\eqref{eq:monitor}, $y_j(t)$ is the logarithm of six-hour H$_2$S at monitor~$j$ (MMF1, MMF2 and MMF9); $\Phi_j(t)$ is the mean of the dispersion field within 300\,m of the monitor, with its own coefficient for each monitor; $\mathbf X_t$ contains weather, window-of-day and seasonal terms with coefficients for each monitor; $u(t)$ is the shared source state and $l_j(t)$ each monitor's own first-order autoregressive memory. For Fig.~\ref{fig:3}c the dispersion field is taken at a lead time of zero and the model forecasts one six-hour window ahead. It was fitted by maximum likelihood with the Kalman filter in each of the eight blocks of the block cross-validation, restarting the optimiser until its convergence test was met (all eight blocks converged); a level-dependent Poisson calibration refitted on past verified days (Supplementary Methods equation~\eqref{eq:vol}) then gives daily totals. Daily exposure is the six-hour mean multiplied by 6\,h, summed over the day and the three monitors. Daily gas tiers use the daily peak of the hourly mean at the worst monitor, with boundaries at 7, 30 and 150\,\textmu g\,m$^{-3}$; the lowest is the WHO odour annoyance guideline.

\textbf{Monitor-based source estimate.} For monitor~$i$ in six-hour window~$t$ within period~$p$ (week or month), with H$_2$S excess $H_{it}$ and methane excess $E_{it}$ above background, we fitted
\begin{equation}\label{eq:twogas}
\log H_{it}=q^{\rm H}_{p}+a^{\rm H}_i+\chi_{it}+\varepsilon^{\rm H}_{it},\qquad \log E_{it}=q^{\rm C}_{p}+a^{\rm C}_i+\chi_{it}+\varepsilon^{\rm C}_{it},\qquad \chi_{it}=\log\Phi_i(t)+\mathbf X_t^{\top}\boldsymbol\beta_i ,
\end{equation}
in which the transport term $\chi_{it}$ is shared by the two gases at the same monitor and window. The source pair $(q^{\rm H}_p,q^{\rm C}_p)$ follows a two-dimensional random walk across periods with a full $2\times2$ innovation covariance, estimated by expectation--maximisation and smoothed. Values below a monitor's detection floor enter as censored observations, background comes from windows outside the plume, and a window is in the plume if the modelled plume covers the monitor for at least three of its six hours. A period counts only with at least 24 (week) or 72 (month) in-plume hours at a monitor. The estimate is relative; the reference monitor has no intercept.

\textbf{Variance shares.} With $y_p=\log(1+N_p)$ for the reports in week~$p$, the source estimate $S$ and weekly weather means $W$ (logarithms of boundary-layer height and wind speed, and the sine and cosine of wind direction, from ERA5), the shares of variance are two-block Shapley values,
\begin{equation}\label{eq:shapley}
s_S=\tfrac12\big[R^2(S)+R^2(S,W)-R^2(W)\big],\qquad s_W=\tfrac12\big[R^2(W)+R^2(S,W)-R^2(S)\big],
\end{equation}
where $R^2(\cdot)$ is the coefficient of determination of a linear regression of $y_p$ on the named blocks. Counting noise is $\sum_pN_p/(1+N_p)^2$ divided by $\sum_p(y_p-\bar y)^2$, and the remainder includes changes in residents' readiness to report, together with source and transport variation not captured by the two blocks, and model error. Shares on withheld days use five contiguous blocks with a one-week gap.

\smethod{Comparison models}\label{smeth:comparators}
This section defines the comparison models used in Results (Supplementary Note~\ref{snote:comparators} gives the results).

\textbf{Persistence.} $k$ days ahead, persistence forecasts $\log(1+N)$ for day~$d$ as the value on the last recorded day, $D_{d-k}$. For the same-day estimate, it repeats the last six-hour window, adjusted by the average daily cycle.

\textbf{Seasonal reference.} The reference $C_d$ in Methods equation~\eqref{eq:msss} is the mean of $D_{d'}$ over recorded days $d'<d$ with the same weekday and month, after a burn-in of 120 days.

\textbf{Variants of SENTINEL.} The report-history model is SENTINEL with the dispersion field and the forecast weather removed from its regressors. The model trained by likelihood alone is SENTINEL with $\lambda=0$. The observed-weather model is SENTINEL given the weather and dispersion field at a lead time of zero, a reference run with the observed weather.

\textbf{Wind-direction map.} The division among sectors replaces the dispersion term of Methods equation~\eqref{eq:map} with the alignment of the forecast flow with the bearing of each sector,
\begin{equation}\label{eq:geo}
p^{\rm wind}_{s,t}\propto\exp\big(b_s+\gamma_{\rm w}\cos(\theta_t-\vartheta'_s)+o_{s,d}\big),
\end{equation}
where $\theta_t$ is the direction towards which the forecast wind blows and $\vartheta'_s$ the direction from the landfill to sector~$s$, both as mathematical angles, so that the cosine is $+1$ when the wind carries the plume to the sector. The coefficient $\gamma_{\rm w}$ was fitted by multinomial likelihood on the training windows, with the same residual shares $o_{s,d}$ as SENTINEL. The share maps use the average shares of the training period, or the shares of the last 28 recorded days.

\textbf{Regression on report history and calendar.} A ridge regression, as in Supplementary Methods equation~\eqref{eq:ridge}, of $D_d$ on recent daily totals and calendar terms, refitted on SENTINEL's schedule; adding the forecast weather and the dispersion field gives its weather variant.

\smethod{Verification measures}\label{smeth:verification}
This section gives the formulas for the measures defined in Methods. Unless stated otherwise, every measure is computed on $D_d=\log(1+N_d)$ for the ten-sector daily total over the scored days.

\textbf{Tier agreement.} Days fall into tiers $0$--$3$ at 3, 9 and 27 reports. With $O_{ij}$ the share of days with recorded tier~$i$ and forecast tier~$j$, and $E_{ij}$ the product of the marginal shares,
\begin{equation}\label{eq:kappa}
\kappa_w=1-\frac{\sum_{ij}v_{ij}O_{ij}}{\sum_{ij}v_{ij}E_{ij}},\qquad v_{ij}=\frac{(i-j)^2}{9}.
\end{equation}

\textbf{Skill of the tier probabilities.} The forecast probabilities for each tier come from a negative-binomial distribution with mean $\hat N_d$ and dispersion
\begin{equation}\label{eq:nb}
\hat a_d=\operatorname{clip}\Big(\frac{\sum_{d'}\big[(N_{d'}-\hat N_{d'})^2-\hat N_{d'}\big]}{\sum_{d'}\hat N_{d'}^2},\ 0.02,\ 50\Big),
\end{equation}
estimated from the days $d'$ already recorded at issue (at least 30). With $F_{d,k}$ and $O_{d,k}$ the forecast and recorded cumulative tier probabilities, the ranked probability score is $\mathrm{RPS}_d=\sum_{k=1}^{3}(F_{d,k}-O_{d,k})^2$, and the skill is $1-\sum_d\mathrm{RPS}_d/\sum_d\mathrm{RPS}^{\rm ref}_d$, with the reference given by the tier frequencies of the days recorded before~$d$.

\textbf{Days with many reports.} For days with at least 27 reports, the area under the precision--recall curve of the forecast is divided by the share of such days, so that 1 corresponds to no skill.

\textbf{Division among sectors.} With forecast shares $\hat p_{d,s}$, recorded shares $p_{d,s}$ and training-period shares $\bar p_s$,
\begin{equation}\label{eq:redist}
\mathrm{RS}=1-\frac{\sum_dN_d\sum_s(\hat p_{d,s}-p_{d,s})^2}{\sum_dN_d\sum_s(\bar p_s-p_{d,s})^2},\qquad \cos_d=\frac{\sum_s\tilde p_{d,s}\,\tilde{\hat p}_{d,s}}{\lVert\tilde{\mathbf p}_d\rVert\,\lVert\tilde{\hat{\mathbf p}}_d\rVert},
\end{equation}
where the redistribution skill RS weights days by their report counts, and the cosine agreement $\cos_d$ uses shares per resident ($\tilde p_{d,s}\propto p_{d,s}/n_s$, with $n_s$ the residents of sector~$s$).

\textbf{Economic value.} For a yes/no alert with hits $h$, false alarms $f$ and misses $m$ as fractions of the scored days, base rate $\bar o$ and ratio $\alpha_c$ of the cost of protective action to the loss avoided, the relative economic value is\cite{zhu2002}
\begin{equation}\label{eq:value}
V(\alpha_c)=\frac{\min(\alpha_c,\bar o)-\big[\alpha_c(h+f)+m\big]}{\min(\alpha_c,\bar o)-\bar o\,\alpha_c},
\end{equation}
which is 1 for a perfect forecast and 0 for the better of always and never acting. The value area is the mean of $V$ over the grid of cost--loss ratios, computed by the trapezoidal rule. At an equal number of alerts, the forecast and persistence thresholds are set so that both issue exactly the same number of alerts.

\smethod{Intervals and tests}\label{smeth:intervals}
This section gives the interval methods behind the Methods subsection on validation design; the registered tests and their outcomes are in Supplementary Note~\ref{snote:intervals}.

\textbf{Paired resampling.} Every comparison between two forecasts is a statistic $\hat\theta$ of a paired daily series. A moving-block bootstrap draws blocks of $L$ consecutive scored days with replacement until the series length is reached and recomputes $\hat\theta^{*}_b$, $b=1,\dots,B$. The percentile interval is $[\hat\theta^{*}_{(0.025)},\hat\theta^{*}_{(0.975)}]$; the basic interval is $[2\hat\theta-\hat\theta^{*}_{(0.975)},\,2\hat\theta-\hat\theta^{*}_{(0.025)}]$; the bias-corrected and accelerated interval adjusts the percentiles by the normal quantile of the share of draws below $\hat\theta$ and by an acceleration from a delete-one-block jackknife. Subsampling recomputes the statistic on every window of $b$ consecutive days and gives
\begin{equation}\label{eq:subsample}
\Big[\hat\theta_n-\frac{q_{0.975}}{\sqrt n},\ \hat\theta_n-\frac{q_{0.025}}{\sqrt n}\Big],\qquad r_i=\frac{\sqrt b\,(\hat\theta_{b,i}-\hat\theta_n)}{\sqrt{1-b/n}},
\end{equation}
where $q_p$ is the $p$-quantile of the $r_i$; the factor $\sqrt{1-b/n}$ corrects for the overlap of each window with the full record. The window length was chosen by the minimum-volatility rule over 14--112 days.

\textbf{Choosing the interval method.} For each family of comparisons, candidate intervals (percentile, basic and bias-corrected bootstrap with blocks of 7, 14, 28, 56, 84 and 112 days, and subsampling) were scored by their coverage, measured in two ways. In simulated records, a Gaussian copula with the real marginal distributions and a latent process that is the sum of a slow and a fast first-order autoregression, fitted to the autocorrelation of the real series, reproduces the paired structure of each forecast and its comparison model; the true value is computed on one 200,000-day record, and coverage is the share of 400 simulated records of the real length whose interval contains it. The second check uses no simulated generator: the real paired series are resampled by subsampling. For each family we used the method and block length with the highest measured coverage. No method reached nominal 95\,\% coverage (Methods).

\textbf{Equivalence and multiplicity.} A statement of no credible evidence is supported by a two-one-sided test against a smallest meaningful effect $m$ declared in advance: with $p_{\rm low}=(1+\#\{\hat\theta^{*}_b\le-m\})/(B+1)$ and $p_{\rm high}=(1+\#\{\hat\theta^{*}_b\ge m\})/(B+1)$, equivalence holds when $\max(p_{\rm low},p_{\rm high})<0.05$, that is, when the 90\,\% interval lies inside $(-m,m)$. Holm's step-down adjustment is applied over each stated family. Detectable effects are the effects, added in simulation, that are detected with 80\,\% power in simulated records built in the same way as above.

\smethod{Software}\label{smeth:software}
Analyses used Python~3.12.14 with NumPy~1.26.4, SciPy~1.15.3, pandas~2.2.3, statsmodels~0.14.6 (state-space models and generalised linear models), scikit-learn~1.6.1 and Matplotlib~3.10.5. CALMET and CALPUFF version 5.8.5 (level 151214) were compiled from the distributed source without modification with GNU Fortran 13.3.0 at optimisation level 1 and run on a Linux computing cluster. These versions are those recorded in the release environment file for the cluster, which the code repository provides.

\snote{The odour-report record}\label{snote:records}
This note supports the description of the report data in the main text (Methods; Results section on odour reports and fugitive emissions). The record is the local council's log of the odour reports made to it by residents, each with the date and time at which the odour was noticed, the postcode and an intensity rating from 0 to 6. We removed reports delivered more than once and collapsed repeats with the same minute, postcode sector and rating, which reduced 9,992 rows to 9,893 reports from all postcodes, of which 9,078 came from the ten sectors around the landfill. Residents per sector come from the 2021 Census. Residents can also report to the Environment Agency, which logs reports separately. Its channel received 2.8–4.9 times as many contacts as the council's in each year from 2021 to 2024, partly because it counts every contact, including repeats from the same address. Its reports are not in our data; its published weekly counts are used only as a check. Over January 2024 to April 2025, reports from the ten sectors make up 91.0\,\% of the reports the council published; the remainder are reports from outside the ten sectors (7.9\,\%) and collapsed repeats (1.1\,\%). The data handling is described in Supplementary Methods~\ref{smeth:records}.

Different analyses use different windows (Supplementary Table~\ref{tab:s1}). The map and wind analyses (Fig.~\ref{fig:1}a,b) use January 2024 to March 2025 (7,517 reports). The exposure analyses (Fig.~\ref{fig:3}d) use the 29 months from March 2023 to July 2025 (9,069 reports). The same-day estimate and the forecasts are scored from January 2024 to April 2025: 486 days one day ahead (7,531 reports), falling to 478 days four days ahead, because longer lead times need more days of history at the start. Every count is also given with September--December 2024 withheld.

The report record agrees with independent published figures. The council's published monthly counts match our de-duplicated record to within 4 reports a month, and the council's and the Environment Agency's weekly published figures are closely correlated (Spearman 0.97). Council officers' detections of landfill odour move with the reports (Supplementary Fig.~\ref{fig:s2}d), although their visits are partly prompted by reports.

\begin{table}[tbp]\centering
\caption[Odour reports in each analysis window]{\textbf{Odour reports in each analysis window, by postcode sector.} Residents from the 2021 Census. Counts are de-duplicated reports; ``excl.'' is with September--December 2024 withheld. Whole record, March 2023 to January 2026 (all but nine reports by July 2025); map window, January 2024 to March 2025 (Fig.~\ref{fig:1}a); exposure window, the 29 months March 2023 to July 2025 (Fig.~\ref{fig:3}d); scored days, January 2024 to April 2025, one day ahead (the same-day estimate is scored on the same days). Scored days at longer lead times are given in Supplementary Note~\ref{snote:records}.}\label{tab:s1}
\footnotesize\setlength{\tabcolsep}{3pt}
\begin{tabular}{lrrrrrrrrr}
\toprule
 & & \multicolumn{2}{c}{Whole record} & \multicolumn{2}{c}{Map window} & \multicolumn{2}{c}{Exposure window} & \multicolumn{2}{c}{Scored days} \\
Sector & Residents & all & excl. & all & excl. & all & excl. & all & excl. \\
\midrule
A & 9,499 & 303 & 185 & 280 & 162 & 303 & 185 & 280 & 162 \\
B & 3,305 & 250 & 159 & 209 & 118 & 250 & 159 & 210 & 119 \\
C & 9,059 & 3,034 & 2,400 & 2,347 & 1,713 & 3,032 & 2,398 & 2,356 & 1,722 \\
D & 9,684 & 955 & 724 & 783 & 552 & 954 & 723 & 783 & 552 \\
E & 6,103 & 300 & 209 & 275 & 184 & 299 & 208 & 275 & 184 \\
F & 5,042 & 55 & 40 & 52 & 37 & 54 & 39 & 52 & 37 \\
G & 9,556 & 2,920 & 2,208 & 2,433 & 1,721 & 2,917 & 2,205 & 2,434 & 1,722 \\
H & 11,209 & 209 & 134 & 187 & 112 & 209 & 134 & 188 & 113 \\
I & 10,775 & 236 & 133 & 222 & 119 & 236 & 133 & 222 & 119 \\
J & 8,056 & 816 & 517 & 729 & 430 & 815 & 516 & 731 & 432 \\
\midrule
All & 82,288 & 9,078 & 6,709 & 7,517 & 5,148 & 9,069 & 6,700 & 7,531 & 5,162 \\
\bottomrule
\end{tabular}
\end{table}

\snote{The earlier filing-dated record (2021--23)}\label{snote:legacy}
This note explains why the earlier record is analysed separately (Methods; first Results section). Before March 2023, reports were logged under an older collection system on a separate sheet, which holds 25,575 reports from all postcodes after removing duplicates, 22,915 of them from the ten sectors, most of them filed in 2021 (19,014). The sheet records when a report was filed, not when the odour occurred; the median delay is about 15\,h, and 18\,\% of reports fall in a later week than the odour. It is therefore a different instrument. We use it only for weekly and monthly comparisons (Supplementary Notes~\ref{snote:behaviour} and~\ref{snote:beyondweather}), never pool it with the current record, and never treat its results as a replication of the daily analyses. The UK Health Security Agency's published monthly totals for 2021--2023 are upper bounds on the corresponding report counts (Data availability).

\snote{Monitors and gas data}\label{snote:monitors}
This note describes the gas measurements used to evaluate the results (Methods). The regulator's network measures H$_2$S (\textmu g\,m$^{-3}$) and methane (mg\,m$^{-3}$) every 15 minutes; the instruments are described in earlier work at this site\cite{Pearce2026}. The H$_2$S analysers are Teledyne API T101 real-time analysers, which use UV fluorescence and are calibrated every fortnight with traceable gas standards; methane is measured with Los Gatos Research analysers, which use off-axis integrated cavity output spectroscopy (OA-ICOS). We applied the regulator's published adjustment of September 2023, then averaged to hourly and six-hour means. The monitors lie 229–772\,m apart. Sector~C is represented by MMF9. Sector~G is represented by MMF1 and MMF2 until September 2024, when MMF1's record ends, and by MMF2 and MMF1A afterwards; its measured exposure differs by a factor of about two across this change, so results for sector~G are also shown on the unchanged monitors (Supplementary Note~\ref{snote:exposure}). The monitors are used for validation only and never enter SENTINEL. The processing of the gas data is described in Supplementary Methods~\ref{smeth:records}.

\snote{Source attribution}\label{snote:attribution}
This note gives the methods and checks behind the first Results section (Fig.~\ref{fig:1}; Supplementary Fig.~\ref{fig:s1}). The source estimate and the variance shares are specified in Supplementary Methods~\ref{smeth:monitor}.

\textbf{Wind at report times.} For each sector with at least 150 reports, we computed the mean wind direction at report times and its circular correlation with the direction from the landfill to the sector's centre. The null distribution comes from copies of the wind record shifted circularly in time, which preserve its persistence: with 200 shifted copies the one-sided permutation $P$ was 0.005, the smallest attainable, and with 10,000 it was \ensuremath{\approx}~0.001. The result holds for ERA5 and on-site wind, and for population-weighted and geometric sector centres (Supplementary Fig.~\ref{fig:s1}a).

\textbf{Source location.} We estimated a source position from the reports in each month and tested the declared landfill against the mean of these monthly estimates with Hotelling's test (Fig.~\ref{fig:1}a).

\textbf{The 2025 source intervention.} Council figures show that reports fell at the same time as the share of half-hours above 7\,\textmu g\,m$^{-3}$ at every monitor (Supplementary Fig.~\ref{fig:s1}b,c). Methane peaks measured on the same analysers showed no detectable change, while H$_2$S peaks fell sharply (Supplementary Fig.~\ref{fig:s1}d); the instruments and the delivery of landfill gas did not change detectably, which is consistent with a fall in the H$_2$S source; the methane interval cannot exclude a fall of up to about half. The cover records show the capping works over the remaining open ground (Supplementary Fig.~\ref{fig:s1}e). The gas and the reports each have a single step date; the order of the two steps cannot be determined from these data (Supplementary Fig.~\ref{fig:s1}k).

\textbf{Variance shares.} The share of week-to-week variance in reports accounted for by the source estimate is stable across aggregation periods, reporting channels and scopes (Supplementary Fig.~\ref{fig:s1}f) and holds in withheld weeks and against a time-shifted source (Supplementary Fig.~\ref{fig:s1}i,j). The source estimate is built from measured gas and therefore includes the transport realised at the monitors (Supplementary Note~\ref{snote:monitormodel}). Before the 2025 source intervention, and with September--December 2024 withheld, weather accounted for more of the variance than the source (0.50 against 0.35; 37~weeks).

\textbf{Nearest monitor.} Each sector's monthly reports correlate with H$_2$S at its nearest monitor, but a single event in January 2025 dominates these correlations, so the correlations reflect a time course shared by all sectors, not a separate relation between each sector's reports and its own monitor (Supplementary Fig.~\ref{fig:s1}g,h).

\sifigpart{\includegraphics[width=\textwidth,height=0.9\textheight,keepaspectratio]{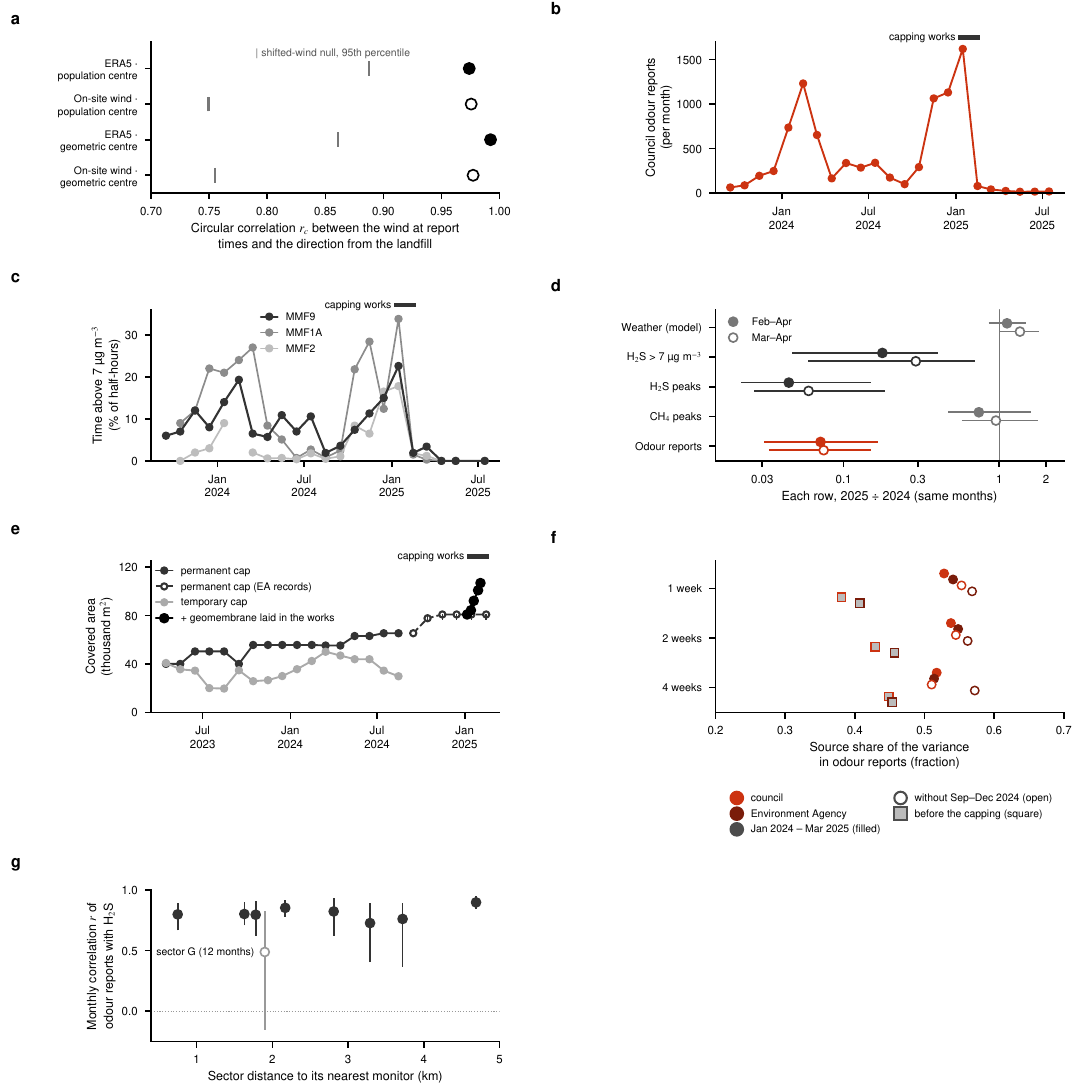}}{fig:s1}
\sifig{\includegraphics[width=\textwidth,height=0.9\textheight,keepaspectratio]{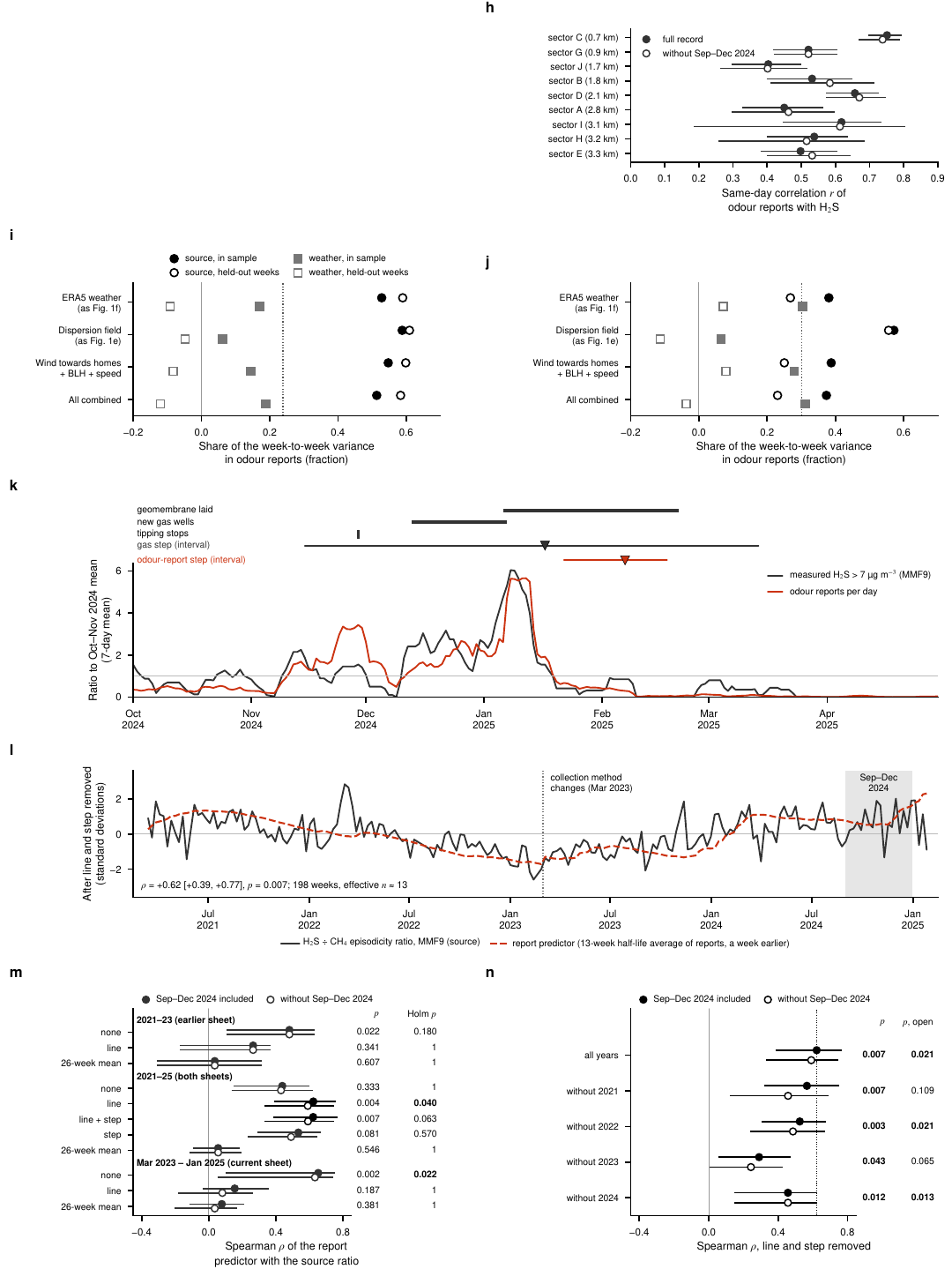}}{Source attribution and the 2025 source intervention}{fig:s1}{\input{legend_SI1}}

\snote{Reporting behaviour}\label{snote:behaviour}
This note addresses two alternative explanations for the link between reports and gas: that reports cluster because residents prompt one another, and that the link reflects a slow trend.

\textbf{Clustering.} Removing reports that arrive in bursts weakens the correlation between reports and H$_2$S more than removing the same number of reports at random, in 56 of 72 declustering variants and in the pre-specified test at MMF9 and MMF2 (Supplementary Fig.~\ref{fig:s2}f,i). Reports later in a burst, including those from other sectors, are almost as well aligned with the wind from the landfill as the first report, and reports from other sectors inside the plume arrive within 15 minutes no more often than exposure predicts (Supplementary Fig.~\ref{fig:s2}i). These results are consistent with clusters of reports arising when the plume passes over a sector. Contagion inside the plume was not tested, and the bound on cross-sector reporting that is unrelated to the wind is weak, so these results do not rule out contagion within the plume.

\snote{The odour-impact factors and the psychometric reporting model}\label{snote:fidol}
This note supports the Results section on the odour-impact factors (Fig.~\ref{fig:2}; Supplementary Fig.~\ref{fig:s2}). The fitting of the psychometric reporting model is described in Supplementary Methods~\ref{smeth:psych}.

\textbf{Definitions.} Regulators define offensiveness by the hedonic tone of an odour and location by the sensitivity of the receiving land\cite{EAH4_2011,IAQM2018}. Here, offensiveness is represented by the community's reporting threshold relative to published odour thresholds, and location by where residents were when they reported; both are proxies.

\textbf{Frequency definitions.} We specified 84 monthly definitions of an odour episode before testing them. The definition shown in Figs~\ref{fig:1}c and~\ref{fig:2}a is the strongest. Its 99\,\% interval from resampling months, 0.82–0.99, is narrower than an interval that treats the 15 months as independent, [0.69,~0.98]; because resampling autocorrelated months would normally widen an interval, the wider one is the more cautious summary. At the daily scale, all 32 combinations of four regulatory frequency definitions, two gas definitions and two windows give correlations with lower bounds above zero (Supplementary Fig.~\ref{fig:s2}a), and frequency and duration rise steadily with the number of hours above 7\,\textmu g\,m$^{-3}$ in a day (Supplementary Fig.~\ref{fig:s2}b,c). Council officers' detection of landfill odour moves with the reports, but their visits are partly prompted by reports, so this is co-movement rather than independent confirmation (Supplementary Fig.~\ref{fig:s2}d).

\textbf{Psychometric reporting model.} The complementary log--log link fits slightly better than the cumulative normal and moves the midpoint on the six-hour mean (Supplementary Fig.~\ref{fig:s2}e). The threshold on the six-hour peak comes from a joint profile over two periods with the spread estimated, which gives the interval quoted in the main text. In the separate profiles the spread was fixed at 0.66; in one of them (September--December 2023) the criterion we had set in advance for abandoning the threshold estimate, an upper bound above 20\,\textmu g\,m$^{-3}$, was exceeded by 1.1\,\textmu g\,m$^{-3}$, which we treat as a sensitivity to the design threshold. Outside the main window, in March--August and September--December 2023, the six-hourly psychometric reporting model is consistent with the WHO guideline and not with 56\,\textmu g\,m$^{-3}$ (the historical mean detection threshold multiplied by the detection-to-annoyance factor), and a model with its threshold fixed at 7\,\textmu g\,m$^{-3}$ retains almost all of the gain of the free fit. The detection baselines are the historical mean threshold of 8\,ppb (Amoore, through the California Office of Environmental Health Hazard Assessment) and a modern value of 0.41\,ppb\cite{Nagata2003}, converted at 1.417\,\textmu g\,m$^{-3}$ per ppb (20\,$^\circ$C); expected annoyance lies 4--5 times above detection\cite{Amoore1985}.

\textbf{Comparison with time above the guideline.} The psychometric reporting model predicted daily reports better than the daily share of time above 7\,\textmu g\,m$^{-3}$ (difference in mean squared error \ensuremath{-}0.22~[\ensuremath{-}0.33,~\ensuremath{-}0.12]; 591~days).

\textbf{Detection thresholds.} Translated into the exposure of residents within the plume, reporting began at 0.30~[0.19,~0.60]\,\textmu g\,m$^{-3}$, within the range of human odour detection. Whether the community is more or less sensitive than expected depends on which detection threshold is taken as the baseline: its threshold is 4–6\,times below the expected annoyance level with the historical mean detection threshold, and 3–4\,times above it with the modern value of 0.58\,\textmu g\,m$^{-3}$. The regulatory definitions of offensiveness and location concern the hedonic tone of the odour and the sensitivity of the land that receives it\cite{EAH4_2011,IAQM2018}.

\textbf{Gas tiers.} The daily report tiers are not interchangeable with gas-concentration tiers: one day ahead, forecast tiers agree with gas tiers with $\kappa=0.42$, close to the value of 0.41 for persistence, and the top tier has not been checked against gas outside the days used to select it. On the same day, recorded report tiers (3, 9 and 27 reports a day) agree moderately with gas tiers anchored at the WHO odour annoyance guideline; forecast tiers one day ahead, however, agree with gas tiers no better than persistence does. An equal-weight average of the monitors is a poor measure of exposure, because each monitor sits in its own microclimate; In March--August 2023, monthly report frequency was not detectably correlated with H$_2$S at MMF1 and MMF2 alone.

\sifig{\includegraphics[width=\textwidth,height=0.9\textheight,keepaspectratio]{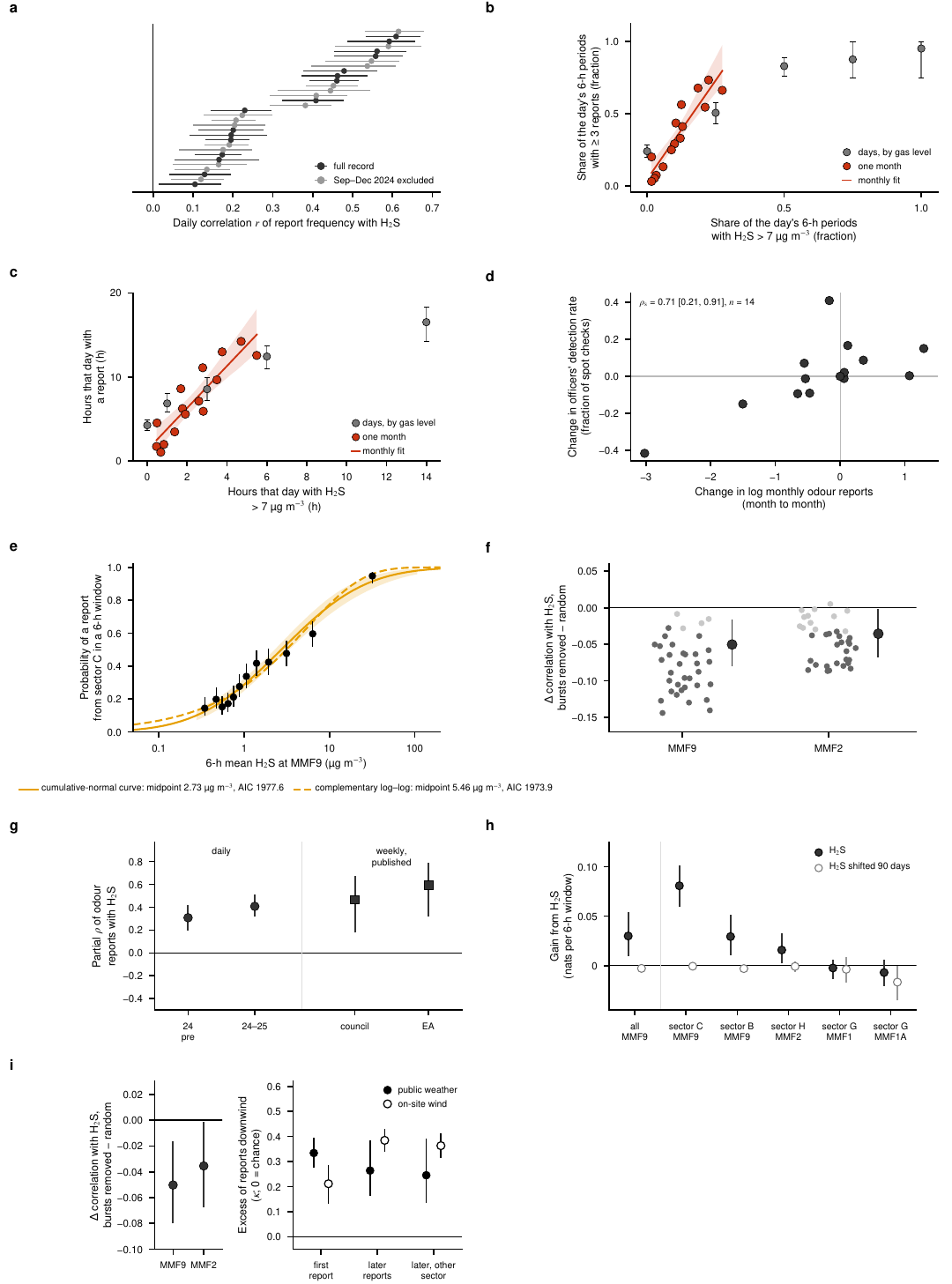}}{Odour-impact factors, the psychometric reporting model and reporting behaviour}{fig:s2}{\input{legend_SI2}}

\snote{Information in the reports beyond weather}\label{snote:beyondweather}
This note supports the statement that daily reports were correlated with H$_2$S beyond season, weekday, weather and recent reporting (Results section on the odour-impact factors). We added H$_2$S to a calibrated model of reports that already included these terms and measured the gain in log-likelihood on withheld days. The gain is positive over the forecast window for all sectors with MMF9 and larger for sector~C alone, but absent for the sector~G pairs; with H$_2$S shifted by 90 days it disappears (Supplementary Fig.~\ref{fig:s2}h). The partial correlation of daily reports with H$_2$S beyond the same terms is positive in 2024 before the 2025 source intervention, over 2024--25, and in the council's and the Environment Agency's published weekly figures (Supplementary Fig.~\ref{fig:s2}g); it is also positive in the earlier filing-dated record, which we report only as a separate instrument (Supplementary Note~\ref{snote:legacy}).

Three limits apply. The test uses a base model calibrated without the days it is tested on. Before the 2025 source intervention, the six-hourly gain over the calibrated base is small and its time-shifted control is not clean, which we treat as a sensitivity. In 2021, the association does not survive controlling for methane; because methane travels in the same plume, this control partly removes the exposure itself.

\snote{Weather and the dispersion chain}\label{snote:chain}
This note describes the dispersion modelling behind Fig.~\ref{fig:3}a,b and how it compares with the monitors (Supplementary Fig.~\ref{fig:s3}). The Met Office UKV forecast, issued at midnight, provides the meteorological input to CALMET, which is run without observations; CALPUFF then releases a constant unit emission from the landfill. For each six-hour window, the dispersion field comes from the shortest available lead time at or beyond the window's time since issue (Supplementary Methods~\ref{smeth:chain}), so one day ahead uses lead times of 6 to 24\,h. Concentrations are averaged over the receptors in each sector (the upper quartile and maximum are checks) and used on a logarithmic scale with a small floor. The configuration of the models is given in Supplementary Methods~\ref{smeth:chain} and Supplementary Table~\ref{tab:chain}.

Against the monitors, the modelled wind direction at MMF9 has a median error of 16$^\circ$, larger on calm, stable nights, and the modelled temperature is within about 1--1.5\,K (Supplementary Fig.~\ref{fig:s3}a,b). The model reproduces the timing of the daily methane cycle but underestimates its amplitude (Supplementary Fig.~\ref{fig:s3}c). Measured gas is several times higher when the modelled plume is over a monitor, except in calm air, which CALPUFF does not represent (Supplementary Fig.~\ref{fig:s3}d). Agreement is closer for methane than for H$_2$S at every monitor, and one possible explanation is a shortfall in the modelled H$_2$S source (Supplementary Fig.~\ref{fig:s3}e). With nothing fitted, the dispersion field agrees with each monitor's H$_2$S, but its skill differs between monitors by a factor of about 1.8--2.3 under identical physics; one possible explanation is local effects that a 100\,m grid cannot resolve (Supplementary Fig.~\ref{fig:s3}f). At three of the four monitors, adding the dispersion field of the monitor's sector to the state-space model of reports improved its agreement with that monitor's H$_2$S, with no monitor data used (Supplementary Fig.~\ref{fig:s3}g). Nights with pooling of cold air can be identified from the public weather analysis without the monitors. Supplementary Movie~1 shows the hourly dispersion field together with the reports and the monitored gas over September 2024 to April 2025.

\sifig{\includegraphics[width=\textwidth,height=0.9\textheight,keepaspectratio]{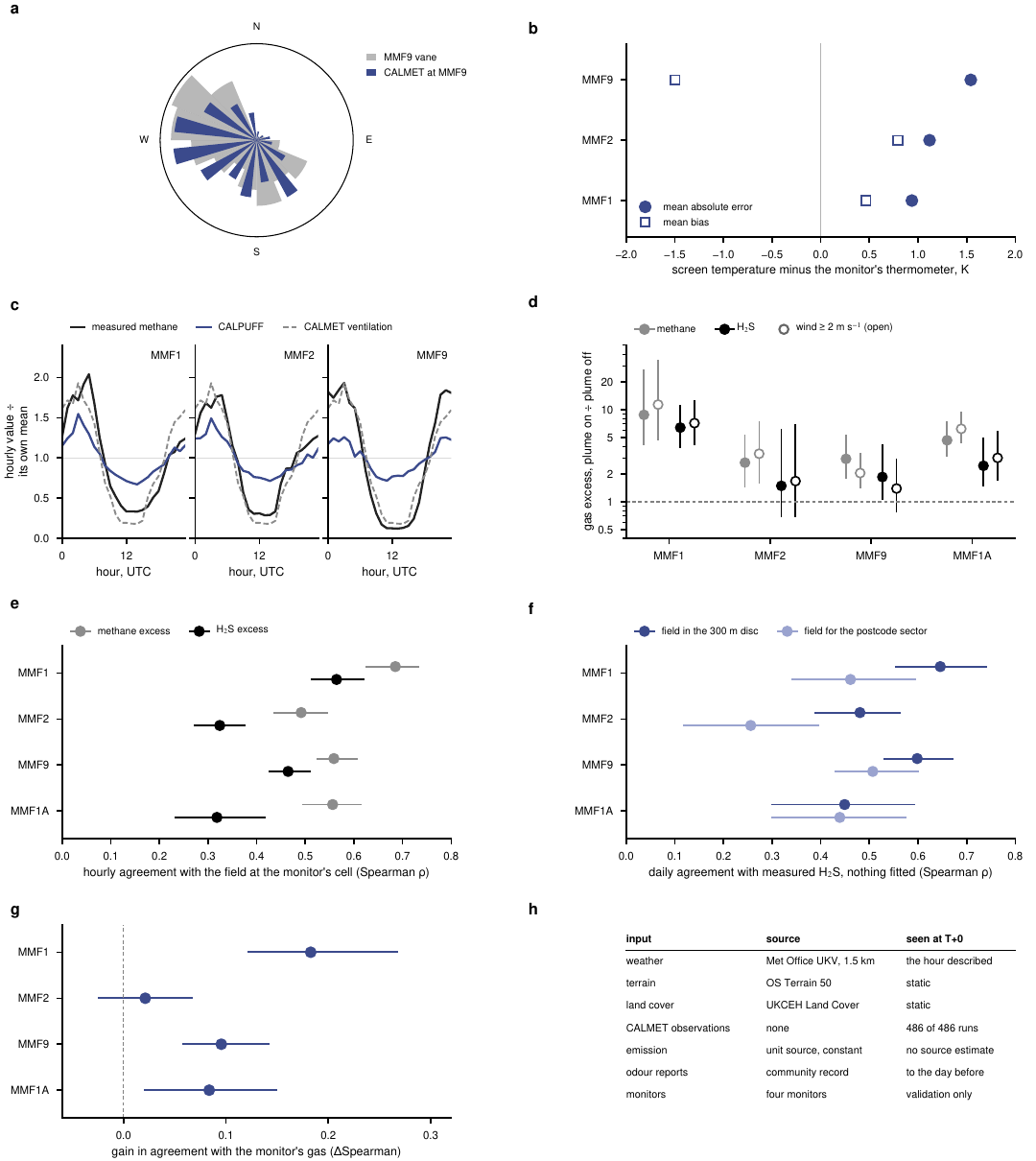}}{Dispersion modelling compared with the monitors}{fig:s3}{\input{legend_SI3}}

\snote{The exposure estimate}\label{snote:exposure}
This note gives the details behind the Results section on exposure estimated from the reports (Fig.~\ref{fig:3}d; Extended Data Fig.~\ref{fig:ed2}). The source state and the exposure calculation are specified in Supplementary Methods~\ref{smeth:psych}.

\textbf{Source state from reports.} The latent source state is estimated by an exact filter on a grid of states, fitted first with a constant source and then with its dynamics. On simulated reports generated from a known source, the filter recovers that source.

\textbf{Gas estimated from reports.} A joint inversion of the sector reports reproduces the gas measured at MMF9 and, more weakly, at MMF2; MMF1 and MMF1A were tested only one sector at a time, and a simple count of the sectors reporting performs as well.

\textbf{Agreement with measured exposure.} For weekly values, the Spearman correlations between report rates and measured exposure were 0.875 in sector~C and 0.712 in sector~G (weekly levels over 52~weeks; the week-on-week changes in the main text are a different statistic). The estimates varied with residents' readiness to report: in the worst month, the estimate was 4.1 times too low in sector~C and 4.9 times too high in sector~G.

\textbf{Calibration.} The main estimate places the WHO guideline at the midpoint of the psychometric reporting model and uses the published spread of 0.55. When we calibrated instead to human detection thresholds, the estimates fell to 0.18 and 0.41 times the measured level; placing the midpoint at a laboratory annoyance level of five times the detection threshold gives levels several times too low. We expect a community's spread to be at least as wide as published spreads, which would make the uncertainty in level mainly one-sided. Summed over the three monitors, the estimate was 1.29~[0.96,~1.88] times the measured exposure, but this total hides errors of opposite sign in the two postcodes (0.79~[0.54,~1.42] in sector~C and 1.76~[1.29,~2.57] in sector~G on all of its monitors; 1.70\ensuremath{\times} on its unchanged monitor MMF2).

\textbf{Between postcodes.} The dispersion model ranks the two monitored postcodes in the opposite order to their measured exposure (Extended Data Fig.~\ref{fig:ed2}d). Month-to-month changes in the difference between the two postcodes' report rates are correlated with the monitors' difference on the full record (Spearman 0.55~[0.09,~0.85], 28 months), but not with September--December 2024 withheld or before sector~G's monitors changed; we report it as a comparison only.

\textbf{After the 2025 source intervention.} The estimate fell further than the measured exposure in both postcodes (0.42~[0.25,~0.72],~p~0.014 in sector~C and 0.38~[0.22,~0.70],~p~0.022 in sector~G), and at sector~G's unchanged monitors the ratio was 0.59~[0.31,~1.16],~p~0.143, an interval that includes 1.

\textbf{Readiness to report.} The Environment Agency's share of reports varies more than counting noise would produce (main text), so residents' readiness to report changed over time; we report its direction only. Ranked by distance from the source alone, the ten sectors are ordered much as their report rates are, because the long-term pattern of the dispersion field is set mainly by distance and bearing. Week-to-week changes in where reports come from, beyond this geography, could not be estimated reliably at this sample size.

\snote{The monitor model and the source estimate}\label{snote:monitormodel}
This note supports the check of the dispersion field at the monitors (Fig.~\ref{fig:3}c) and the source estimate used in Fig.~\ref{fig:1}f. The monitor model uses each monitor's own past H$_2$S, the dispersion field at its grid cell and weather terms, with a shared source state and a local state for each monitor (Methods). No odour reports enter it. It tests whether the dispersion field adds information near the source; it is not designed as a better same-day estimate of the gas, and on the day six-hour persistence of the monitors' own readings is more accurate (skill 0.806 against 0.716). The source estimate treats H$_2$S and methane at each monitor and period as sharing one transport term and models the two gases as a random walk across periods. Because it is built from measured gas, it includes the transport realised at the monitors. Both models are specified in Supplementary Methods~\ref{smeth:monitor}.

\snote{The same-day estimate}\label{snote:nowcast}
This note gives the details behind the same-day estimate results (Fig.~\ref{fig:4}; Extended Data Fig.~\ref{fig:ed3}). The same-day estimate is the six-hour state-space model run with the observed weather and trained on earlier days. Its sector map blends the dispersion model's shares with the reports of the previous six hours; the blend weight is trained on earlier days, is zero in January 2024 when there are no training days, and is at most 0.30 thereafter. Six-hour estimates are more closely correlated with the preceding period than with the target period (Extended Data Fig.~\ref{fig:ed3}b), so the six-hour gains over persistence describe periods already observed. At the daily scale the same-day estimate and persistence have similar skill in all four settings, and, judged by how it ranks the sectors each day, the blended map is less accurate than the previous six hours' reports, because its model part is updated only with reports from earlier days; by daily cosine agreement it is more accurate (Results). The model and the comparison models are specified in Supplementary Methods~\ref{smeth:sentinel} and~\ref{smeth:comparators}.

\snote{Choosing SENTINEL}\label{snote:selection}
This note describes how SENTINEL was selected (Methods; Results sections on the same-day estimates and the forecasts; Supplementary Fig.~\ref{fig:s4}). The candidates were 226 configurations formed from 12 state-space-model bases, several half-lives of the error correction and several volume calibrations. The rule, set before scoring, was to take the configuration with the highest tier agreement among those whose correlation with reports peaked on the target day on the first half of the days, and then to confirm it on the second half, on all days and with training on earlier days only. The selected model is trained on day-to-day change with a weight of 1.5 and a 15-day half-life and has no volume calibration; adding a volume calibration moved the correlation peak back to the previous day in most configurations. The model trained on daily levels with the same settings scores slightly higher tier agreement but peaks on the previous day. The design of the model and the full search are in Supplementary Methods~\ref{smeth:sentinel} and Supplementary Table~\ref{tab:search}.

The selection used the scored days, and we made no adjustment for it; a selection-adjusted one-day margin, computed with short blocks, has a calibrated $p$ of 0.024. The advantage of training on day-to-day change depends on the report history being up to date: when the history is closed 24 hours before issue, the timing advantage over the model trained on daily levels falls close to zero (Supplementary Fig.~\ref{fig:s4}c). Beyond one day ahead, neither the timing statistic nor the paired advantage distinguishes the two training methods. Shortening the report history available to the model trained on daily levels improved it almost as much as training on day-to-day changes did, so this comparison gives no separate evidence for training on changes. The month-by-month ratio of forecast to recorded reports and the agreement of the forecast sector map are shown in Supplementary Fig.~\ref{fig:s4}b,e.

\sifig{\includegraphics[width=\textwidth,height=0.9\textheight,keepaspectratio]{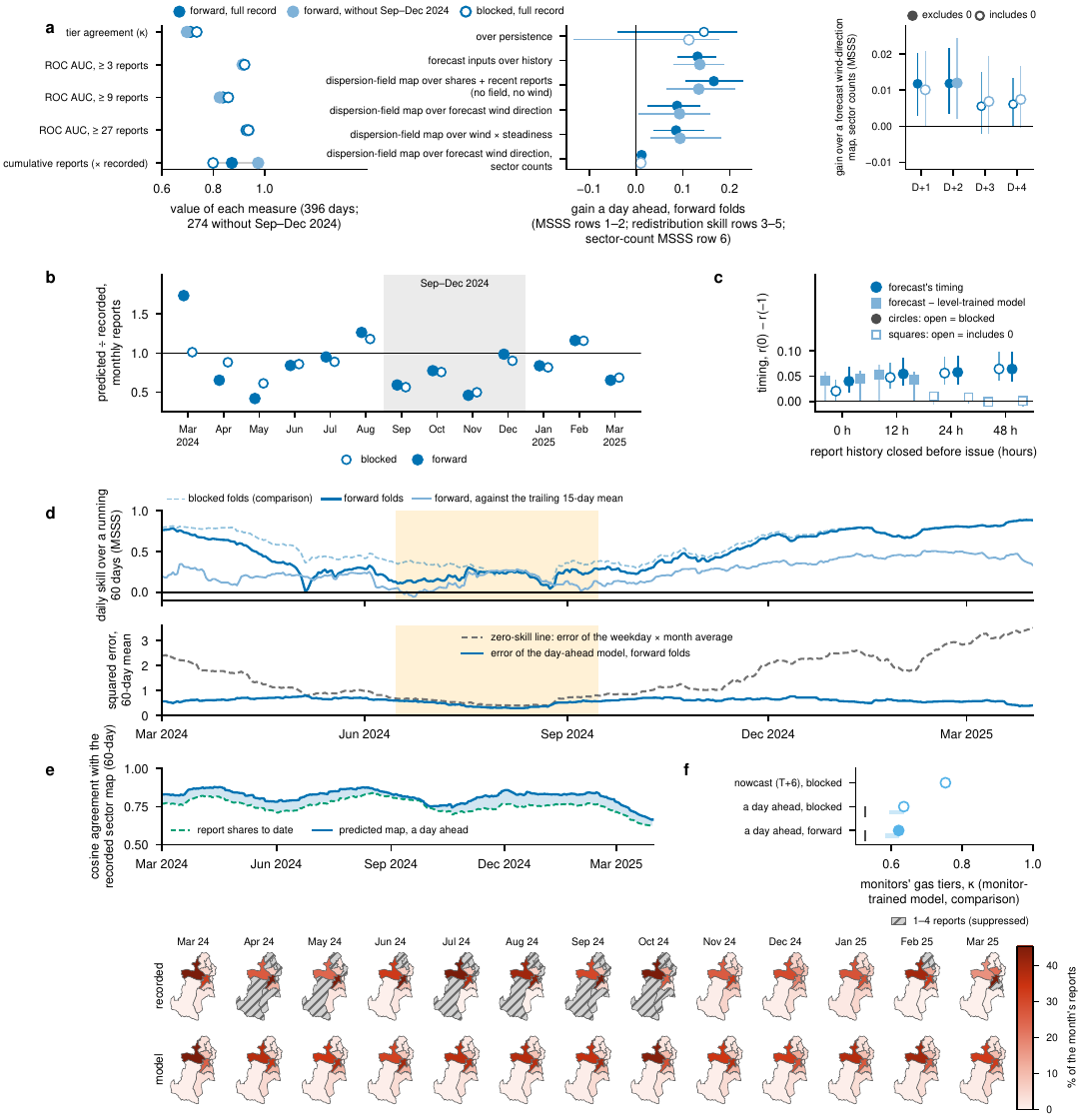}}{One-day-ahead forecasts in three evaluation settings}{fig:s4}{\input{legend_SI5}}

\snote{Comparison models}\label{snote:comparators}
This note describes the comparison models behind the fifth Results section (Fig.~\ref{fig:6}; Supplementary Fig.~\ref{fig:s5}). Persistence repeats the last recorded day; for the same-day estimate it repeats the last six hours with their average daily cycle. The report-history model is SENTINEL without the weather forecast or the dispersion field. Skill against a trailing 7- or 15-day mean, instead of the weekday and month average, gives the same ordering of models (Supplementary Fig.~\ref{fig:s5}d). The comparison models are defined in Supplementary Methods~\ref{smeth:comparators}.

SENTINEL was compared with a regression on report history and calendar. When refitted on SENTINEL's schedule, the regression had lower daily skill than SENTINEL in 14~of~16 combinations of lead time and setting. When refitted every day, which gives it access to more recent data than SENTINEL uses, it did not differ from SENTINEL in 5~of~16. Adding the weather forecast and the dispersion field to the regression increased its skill one and two days ahead in six~of~eight combinations and in none three and four days ahead. A regression given the same weather had higher one-day skill when trained on earlier days only, but its interval included zero and its correlation was not highest on the target day, so it is reported as a comparison only. In a factorial comparison, the dispersion field and the weather each add skill to the report history, and the dispersion field improves the allocation of reports among sectors (Supplementary Fig.~\ref{fig:s5}h). For the ten-sector daily total, the dispersion field added at most 0.03 in daily skill beyond the weather forecast (upper bound; equivalence at 0.02 not shown), whereas the weather adds 0.01–0.03 beyond the dispersion field. When the allocation uses only each sector's recent reports and no weather, adding the dispersion field improves the division of each day's reports among the ten sectors at every lead time, and nine of ten sectors gain.

After the 2025 source intervention, forecast report counts fell in line with the lower recorded counts, within a factor of two of the recorded level in each of February, March and April 2025 (Supplementary Fig.~\ref{fig:s4}b). Persistence and the model trained on daily levels, which also use the report history, fell in the same way, so the forecast's response to the intervention comes from using the report history. Whether the fall reflects a change in emissions or in reporting cannot be determined from these data: weekly comparisons that remove the weather do not separate the two.

\sifigpart{\includegraphics[width=\textwidth,height=0.9\textheight,keepaspectratio]{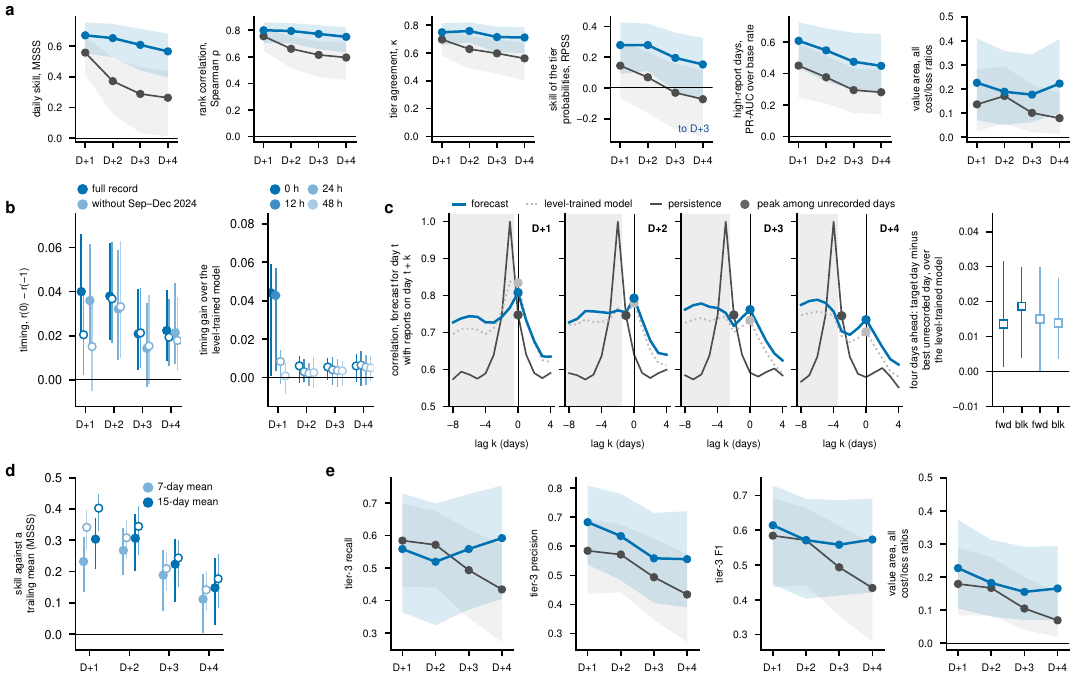}}{fig:s5}
\sifig{\includegraphics[width=\textwidth,height=0.9\textheight,keepaspectratio]{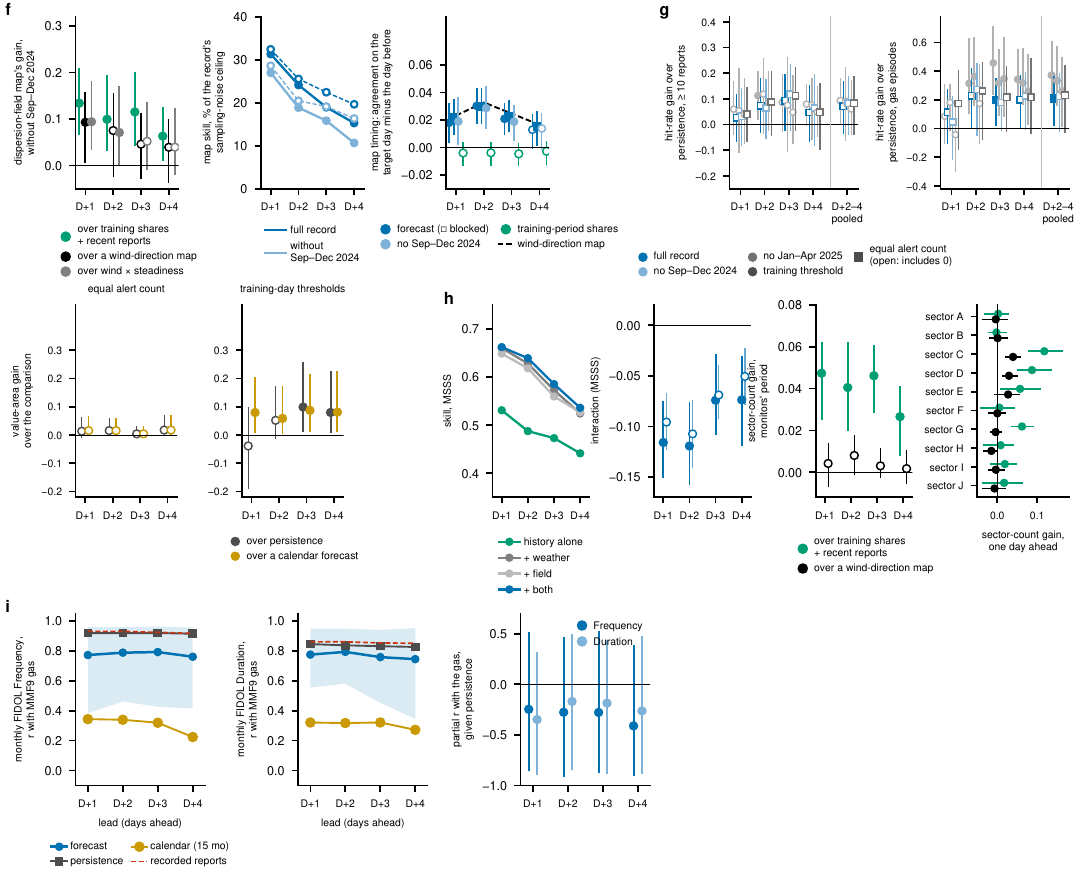}}{Forecast performance by lead time, comparison models, sector allocation and alerts}{fig:s5}{\input{legend_SI6}}

\textbf{Sector division two days ahead and timing.} For the division of reports among sectors, the gain over the shares of the last 28 recorded days was +0.112~[+0.026,~+0.221] three days ahead. Two days ahead, the interval method for this comparison did not reach acceptable coverage in simulations with a known effect, so the value in the Results (+0.135~[+0.044,~+0.240]) carries that limit. The dispersion-model and wind-direction forecasts were equally well timed, except in one of 16 comparisons, in which the wind-direction forecast was slightly better timed.

\snote{Alerts and economic value}\label{snote:alerts}
We do not rely on the gain in identifying days of high gas concentration at an equal number of alerts, because it is not robust to the training scheme or to the period withheld. The intended uses, with the evidence and limit of each, are in Supplementary Table~\ref{tab:uses}.

\begin{table}[tbp]\centering\scriptsize
\caption[Intended uses of the forecast and the exposure estimates]{\textbf{Intended uses of the forecast and the exposure estimates.} For each use: the user and decision, the product and its lead, the verification result and the limit. Timing within a monitored postcode and relative exposure across months come from the reports; an absolute level, and the size of an intervention's effect, need a monitor in the sector concerned.}\label{tab:uses}
\begin{tabular}{p{1.6cm}p{2.4cm}p{2.0cm}p{1.2cm}p{3.6cm}p{3.0cm}}\toprule
User & Decision & Product & Lead & Verification & Limit \\\midrule
Site operator & Schedule odour-generating work (waste handling, capping) away from high-impact days & Daily report count and alert & 1--4 days & Daily skill 0.662 (D+1) to 0.536 (D+4); 0.642 / 0.500 with operational data delays; gain over the climatology--persistence blend +0.23 [+0.12, +0.34] at D+4 (trained on earlier days) & Most alerts are false; no hit-rate gain at a single lead is robust \\
Site operator or regulator & Which sectors to protect or inspect & Map of the division of reports among sectors & 1 day (2--3 days: intervals include zero) & Gain over a forecast wind-direction map +0.088 [+0.026, +0.136]; over a Gaussian plume from the same forecast +0.125 [+0.079, +0.170] & Two days ahead the gain over the wind map has an interval that includes zero \\
Regulator & Track odour impact against permit conditions (frequency and duration) & Monthly frequency and duration from reports & Monthly, after the fact & Monthly r 0.92 (strongest of 84 definitions; 0.85 fixed in advance) and 0.90 with H$_2$S at MMF9 (0.90 and 0.89 on a 30-minute basis) & Intensity, offensiveness and location only approximated \\
Environmental-health team & Exploratory exposure--health studies & Weekly or monthly exposure by postcode & After the fact & Monthly Spearman 0.87 (C) and 0.77 (G) with measured H$_2$S; level 0.38--1.33 (sector C) and 0.69--2.40 (sector G) times measured, across conversions & Relative without a monitor; timing better than level; intervention effects need a monitor in the sector concerned \\
\bottomrule\end{tabular}
\end{table}

This note gives the alert and value results summarised in the Results section on the forecasts (Extended Data Fig.~\ref{fig:ed6}b; Supplementary Fig.~\ref{fig:s5}e,g). Economic value is the relative value of acting on a forecast over all ratios of the cost of protective action to the loss avoided\cite{zhu2002}. The formulas for economic value and the alert comparison are in Supplementary Methods~\ref{smeth:verification}.

At an exactly equal number of alerts, forecast alerts issued two to four days ahead identified a larger share of days with gas episodes than persistence alerts (difference 0.210, $p=0.039$), although most alerts of both kinds were false. This difference was not consistent across training schemes or periods: under block cross-validation and with September--December 2024 withheld it was statistically indistinguishable from zero, and after Holm correction over the pre-specified events and the equal-count comparisons no difference remained. At equal alert counts we found no credible evidence that acting on the forecast is worth more than acting on persistence; a value advantage appeared only when the forecast issued about 1.5 times as many alerts as persistence. With thresholds set on training days, as a real-time service would set them, forecast alerts identified more episode days than 500 random sets of alerts of the same size in all four settings, whereas persistence did so only under block cross-validation (Supplementary Fig.~\ref{fig:s5}g).

\snote{Monthly Frequency and Duration a day ahead}\label{snote:monthly}
This note supports the statement about monthly factors forecast a day ahead (Extended Data Fig.~\ref{fig:ed4}a; Supplementary Fig.~\ref{fig:s5}i). Monthly frequency and duration forecast one day ahead were correlated with monitored H$_2$S ($r=0.77$ and 0.78), as were persistence forecasts over the same months (0.92 and 0.85). The forecast was more closely correlated with the gas than a calendar forecast (Williams test, which assumes independent months), but after Holm correction over the eight comparisons no difference remained. After allowing for persistence, the partial correlation of the forecast with the gas has an interval that includes zero, and a monthly gain beyond persistence could have been detected only above about 0.7. Persistence's advantage is shown for frequency but not for duration. Without September--December 2024, the forecast remains more closely correlated with the gas than the calendar forecast under block cross-validation, and the difference is borderline when trained on earlier days only. The monthly relation therefore reflects the slowly varying level of reporting, which is also present in the forecast; the daily relation, unlike the monthly one, has its own timing test (Methods).

\snote{Local instruments and the forecast}\label{snote:instruments}
This note supports the statement that adding the local measurements tested did not detectably improve the forecast (Extended Data Fig.~\ref{fig:ed1}). We added each of 40 single local measurements, known at issue, to SENTINEL: monitor gas, source estimates and source terms, monitor wind, a forecast wind calibrated to the monitor, the target day's measured weather (a reference that is not known at issue) and network anchors, the mean of the monitor network over the previous 28 or 84 days. For all 40 forms, a gain in daily skill above 0.02 is excluded one and two days ahead (about an 87\,\% one-sided bound); for the 84-day network anchor trained on earlier days, the upper bound one day ahead is 0.027. Month by month, a reduction in monthly level error of more than about 0.014--0.025 is excluded one day ahead on the full record. The monthly test's 80\,\%-power detectable effect is 0.036 (up to 0.048). Drift in residents' readiness to report over several years remains untested. The monitors remain necessary for validation, for the units of exposure, for defining gas episodes and for attributing a change in reports to the source (main text). The equivalence tests are described in Supplementary Methods~\ref{smeth:intervals}.

\snote{Intervals, coverage and registered tests}\label{snote:intervals}
Each four-month period of the scored window was withheld in turn: the correlations, skill and comparator gains of the summary paragraph were re-estimated without it, and the forecast was trained without it and scored on it (Supplementary Table~\ref{tab:periods}).

\begin{sidewaystable}\centering\scriptsize
\caption[Each four-month period withheld]{\textbf{Each four-month period withheld.} Upper part: the correlations, skill and comparator gains of the summary paragraph and the main forecast measures re-estimated with each four-month period of the scored window withheld in turn (jackknife; the worst period beside each). Lower part: the forecast trained with each period withheld and scored on it (published configuration; 7-day block 95\,\% intervals; information beyond recorded days, 56-day blocks). The four periods confound season with year; neither check is an unseen period or a second site.}\label{tab:periods}
\begin{tabular}{p{8.0cm}rrrrrl}\toprule
Quantity & All & \multicolumn{4}{c}{Period withheld} & Worst \\
 & & Jan--Apr 2024 & May--Aug 2024 & Sep--Dec 2024 & Jan--Apr 2025 & \\\midrule
Monthly frequency relation, $r$ (Fig.~\ref{fig:1}c) & 0.92 & 0.91 & 0.93 & 0.94 & 0.91 & 0.91 (Jan--Apr 2025) \\
Monthly duration relation, $r$ & 0.90 & 0.90 & 0.92 & 0.90 & 0.87 & 0.87 (Jan--Apr 2025) \\
Partial $r$, daily reports with H$_2$S beyond weather & 0.41 & 0.39 & 0.47 & 0.39 & 0.31 & 0.31 (Jan--Apr 2025) \\
Monthly Spearman, report rate with measured H$_2$S, sector~C & 0.87 & 0.83 & 0.91 & 0.85 & 0.84 & 0.83 (Jan--Apr 2024) \\
Reporting threshold, six-hour peak of 30-min means (\textmu g\,m$^{-3}$) & 11.46 & 8.29 & 9.74 & 11.20 & 12.51 & 8.29 (Jan--Apr 2024) \\
Daily skill one day ahead (trained on earlier days) & 0.66 & 0.68 & 0.71 & 0.67 & 0.56 & 0.56 (Jan--Apr 2025) \\
Daily skill four days ahead (trained on earlier days) & 0.54 & 0.54 & 0.60 & 0.57 & 0.41 & 0.41 (Jan--Apr 2025) \\
Information beyond recorded days, one day ahead (trained on earlier days) & 0.48 & 0.44 & 0.51 & 0.47 & 0.50 & 0.44 (Jan--Apr 2024) \\
Gain over damped persistence, four days ahead (trained on earlier days) & 0.21 & 0.22 & 0.25 & 0.19 & 0.15 & 0.15 (Jan--Apr 2025) \\
Gain over the climatology--persistence blend, four days ahead & 0.23 & 0.19 & 0.27 & 0.28 & 0.18 & 0.18 (Jan--Apr 2025) \\
\midrule
\multicolumn{7}{l}{\emph{Trained without the period, scored on it}} \\
Daily skill on the withheld period, one day ahead & 0.71 & 0.72 [0.59, 0.80] & 0.35 [0.06, 0.45] & 0.67 [0.51, 0.76] & 0.83 [0.74, 0.89] & \\
Daily skill on the withheld period, four days ahead & 0.55 & 0.52 [0.26, 0.69] & 0.07 [$-$0.25, 0.17] & 0.47 [0.19, 0.63] & 0.75 [0.58, 0.86] & \\
Information beyond recorded days on the withheld period, one day ahead & 0.52 & 0.48 [0.05, 1.38] & 0.39 [0.20, 0.73] & 0.57 [0.43, 0.76] & 0.55 [0.43, 0.64] & \\
\bottomrule\end{tabular}
\end{sidewaystable}

This note supports every interval in the paper (Methods). For each family of paired comparisons we measured the coverage of candidate interval methods in two ways: with a known effect added to the real data in simulation, and with a check that uses no generating model. The candidates were moving-block percentile, basic and bias-corrected bootstrap intervals with blocks of 7 to 112 days, and subsampling. We chose, before any result, the method with the highest minimum coverage under the two checks. No method reached 0.94 coverage on these records; measured coverage ranged from 0.63 for sector allocation to 0.91 for paired timing, and the coverage of each family is given with its intervals in the figure legends and Methods. Two results were computed with short blocks and were not recalculated with the chosen method: the difference between lead times and the selection-adjusted one-day margin (Methods). The interval methods and the coverage checks are described in Supplementary Methods~\ref{smeth:intervals}.

Nine further tests were specified after the first results had been reviewed, each with its prediction and before it was run. No family-wise correction is applied except where a Holm adjustment over a stated family is given, and results whose lower bound lies within 0.01 of zero are marked.

\section*{Supplementary Data}\addcontentsline{toc}{section}{Supplementary Data}
\noindent\textbf{Supplementary Movie 1 (separate file, MP4). Hourly dispersion field, weather, odour reports and monitored H$_2$S, September 2024 to April 2025.}\label{movie:s1} Each frame shows one hour of the dispersion field $\Phi$ from the public UKV forecast (unit emission, nothing fitted) over the ten postcode sectors, with the forecast wind, the odour reports made in each sector and the H$_2$S measured at the monitors, which SENTINEL does not use. The frame for 10 January 2025, 17:00 UTC, the hour with the most reports in January--April 2025, is shown in Fig.~\ref{fig:3}b. The file will be uploaded with the submission or made available at a public repository (Data availability).

\end{document}

%% file: legend_Fig1.tex
\textbf{a}, Postcode sectors A–J shaded by odour reports per 1,000 residents (January 2024 – March 2025). Roses, wind at report times (45\textdegree{} bins); arrows, mean report-time wind (dashed, outside the rotation test); grey bars, direction of the landfill. The source located from the reports lies 141 m from the landfill (\textit{p} = 0.93). \textbf{b}, Mean report-time wind direction against the direction that would carry the landfill's odour to each sector (eight sectors with \ensuremath{\geq} 150 reports; filled, ERA5; open, on-site wind): circular correlation \textit{r} = 0.974; one-sided permutation test against 10,000 time-shifted wind records, \textit{P} \ensuremath{\approx} 0.001. \textbf{c}, Monthly share of six-hour periods with H\ensuremath{_2}S above 7 \textmu{}g m\ensuremath{^{-3}} at ambient air quality monitor (AAQM) MMF9 (black) and with at least three reports (crimson), each \ensuremath{\div} its mean; shaded, the capping works; \textit{r} = 0.92 for the strongest of 84 episode definitions (99 \% month-bootstrap interval 0.82–0.99, not adjusted for this selection; 15 months); 0.91–0.94 leaving out any one month; 0.85 for the definition fixed in advance. \textbf{d}, Partial correlation of daily reports with each pollutant at MMF9, beyond weather, season, weekday and recent reports (H\ensuremath{_2}S 0.41; 0.37 on unstable and 0.40 on stable nights); the particle bars are one weather signal. \textbf{e}, Ratio after \ensuremath{\div} before the 2025 source intervention (February–April 2025 \ensuremath{\div} 2024; 95 \% block-bootstrap intervals): reports fell fourteen-fold (6.0–32.4), H\ensuremath{_2}S half-hours six-fold and 30-minute H\ensuremath{_2}S peaks to 0.04 [0.02, 0.14], while 30-minute methane peaks (0.77 [0.47, 1.51]) and modelled transport (1.12 [0.86, 1.48]) showed no detectable change. Squares, H\ensuremath{_2}S \ensuremath{\div} methane peaks, February–April 2025 \ensuremath{\div} October–December 2024: wind from the landfill 0.10 [0.06, 0.18], other hours 0.15 [0.10, 0.25]. \textbf{f}, Shares of week-to-week variance in log(1 + reports), 65 weeks, January 2024 – March 2025 (open, September–December 2024 withheld): the source estimate accounts for the largest share (0.53 [0.32, 0.66]). \textbf{g}, Daily correlation of each sector's reports with H\ensuremath{_2}S at its nearest monitor (nine sectors; colour, distance): for every sector the correlation is highest when reports and H\ensuremath{_2}S are taken on the same day.

%% file: legend_Fig2.tex
\textbf{a}, The five odour-impact factors (FIDOL) against monitor MMF9, January 2024 – March 2025. Frequency: monthly share of six-hour periods with at least 3 reports against the share with H\ensuremath{_2}S above 7 \textmu{}g m\ensuremath{^{-3}}; \textit{r} = 0.92 (99 \% 0.82–0.99; 15 months; band, 95 \% month bootstrap). Intensity: the reporting threshold for each reported intensity score \ensuremath{\div} the threshold for any report (388, 666 and 1,906 sector-windows per plotted score); reported intensity is only weakly related to H\ensuremath{_2}S (Spearman 0.28 [0.22, 0.35]). Duration: monthly hours per day with a report against hours above 7 \textmu{}g m\ensuremath{^{-3}}; \textit{r} = 0.90 (99 \% 0.73–0.98). Offensiveness (a proxy, not hedonic tone): the community threshold and the expected annoyance level as multiples of the historical mean odour threshold. Location (a proxy): where 8,275 reports were made. Frequency, Intensity and Duration are correlated for a near-continuous source and are not independent confirmations. \textbf{b}, The psychometric reporting model: probability of at least one report from sector C per six-hour window against six-hour mean H\ensuremath{_2}S at MMF9 (points, 1,815 windows in twelve equal-count bins, Wilson 95 \% intervals; line, cumulative-normal fit, threshold \ensuremath{\alpha} 2.7 \textmu{}g m\ensuremath{^{-3}} on this basis; band, 3-day block bootstrap; dashed, complementary log–log curve). Strip below, reference values at their own averaging times (vertical position in the strip has no meaning): the WHO odour annoyance guideline, 7 \textmu{}g m\ensuremath{^{-3}}; this sector's threshold \ensuremath{\alpha} on the six-hour peak of hourly means, 10 [6.2–14.0] \textmu{}g m\ensuremath{^{-3}}, and on the guideline's 30-minute basis, 11.5 [5.7, 15.6] \textmu{}g m\ensuremath{^{-3}} averaged over three curve shapes (best-fitting shape 12.9 [10.1, 16.4], which excludes 7), about 1.5–2 times the guideline value and of the same order; the concentration at which the daily dose–response changes slope, 1.04 [0.68, 1.78] \textmu{}g m\ensuremath{^{-3}}; published detection and odour-threshold ranges. The threshold is specific to this sector and monitor.

%% file: legend_Fig3.tex
\textbf{a}, The modelling chain, which uses public inputs only. Dispersion model: CALMET and CALPUFF give a dispersion field \ensuremath{\Phi} for each postcode sector. Odour-report model: a state-space model of each day's total with an error correction, divided among the sectors by \ensuremath{\Phi}. Monitor check (light blue, \ensuremath{\rightarrow} \textbf{c}): each monitor's H\ensuremath{_2}S modelled from its own past with \ensuremath{\Phi}. Exposure model (orange, \ensuremath{\rightarrow} \textbf{d}), which uses no gas data: the psychometric reporting model, with its threshold at the WHO guideline, converts reports into a source state, and \ensuremath{\Phi} divides this among postcodes. The ambient air quality monitors (AAQMs) are used only to evaluate the result. \textbf{b}, \ensuremath{\Phi} at 17:00 UTC on 10 January 2025 (log scale); outlined, sectors C and G; triangles, monitors. \textbf{c}, Monitor check, 396 days, at the three monitors near the source (it checks the dispersion field near the source, not transport to the sectors): modelled against measured daily exposure, \textit{r} = 0.87, slope 0.80 [0.72, 0.86]; cumulative exposure 1.13\ensuremath{\times} [1.02, 1.28] measured. With CALPUFF, the monitor model had higher skill than with a Gaussian plume run with the same forecast (+0.032 [+0.007, +0.060]). ROC per tier of measured daily peak H\ensuremath{_2}S, AUC 0.89–0.95 (dashed, without the dispersion field); tiers \ensuremath{\kappa} = 0.75. \textbf{d}, Monthly H\ensuremath{_2}S exposure by postcode, March 2024 – April 2025: the report-derived estimate (no gas data; orange; band, 95 \%) against the monitor(s) (open, sector G's monitors after September 2024); grey, the September–December 2024 period; hatched, after the 2025 source intervention. Right, each postcode against its own monitor(s): Spearman 0.89 and 0.75 over the 14 months shown (0.87 and 0.77 over 29 months, main text); 14-month level 0.64–0.87\ensuremath{\times} measured in sector C and 1.41–1.87\ensuremath{\times} in sector G on its unchanged monitor MMF2, and 0.38–1.33 and 0.69–2.40 across averaging-time conversions. The calibration was chosen by comparison with these monitors, so the level is not an independent test; the data do not support a difference in level between the two postcodes.

%% file: legend_Fig4.tex
The weather is observed, so this is a reference run rather than a forecast; the same-day estimate is the state-space model with its error correction (Fig. 3a), trained on earlier days. Per six-hour period its skill is higher than that of the last six hours carried forward with their daily cycle (+0.102 [+0.055, +0.151], 1,944 periods), but these gains describe periods already observed (Supplementary Note 11); day by day, its skill does not differ detectably from that of the last six hours carried forward. \textbf{a}, Daily reports, ten sectors: recorded (crimson), raw state model (dotted), same-day estimate. Right, estimate against recorded (396 days): \textit{r} = 0.94, slope 0.80 [0.76, 0.84] (band, 7-day block bootstrap). \textbf{b}, Daily skill, running 60 days (MSSS; 0, the weekday \ensuremath{\times} month average). Right top, whole-record skill (488 days, each model on its own days): same-day estimate 0.89, six-hour persistence 0.87. Right bottom, correlation of the estimate for day \textit{t} with reports on day \textit{t} + \textit{k} (band, 95 \%): highest on the target day (0.949 [0.931, 0.960]). \textbf{c}, Daily cosine agreement with the recorded sector map (60-day running mean): the blended map (the dispersion model's sector shares blended with the last six hours of reports, model weight at most 0.30) 0.908 against 0.893 for the last six hours carried forward (+0.0145 [+0.0081, +0.0223], 486 days); green, report shares to date. The blend is mostly the last six hours of reports (monthly maps: Extended Data Fig. 3a). Right, cumulative reports: 0.93\ensuremath{\times} recorded. \textbf{d}, Daily tiers (3, 9 and 27 reports), 396 days: \ensuremath{\kappa} = 0.87 (carrying forward 0.86; paired gain +0.013 [\ensuremath{-}0.015, +0.026]); ROC AUC 0.97 for each tier.

%% file: legend_Fig5.tex
SENTINEL (Fig. 3a), issued at midnight from the public weather forecast and earlier reports, with no instrument at the site; trained on earlier days. \textbf{a}, What is known at issue and what is scored (the target day's four six-hour periods). \textbf{b}, Daily reports recorded (crimson) and forecast a day ahead (blue), 396 days: \textit{r} = 0.78, slope 0.70 [0.59, 0.76]; tiers \ensuremath{\kappa} = 0.71 (persistence 0.65); ROC AUC 0.92, 0.84 and 0.93 (dashed, persistence); cumulative 0.87\ensuremath{\times} recorded. \textbf{c}, Same-day estimate with observed weather (Fig. 4) to one day ahead, same days: daily skill 0.88 \ensuremath{\rightarrow} 0.63, mostly from losing the day's own reports. \textbf{d}, Information beyond the recorded report history: each forecast's partial correlation with the target day's reports after a regression on the 14 most recent recorded days, the mean of earlier days, weekday and season (one day ahead; 304 days scored with the configuration re-chosen each week from earlier days, July 2024 – April 2025; bands, 95 \%, 56-day blocks). SENTINEL (blue; filled, configuration chosen without the scored days; open, fixed): re-chosen weekly 0.50 [0.41, 0.59], on a withheld window 0.45 [0.35, 0.52], averaged over eligible configurations 0.50 [0.40, 0.59], published 0.48 [0.38, 0.57]. A regression on the same weather and dispersion inputs, 0.52 [0.42, 0.64]; the report history alone, \ensuremath{-}0.02 [\ensuremath{-}0.16, 0.12]; damped persistence \ensuremath{-}0.05 [\ensuremath{-}0.12, 0.05] and the climatology–persistence blend \ensuremath{-}0.05 [\ensuremath{-}0.15, 0.11]. \textbf{e}, Gains one day ahead (circles, trained on earlier days; squares, block cross-validation; open, interval includes zero). Daily skill over persistence +0.145 [\ensuremath{-}0.038, +0.214]; over the report history alone +0.131 (7-day blocks, [+0.089, +0.169]; interval from the chosen method [+0.057, +0.186]). Division of reports among the ten sectors (redistribution skill; coverage 0.63–0.83): over a Gaussian plume from the same forecast +0.125 [+0.079, +0.170]; over training-period shares +0.166 [+0.107, +0.228]; over a forecast wind-direction map +0.088 [+0.026, +0.136]; over wind direction \ensuremath{\times} steadiness +0.085 [+0.039, +0.143]; over the shares of the last recorded day +0.925 (off scale).

%% file: legend_Fig6.tex
SENTINEL is issued at midnight from the public weather forecast and the report record, with no instrument at the site; blue, trained on earlier days (dashed, block cross-validation); grey column, D+0, the same-day estimate with observed weather (Fig. 4). \textbf{a}, Five measures against lead: daily skill (MSSS), rank correlation, tier agreement \ensuremath{\kappa}, skill of the tier probabilities (RPSS) and high-report-day detection (PR-AUC over the base rate). Dark grey, persistence (the last recorded day; at D+0, the last six hours); green, the report history alone; dotted grey, the level-trained model (comparison); daily skill only, damped persistence (dashed, squares) and the climatology–persistence blend (dash-dot, triangles). Daily skill 0.662 [0.569, 0.730] one day and 0.536 [0.403, 0.636] four days ahead (0.642 and 0.500 with operational data delays). Small panels, gain in the same measure (shaded, 95 \%; open, interval includes zero): every measure gains over persistence from three days ahead (daily skill +0.354 [+0.248, +0.441] and +0.323 [+0.145, +0.414]); one day ahead the daily-skill gain includes zero. Four days ahead daily skill exceeds damped persistence by 0.21 [0.10, 0.30] and the blend by 0.23 [0.12, 0.34]. All 20 paired gains over persistence three and four days ahead are positive; trained on earlier days, all 10 exceed zero at the 99.9 \% level (joint-bound coverage about 0.89). Green squares, gain over the report history alone, +0.131 [+0.089, +0.169] one day to +0.095 [+0.051, +0.141] four days ahead. \textbf{b}, Information beyond the recorded report history by lead time (trained on earlier days; bands, 95 \%, 56-day blocks): SENTINEL 0.48 [0.38, 0.56] one day to 0.31 four days ahead, above zero at every lead; damped persistence and the blend \ensuremath{-}0.05 to +0.06.

%% file: legend_ED1.tex
\textbf{a}, Change in daily skill when one local measurement, known at issue, is added to SENTINEL, one to four days ahead, all days (478–486 days). Rows are gas at the ambient air quality monitors (AAQMs) and source terms, source estimates, monitor wind, the target day's measured weather (light grey; a reference that is not known at issue) and the 28- and 84-day network anchors (the mean of the monitor network over the previous 28 or 84 days). Filled markers, trained on earlier days; open, block cross-validation; bars, paired 14-day block-bootstrap 95 \% intervals (arrows, beyond the axis); shading, the pre-declared \ensuremath{\pm}0.02 margin. \textbf{b}, As \textbf{a}, with September–December 2024 withheld (356–364 days). \textbf{c}, Daily skill of SENTINEL, persistence and models without the report history (solid, trained on earlier days; dashed, block cross-validation); intervals are not shown for these levels. Across the 160 comparisons none shows an improvement after Holm adjustment (smallest adjusted \textit{P} = 0.144). A gain above 0.02 is excluded, at about 87 \% one-sided confidence, one and two days ahead for all 40 single-measurement forms; for the 84-day network anchor trained on earlier days, the upper bound one day ahead is 0.027, just above the margin.

%% file: legend_ED2.tex
\textbf{a}, \textbf{b}, Monthly series, March 2023 – July 2025, each on the log scale and standardised: odour reports per 1,000 residents (crimson) and the H\ensuremath{_2}S exposure measured at the postcode's ambient air quality monitor(s) (AAQMs; black; a monitor point standing in for the area). \textbf{a}, Sector C against MMF9. \textbf{b}, Sector G against the monitors reporting (dotted, MMF2 alone); the line breaks where the monitor set changes in September 2024. Shaded, 95 \% intervals: for reports, the exact Poisson interval of the month's count; for H\ensuremath{_2}S, the month's days resampled; grey, September–December 2024. Monthly Spearman 0.87 [0.67, 0.94] and 0.77 [0.34, 0.94] (3-month blocks), 0.89 and 0.78 with the season removed; with each four-month period withheld the lowest is 0.83 (sector C) and 0.70 (sector G). \textbf{c}, The 14-month level per postcode, estimate \ensuremath{\div} measured, against the spread of the psychometric reporting model, with intervals at the calibrated level (bootstrap-t, nominal 99.9 \%; bands: sector C filled, sector G hatched, three monitors grey). At the published spread (dotted, 0.55) sector C 0.79 [0.21, 156] and sector G, on all of its monitors, 1.76 [0.24, 17.1] (1.70 on MMF2 alone, main text): a per-postcode level cannot be bounded usefully with 14 months; summed over the three monitors 1.29 [0.93, 2.03]. Open, spread 0.66, a sensitivity analysis. The calibration was chosen by comparison with these monitors, so the level is not an independent test. \textbf{d}, Week by week, the between-postcode Spearman of the dispersion field's postcode shares with the report shares, against fixed geography (each postcode's time-mean share), all 59 weeks and with September–December 2024 withheld (41 weeks): the dispersion field's shares are no more closely correlated with the report shares than fixed geography is. \textbf{e}, After the 2025 source intervention, the change in each sector estimated by correcting its reports with the monitor in the other sector, divided by the sector's own measured change (green band, 0.8–1.25, agreement): MMF9 applied to sector G 0.79 [0.57, 1.08], MMF2 applied to sector C 1.89 [1.14, 3.20] (month bootstrap; coverage 81–86 \%, so too narrow). Measuring the size of an intervention's effect needs a monitor in the sector concerned.

%% file: legend_ED3.tex
\textbf{a}, Share of each month's odour reports by postcode sector, March 2024 – March 2025 (month and its report count above each column): recorded (top), the last six hours carried forward (middle) and the blended same-day map (bottom); grey, sectors with 1–4 reports in the month (share suppressed); the weather is observed. \textbf{b}, Paired gains of the same-day estimate per six-hour period on five measures, over the last six hours carried forward (dark grey) and over the estimate without weather (green), trained on earlier days or by block cross-validation, each also with September–December 2024 withheld (1,944 and 1,456 periods; 7-day block bootstrap; open, the interval includes zero). Skill gains +0.26 [+0.21, +0.32] and tier probabilities +0.15 [+0.11, +0.20]; weather's share of the skill margin is 0.19 [0.13, 0.26]. These gains describe periods already recorded and are not forecasts, because per period the estimate is more closely correlated with the previous period than with the target period. Right top, gains of the daily total over the last six hours carried forward (+0.019 on skill, +0.013 on tier agreement; both intervals include zero; 486 days). Right bottom, six-hour timing, \textit{r}(target 6 h) \ensuremath{-} \textit{r}(previous 6 h): same-day estimate \ensuremath{-}0.216 [\ensuremath{-}0.273, \ensuremath{-}0.175], against the estimate without weather and the last six hours carried forward.

%% file: legend_ED4.tex
\textbf{a}, Monthly frequency and duration from SENTINEL's forecast one day ahead (blue) and from the recorded reports (crimson) against H\ensuremath{_2}S at MMF9, 16 months; lines, least-squares fits to the forecast; bands, 95 \% intervals resampling the months. Forecast \textit{r} is 0.77 for frequency and 0.78 for duration (99 \% intervals 0.38–0.96 and 0.56–0.95; effective \textit{n} about 11); frequency from persistence is more closely correlated with H\ensuremath{_2}S (0.92). Before correction the forecast was more closely correlated with H\ensuremath{_2}S than a calendar forecast was; after Holm adjustment over eight comparisons it was not (smallest adjusted \textit{P} = 0.25). \textbf{b}, After the 2025 source intervention, each month's predicted \ensuremath{\div} recorded reports with intervals that allow for autocorrelation (HAC; shaded, within a factor of two): the forecast (blue), persistence (dark grey) and the model trained on daily levels (light grey) all use the report history, and their predictions fell with the recorded counts; a model with a fixed emission and no report history (triangles) predicted too many reports. \textbf{c}, Why the report-based exposure estimate fell further than measured exposure (ratio, after \ensuremath{\div} before; month bootstrap, coverage 81–86 \%, so too narrow): estimate \ensuremath{\div} measured 0.42 [0.25, 0.72] in sector C and 0.38 [0.22, 0.71] in sector G; reports per unit of measured exposure 0.69 [0.44, 1.12] and 0.14 [0.03, 0.65] (at the point estimate, about two thirds of the excess fall in sector C's estimate); the dispersion field's monthly mean 0.58 and 1.09. \textbf{d}, Reports per unit of exposure, three months after \ensuremath{\div} three before, at the intervention (filled) and at every other month boundary (open): within the range at other month boundaries in sector C (0.80) and lower than at any other boundary in sector G (0.20).

%% file: legend_ED5.tex
 \textbf{a}– \textbf{e}, Correlation of the forecast for day \textit{t}, issued zero (\textbf{a}) to four (\textbf{e}) days ahead, with recorded reports on day \textit{t} + \textit{k} (solid, trained on earlier days, with its 95 \% interval shaded, 7-day blocks of day pairs; dashed, block cross-validation; grey, days already recorded at issue; star, the target day; \textbf{a} is the same-day estimate with observed weather). Target-day correlation 0.949 [0.931, 0.960] in \textbf{a} and 0.808 [0.748, 0.851] in \textbf{b}. Timing, a diagnostic of the model family rather than evidence of skill: \textit{r}(target day) \ensuremath{-} \textit{r}(day before) +0.040 [+0.020, +0.064] one day ahead (\textbf{b}) and +0.163 [+0.126, +0.220] for the same-day estimate (\textbf{a}); the target day against the best day already recorded at issue +0.030 [\ensuremath{-}0.008, +0.052] (\textbf{c}), \ensuremath{-}0.023 [\ensuremath{-}0.063, +0.010] (\textbf{d}) and \ensuremath{-}0.061 [\ensuremath{-}0.107, \ensuremath{-}0.030] (\textbf{e}), so four days ahead part of the skill comes from report levels known at issue. \textbf{f}– \textbf{j}, Checks on the days scored with the configuration re-chosen each week from earlier days, July 2024 – April 2025, all ten sectors. \textbf{f}, \textbf{g}, Every combination of design choices scored as its own forecast without selection (7,232 one-day specifications: model base, half-life, volume calibration, burn-in, start date, retraining schedule; blue, trained on day-to-day change; grey, on levels). \textbf{f}, Partial correlation with the target day beyond the recorded report history: at least 0.33 in every specification. \textbf{g}, Target-day timing: above zero in 93 \% of change-trained and 12 \% of level-trained specifications. \textbf{h}, Weekly skill with the configuration chosen on the evaluated week (blue) or with that week left out (grey, dashed), minus skill when chosen on a 28-day window withheld from training (symmetric-log axis): choosing on the scored week scores higher in 42 of 44 weeks (pooled +0.090 [+0.048, +0.137]). \textbf{i}, Target-day timing of these forecasts (bands, 95 \%, 28-day blocks; daily skill in brackets; filled, re-chosen; open, fixed): re-chosen weekly +0.030 [\ensuremath{-}0.004, +0.047], published configuration +0.035 [\ensuremath{-}0.002, +0.051]; every interval includes zero. \textbf{j}, Timing within sectors, averaged: absolute (grey), \ensuremath{-}0.007 to +0.013, every interval including zero; margin over the same model trained on levels (blue), +0.019 to +0.039, every interval above zero. Intervals cover the true value less often than their nominal level (Methods).

%% file: legend_ED6.tex
All panels are trained on earlier days; open markers, the interval includes zero. \textbf{a}, Gain in dividing each day's reports among the ten sectors (redistribution skill) over a forecast wind-direction map, one to three days ahead (14-day block bootstrap; coverage 0.63–0.83): +0.088 [+0.026, +0.136] one day ahead, with intervals including zero from two days ahead; four days ahead the division among sectors is not resolved. \textbf{b}, Extra share of gas-episode days identified by forecast alerts, over persistence alerts, at an equal number of alerts, by lead (14-day blocks; coverage 0.93). Pooled over two to four days ahead, forecast alerts identified 0.29 of 105 gas-episode days (counted once per lead time), against 0.08 for persistence alerts (+0.21, \textit{P} = 0.039), with 166 alerts each, of which 136 and 158 were false; the result is not robust to the training scheme or the period withheld (\textit{P} = 0.165 and 0.070). \textbf{c}, Gain from the dispersion field in forecasting the monitors' measured daily H\ensuremath{_2}S beyond the forecast weather, December 2023 – April 2025 (line, the mean over three monitors; shaded, the range across them, not an interval): +0.18 to +0.20 one to three days ahead and +0.11 four days ahead. Adding the dispersion field did not detectably improve the forecast ten-sector daily total of reports beyond the weather: gains above 0.03 were excluded, but equivalence at the pre-declared 0.02 margin was not shown.

%% file: table_model_search.tex
\setlength{\tabcolsep}{1.9pt}
\begin{tabular}{@{}lll rrrr rrr@{}}
\toprule
 & & & \multicolumn{4}{c}{block cross-validation} & \multicolumn{3}{c}{trained on earlier days} \\
\cmidrule(lr){4-7}\cmidrule(l){8-10}
training & offset (d) & volume & $\kappa$, 1st half & $\kappa$, 2nd half & skill & $T$ & $\kappa$ & skill & $T$ \\
\midrule
\textbf{changes, $\lambda=1.5$} & \textbf{15} & \textbf{none} & \textbf{0.652} & \textbf{0.836} & \textbf{0.710} & \textbf{0.020} & \textbf{0.744} & \textbf{0.662} & \textbf{0.040} \\
changes, $\lambda=1$ & 30 & none & 0.647 & 0.832 & 0.714 & 0.022 & 0.751 & 0.657 & 0.043 \\
changes, $\lambda=1.5$ & 20 & none & 0.640 & 0.838 & 0.707 & 0.024 & 0.747 & 0.652 & 0.044 \\
changes, 4 par., $\lambda=2$ & 15 & none & 0.639 & 0.835 & 0.702 & 0.023 & 0.744 & 0.650 & 0.041 \\
changes, $\lambda=2$ & 15 & none & 0.636 & 0.838 & 0.703 & 0.024 & 0.738 & 0.652 & 0.042 \\
changes, $\lambda=0.75$ & 90 & refit & 0.634 & 0.823 & 0.692 & 0.018 & 0.705 & 0.568 & 0.041 \\
changes, $\lambda=1$ & 45 & none & 0.628 & 0.821 & 0.710 & 0.025 & 0.741 & 0.648 & 0.047 \\
changes, $\lambda=1$ & 60 & refit & 0.628 & 0.829 & 0.691 & 0.022 & 0.694 & 0.561 & 0.042 \\
changes, 4 par., $\lambda=1$ & 60 & refit & 0.628 & 0.829 & 0.690 & 0.021 & 0.698 & 0.558 & 0.040 \\
changes, $\lambda=2$ & 20 & none & 0.627 & 0.835 & 0.698 & 0.028 & 0.743 & 0.640 & 0.046 \\
\midrule
\multicolumn{10}{@{}l}{\textit{best configuration of each base trained by likelihood alone}} \\
likelihood + day index & 45 & none & 0.668 & 0.843 & 0.737 & $-$0.018 & 0.775 & 0.711 & $-$0.002 \\
likelihood, no index & 60 & refit & 0.638 & 0.853 & 0.703 & $-$0.039 & 0.775 & 0.702 & $-$0.029 \\
\bottomrule
\end{tabular}

%% file: legend_SI1.tex
\textbf{a}, Circular correlation between each sector's mean wind direction at report times and the direction from the landfill, eight sectors, for ERA5 (filled) and on-site wind (open) and for population-weighted or geometric sector centres; ticks, the 95th percentile of 200 circularly shifted winds. Circular \textit{r} is 0.974–0.992, against null 95th percentiles of 0.75–0.89 (permutation \textit{P} = 0.005 in all four, the smallest attainable with 200 shifts; with 10,000 shifts \textit{P} \ensuremath{\approx} 0.001, main text). \textbf{b}, Odour reports to the council per month (published figures), September 2023 – July 2025: 1,620 in January 2025, 79 in February and 15–24 a month from April 2025; bar, the capping works. \textbf{c}, Share of half-hours with H\ensuremath{_2}S above 7 \textmu{}g m\ensuremath{^{-3}} at each monitor (council figures): 17.8–33.8 \% in January 2025, 1.5–1.9 \% in February and zero at every reporting monitor from April 2025, so every monitor fell at the same time. \textbf{d}, Each quantity in 2025 \ensuremath{\div} the same months of 2024 (filled, February–April; open, March–April; 95 \% block-bootstrap intervals over calendar days). Methane peaks on the same analysers changed by \ensuremath{\times}0.74 [0.47, 1.61], no change detected, while H\ensuremath{_2}S peaks fell to \ensuremath{\times}0.04 [0.02, 0.15] and odour reports to \ensuremath{\times}0.07. The instruments and the delivery of landfill gas did not change detectably, which is consistent with a fall in the H\ensuremath{_2}S source; the methane interval cannot exclude a fall of up to about half. \textbf{e}, Landfill cover from the operator's published figures and Environment Agency records: permanent cap (filled; open, the Agency's records), temporary cap (grey) and the geomembrane laid in the capping works (black, stacked on the permanent cap, a lower bound). During the capping works, about 26,000 m\ensuremath{^{2}} was added over the remaining open ground, on top of 80,900 m\ensuremath{^{2}} of permanent cap, after two years in which the capped share swung between 57 and 83 \%. \textbf{f}, Share of the variance of log(1 + odour reports) accounted for by the monitor-based source estimate, aggregated over 1, 2 and 4 weeks (crimson, council; dark crimson, Environment Agency; filled, January 2024 – March 2025; open, without September–December 2024; squares, before the source intervention; point values, no intervals). The share is 0.51–0.57 in every aggregation, channel and scope, and 0.38–0.46 before the source intervention. \textbf{g}, Monthly correlation of each sector's odour reports with H\ensuremath{_2}S at its nearest monitor, against the distance to it (12–23 months; month-resampling 95 \% intervals). \textit{r} is 0.73–0.90 in eight of nine sectors; sector G, which surrounds the landfill, has 0.49 [\ensuremath{-}0.15, 0.82]. One shared event, in January 2025, dominates, so the panel shows a time course shared by all sectors, not a separate relation between each sector's reports and its own monitor. \textbf{h}, Same-day correlation per sector with and without September–December 2024: within 0.06 of the full record everywhere (0.40–0.74 without); for eight of nine sectors the correlation is still highest on the same day. \textbf{i}, Shares of the week-to-week variance of odour reports accounted for by the source estimate (circles) and the weather (squares), for four ways of representing transport, January 2024 – March 2025 (filled, weeks used in fitting; open, withheld weeks; dotted, the 95th percentile of the source share with the source shifted in time). The source estimate accounts for 0.51–0.59 in fitted weeks and 0.58–0.61 in withheld weeks; the weather for 0.06–0.19 in fitted weeks and \ensuremath{-}0.12 to \ensuremath{-}0.05 in withheld weeks. \textbf{j}, As \textbf{i} before the source intervention (January 2024 – January 2025). On withheld weeks the source estimate accounted for more variance than the weather (0.23–0.56 against \ensuremath{-}0.11 to 0.08), and more than a time-shifted source estimate only when transport was represented by the dispersion field (0.56). The source estimate is built from measured gas, so it includes the transport realised at the monitors. \textbf{k}, Seven-day rolling share of half-hours with H\ensuremath{_2}S above 7 \textmu{}g m\ensuremath{^{-3}} at MMF9 and odour reports per day, each relative to its October–November 2024 mean, with the capping works and each series' single step date (triangles; 99 \% intervals). The gas stepped on 17 January 2025 [15 November – 14 March] and the reports on 7 February [22 January – 18 February], so their order is not determined; neither fell at the end of tipping.

%% file: legend_SI2.tex
\textbf{a}, Daily correlation of report frequency with H\ensuremath{_2}S exposure for all 32 pre-specified combinations of four regulatory frequency definitions, two gas definitions and two windows (black, full record; grey, without September–December 2024; 95 \% intervals). \textit{r} is 0.10–0.62, with every lower bound above zero. \textbf{b}, Share of each day's six-hour periods with at least three reports. Grey, days grouped by how many periods had H\ensuremath{_2}S above 7 \textmu{}g m\ensuremath{^{-3}} (95 \% block-bootstrap intervals); crimson, months and their fit, as in Fig. 2a. The share is 0.24 on days with no exceedance (317 days), 0.51 with one exceeding period and 0.83–0.95 with two or more; the monthly line covers only the low end. \textbf{c}, As \textbf{b} for duration: hours with a report rise from 4.2 h on days with no exceedance (244 days) to 16.5 h on days with more than 8 h above the threshold. \textbf{d}, Month-to-month change in the council officers' landfill-odour detection rate against the change in odour reports (14 changes): Spearman 0.71 [0.21, 0.91]. Officer visits are partly triggered by reports, so this is co-movement, not independent confirmation. \textbf{e}, The psychometric reporting model for sector C against MMF9 (1,816 six-hour windows, 659 with a report). The cumulative-normal curve (solid, as Fig. 2b; band, 3-day block bootstrap) is set against a complementary log–log link (dashed), which fits slightly better (difference in Akaike information criterion, \ensuremath{\Delta}AIC, 3.7) and moves the midpoint from 2.7 to 5.5 \textmu{}g m\ensuremath{^{-3}}. \textbf{f}, Change in the report–H\ensuremath{_2}S correlation when report bursts are removed, minus the change when the same number of reports is removed at random, for 72 declustering variants: 56 below zero, 16 spanning it and none above. Large points, the pre-specified test. \textbf{g}, Partial correlation of daily odour reports with H\ensuremath{_2}S beyond season, weekday, weather and the past week's reports (95 \% block-bootstrap intervals): 0.31 in 2024 before the source intervention and 0.41 over 2024–25; 0.47 and 0.60 in the council's and the Environment Agency's weekly published figures. \textbf{h}, Gain in log-likelihood on withheld days from adding H\ensuremath{_2}S, per sector and monitor (open, H\ensuremath{_2}S shifted by 90 days): 0.030 [0.009, 0.054] nats (natural-log units) per window for all sectors with MMF9 and 0.081 [0.060, 0.101] for sector C; the sector G pairs show none. \textbf{i}, Left, the pre-specified declustering test: MMF9 \ensuremath{-}0.050 [\ensuremath{-}0.080, \ensuremath{-}0.016], MMF2 \ensuremath{-}0.035 [\ensuremath{-}0.067, \ensuremath{-}0.002]. Right, the excess of reports downwind (\ensuremath{\kappa}; 0 = chance) for the first and later reports of bursts (filled, public weather; open, on-site wind); later reports from other sectors keep 0.73 of the first report's wind alignment. The earlier filing-dated report sheet (2021–23) is not used for day-level results.

%% file: legend_SI3.tex
\textbf{a}, Wind direction at MMF9 from the monitor's vane (grey) and from CALMET at the monitor's cell (navy), July and November 2024, 1,258 hours with wind above 0.5 m s\ensuremath{^{-1}}. The median direction error is 16\textdegree{} (37\textdegree{} on calm, stable nights) and the median speed ratio 1.06; only MMF9 was checked. \textbf{b}, The weather analysis against each monitor's thermometer on the nights of July and November 2024 identified as having pooled cold air (filled, mean absolute error; open, mean bias): mean absolute error 0.9, 1.1 and 1.5 K. \textbf{c}, Mean hour-of-day profile, each series divided by its own mean, September 2023 – November 2024: measured methane excess (black), CALPUFF at the monitor (navy) and CALMET's ventilation (grey). CALPUFF reproduces the timing of the daily cycle but underestimates its amplitude: the median daily Spearman with the gas is 0.59, 0.52 and 0.47 for CALPUFF and 0.57, 0.62 and 0.53 for CALMET's ventilation (MMF1, MMF2, MMF9). \textbf{d}, Measured gas excess in hours with a CALPUFF puff at the monitor's cell \ensuremath{\div} hours without, September 2023 – April 2025 (black, H\ensuremath{_2}S; grey, methane; filled, all hours; open, wind at least 2 m s\ensuremath{^{-1}}; 7-day block intervals). The gas is several times higher when the plume is over the monitor. In calm air the methane ratio falls to 1.1–2.2, because CALPUFF does not represent calm-air accumulation. \textbf{e}, Hourly agreement of the dispersion field at the monitor's cell with measured excess (grey, methane; black, H\ensuremath{_2}S): agreement is better with methane at every monitor; one possible explanation is a shortfall in the modelled H\ensuremath{_2}S source. \textbf{f}, Daily agreement of the dispersion field alone with measured H\ensuremath{_2}S, with nothing fitted (dark, the 300 m disc around the monitor; light, the postcode sector): every interval lies above zero. The skill of the dispersion field alone differs between monitors under identical physics: MMF1 +0.381, MMF9 +0.298, MMF1A +0.370 and MMF2 +0.164. \textbf{g}, Gain in daily agreement with each monitor's H\ensuremath{_2}S when the sector's dispersion field is added to the odour-report model, with no monitor data in any fit (7-day block intervals): MMF1 +0.18, MMF9 +0.10 and MMF1A +0.08; MMF2 +0.02, whose interval includes zero. \textbf{h}, The inputs and the data available at issue: no local instrument is used, and the monitors are used only to evaluate the results.

%% file: legend_SI5.tex
SENTINEL is trained on earlier days unless stated. Blue, the forecast; light blue, without September–December 2024; open markers, block cross-validation or an interval that includes zero, as keyed. \textbf{a}, Left, each measure on the Fig. 5 window in three evaluation settings (trained on earlier days with the full record and without September–December 2024, and under block cross-validation): tier agreement 0.71, 0.70 and 0.74; ROC AUC 0.92, 0.91 and 0.92 (at least 3 reports), 0.84, 0.83 and 0.86 (at least 9) and 0.93, 0.94 and 0.94 (at least 27); cumulative reports 0.87, 0.97 and 0.80 \ensuremath{\times} recorded (396 days; 274 without September–December 2024). Middle, gains one day ahead with and without September–December 2024. Over persistence the gain is +0.145 [\ensuremath{-}0.038, +0.214] and +0.113 [\ensuremath{-}0.131, +0.176], so no gain is shown. The dispersion-field map gains over a forecast wind-direction map by +0.088 [+0.026, +0.136] and +0.093 [+0.006, +0.157]. Right, the dispersion-field map's gain over the wind-direction map in the ten sectors' counts, by lead: +0.012 [+0.003, +0.020] and +0.012 [+0.004, +0.021] one and two days ahead; from three days ahead the intervals include zero. \textbf{b}, Each month's predicted \ensuremath{\div} recorded reports (filled, trained on earlier days; open, block cross-validation). \textbf{c}, Target-day timing r(0) \ensuremath{-} r(\ensuremath{-}1) with the report history closed 0, 12, 24 or 48 hours before issue (circles), and the forecast's advantage over the same model trained on levels (squares). With no gap the advantage is +0.044 [+0.001, +0.059]; with the history closed 24 hours earlier it is +0.008 [\ensuremath{-}0.001, +0.015]. \textbf{d}, Daily skill over a running 60 days against the weekday \ensuremath{\times} month average (dark blue, trained on earlier days; dashed, block cross-validation) and against the trailing 15-day mean (light blue); below, the squared errors behind them. \textbf{e}, Daily agreement of the forecast sector map with the recorded map (blue) and of the report shares to date (green). Below, each month's share of reports by sector, recorded and forecast (hatched, 1–4 reports, share suppressed; 35 of 130 sector-months). \textbf{f}, For comparison, the agreement of the monitors' gas tiers with a monitor-trained model (pale bars, the configurations not selected; ticks, persistence). \ensuremath{\kappa} is 0.75 for the same-day estimate and 0.64 and 0.62 one day ahead under block cross-validation and trained on earlier days. The day-ahead timing r(0) \ensuremath{-} r(\ensuremath{-}1) is +0.026 [\ensuremath{-}0.024, +0.076] and +0.102 [+0.033, +0.173] respectively.

%% file: legend_SI6.tex
Blue, the forecast (light blue, without September–December 2024); dark grey, persistence; shaded, 95 \% intervals; open markers, block cross-validation or an interval that includes zero, as keyed. \textbf{a}, Six measures against lead time with persistence: daily skill, rank correlation, tier agreement, skill of the tier probabilities, high-report-day detection, and the value area over all cost/loss ratios (without September–December 2024). \textbf{b}, Left, target-day timing r(0) \ensuremath{-} r(\ensuremath{-}1) with and without September–December 2024 (filled, trained on earlier days; open, block cross-validation). Right, the timing advantage over the level-trained model with the report history closed 0–48 hours before issue. One day ahead with no gap it is +0.044 [+0.001, +0.059]; beyond one day ahead every variant's interval includes zero. \textbf{c}, Lag profiles one to four days ahead for the forecast, the level-trained model (dotted) and persistence; circles, the peak among days not yet recorded at issue. Right, four days ahead, the target day minus the best unrecorded day, over the level-trained model, in four variants: +0.014 [+0.001, +0.032] trained on earlier days. \textbf{d}, Skill against a trailing 7-day or 15-day mean instead of the weekday \ensuremath{\times} month average. \textbf{e}, Recall, precision and F1 for days with at least 27 reports, and the value area, against lead with persistence. Four days ahead the value area is 0.165 against persistence's 0.069, a gain of +0.096 [+0.018, +0.183]; one day ahead the gain is +0.048 [\ensuremath{-}0.061, +0.143]. \textbf{f}, Where reports fall. Left, the dispersion-field map's gain without September–December 2024 over the training-period shares with recent reports (green), a forecast wind-direction map (black) and wind direction \ensuremath{\times} the day's steadiness (grey). Middle, map skill as a share of the maximum attainable given sampling noise in the data: 31 \% one day ahead and 15 \% [7, 23] four days ahead trained on earlier days (32 \% and 20 \% under block cross-validation). Right, the map's timing (agreement on the target day minus the day before) against the training-period shares and a wind-direction map. The forecast map was not better timed than the wind-direction map: \ensuremath{-}0.005 [\ensuremath{-}0.011, +0.000] one day ahead and \ensuremath{-}0.004 [\ensuremath{-}0.007, \ensuremath{-}0.001] four days ahead. \textbf{g}, Hit-rate gains over persistence for days with at least 10 reports (left) and for gas episodes (right), by lead and pooled two to four days ahead. Variants: blue, full record; light blue, without September–December 2024; greys, without January–April 2025 and with training-day thresholds; squares, an equal number of alerts. Pooled gas episodes at an equal number of alerts: 0.29 against 0.08 (+0.21 [+0.014, +0.369], \textit{p} = 0.039). The gain does not hold without September–December 2024 (\textit{p} = 0.070) or under block cross-validation (\textit{p} = 0.165). With thresholds set on training days, forecast alerts identified 0.59 of gas-episode days against 0.22 for persistence alerts (gain +0.37, lower bound +0.19). Below, the value-area gain over persistence and over a calendar forecast, at an equal number of alerts (53–57 a lead) and with training-day thresholds (forecast 60–69 alerts, persistence 39–42, calendar 93–95). \textbf{h}, The factorial comparison: daily skill of the report history alone (green), with the weather, with the dispersion field and with both (blue), and the interaction of weather and field. Then the sector-count gain over the training-period shares with recent reports (green) and over a wind-direction map (black) in the monitors' period, and, one day ahead, per sector: nine of ten sectors gain over the training-period shares and five of ten over the wind-direction map. \textbf{i}, Correlation of monthly frequency and duration one day ahead with H\ensuremath{_2}S at MMF9 (blue; 99 \% month-resampling band), against persistence (dark grey), a calendar forecast (gold) and the recorded reports' own correlation (crimson, dashed). The forecast reaches 0.77 and 0.78 one day ahead and persistence 0.92 and 0.85; the forecast lies above the calendar forecast (Williams test, \textit{P} = 0.011–0.034). Right, partial correlation with the gas given persistence: every interval includes zero.

%% file: sentinel_arxiv.bbl
\begin{thebibliography}{68}
\ifx \bisbn   \undefined \def \bisbn  #1{ISBN #1}\fi
\ifx \binits  \undefined \def \binits#1{#1}\fi
\ifx \bauthor  \undefined \def \bauthor#1{#1}\fi
\ifx \batitle  \undefined \def \batitle#1{#1}\fi
\ifx \bjtitle  \undefined \def \bjtitle#1{#1}\fi
\ifx \bvolume  \undefined \def \bvolume#1{\textbf{#1}}\fi
\ifx \byear  \undefined \def \byear#1{#1}\fi
\ifx \bissue  \undefined \def \bissue#1{#1}\fi
\ifx \bfpage  \undefined \def \bfpage#1{#1}\fi
\ifx \blpage  \undefined \def \blpage #1{#1}\fi
\ifx \burl  \undefined \def \burl#1{\textsf{#1}}\fi
\ifx \doiurl  \undefined \def \doiurl#1{\url{https://doi.org/#1}}\fi
\ifx \betal  \undefined \def \betal{\textit{et al.}}\fi
\ifx \binstitute  \undefined \def \binstitute#1{#1}\fi
\ifx \binstitutionaled  \undefined \def \binstitutionaled#1{#1}\fi
\ifx \bctitle  \undefined \def \bctitle#1{#1}\fi
\ifx \beditor  \undefined \def \beditor#1{#1}\fi
\ifx \bpublisher  \undefined \def \bpublisher#1{#1}\fi
\ifx \bbtitle  \undefined \def \bbtitle#1{#1}\fi
\ifx \bedition  \undefined \def \bedition#1{#1}\fi
\ifx \bseriesno  \undefined \def \bseriesno#1{#1}\fi
\ifx \blocation  \undefined \def \blocation#1{#1}\fi
\ifx \bsertitle  \undefined \def \bsertitle#1{#1}\fi
\ifx \bsnm \undefined \def \bsnm#1{#1}\fi
\ifx \bsuffix \undefined \def \bsuffix#1{#1}\fi
\ifx \bparticle \undefined \def \bparticle#1{#1}\fi
\ifx \barticle \undefined \def \barticle#1{#1}\fi
\bibcommenthead
\ifx \bconfdate \undefined \def \bconfdate #1{#1}\fi
\ifx \botherref \undefined \def \botherref #1{#1}\fi
\ifx \url \undefined \def \url#1{\textsf{#1}}\fi
\ifx \bchapter \undefined \def \bchapter#1{#1}\fi
\ifx \bbook \undefined \def \bbook#1{#1}\fi
\ifx \bcomment \undefined \def \bcomment#1{#1}\fi
\ifx \oauthor \undefined \def \oauthor#1{#1}\fi
\ifx \citeauthoryear \undefined \def \citeauthoryear#1{#1}\fi
\ifx \endbibitem  \undefined \def \endbibitem {}\fi
\ifx \bconflocation  \undefined \def \bconflocation#1{#1}\fi
\ifx \arxivurl  \undefined \def \arxivurl#1{\textsf{#1}}\fi
\csname PreBibitemsHook\endcsname

\bibitem[\protect\citeauthoryear{Bokowa et~al.}{2021}]{bokowa2021}
\begin{barticle}
\bauthor{\bsnm{Bokowa}, \binits{A.}},
\bauthor{\bsnm{Diaz}, \binits{C.}},
\bauthor{\bsnm{Koziel}, \binits{J.A.}},
\bauthor{\bsnm{McGinley}, \binits{M.}},
\bauthor{\bsnm{Barclay}, \binits{J.}},
\bauthor{\bsnm{Schauberger}, \binits{G.}},
\bauthor{\bsnm{Guillot}, \binits{J.-M.}},
\bauthor{\bsnm{Sneath}, \binits{R.}},
\bauthor{\bsnm{Capelli}, \binits{L.}},
\bauthor{\bsnm{Zorich}, \binits{V.}},
\bauthor{\bsnm{Izquierdo}, \binits{C.}},
\bauthor{\bsnm{Bilsen}, \binits{I.}},
\bauthor{\bsnm{Romain}, \binits{A.-C.}},
\bauthor{\bsnm{Carmen~Cabeza}, \binits{M.}},
\bauthor{\bsnm{Liu}, \binits{D.}},
\bauthor{\bsnm{Both}, \binits{R.}},
\bauthor{\bsnm{Van~Belois}, \binits{H.}},
\bauthor{\bsnm{Higuchi}, \binits{T.}},
\bauthor{\bsnm{Wahe}, \binits{L.}}:
\batitle{Summary and overview of the odour regulations worldwide}.
\bjtitle{Atmosphere}
\bvolume{12}(\bissue{2}),
\bfpage{206}
(\byear{2021})
\doiurl{10.3390/atmos12020206}
\end{barticle}
\endbibitem

\bibitem[\protect\citeauthoryear{Nicell}{2009}]{nicell2009}
\begin{barticle}
\bauthor{\bsnm{Nicell}, \binits{J.A.}}:
\batitle{Assessment and regulation of odour impacts}.
\bjtitle{Atmospheric Environment}
\bvolume{43}(\bissue{1}),
\bfpage{196}--\blpage{206}
(\byear{2009})
\doiurl{10.1016/j.atmosenv.2008.09.033}
\end{barticle}
\endbibitem

\bibitem[\protect\citeauthoryear{Brancher et~al.}{2017}]{brancher2017}
\begin{barticle}
\bauthor{\bsnm{Brancher}, \binits{M.}},
\bauthor{\bsnm{Griffiths}, \binits{K.D.}},
\bauthor{\bsnm{Franco}, \binits{D.}},
\bauthor{\bsnm{Melo~Lisboa}, \binits{H.}}:
\batitle{A review of odour impact criteria in selected countries around the
  world}.
\bjtitle{Chemosphere}
\bvolume{168},
\bfpage{1531}--\blpage{1570}
(\byear{2017})
\doiurl{10.1016/j.chemosphere.2016.11.160}
\end{barticle}
\endbibitem

\bibitem[\protect\citeauthoryear{Wald et~al.}{2012}]{wald2012}
\begin{botherref}
\oauthor{\bsnm{Wald}, \binits{D.J.}},
\oauthor{\bsnm{Quitoriano}, \binits{V.}},
\oauthor{\bsnm{Worden}, \binits{C.B.}},
\oauthor{\bsnm{Hopper}, \binits{M.}},
\oauthor{\bsnm{Dewey}, \binits{J.W.}}:
Usgs “did you feel it?” internet-based macroseismic intensity maps.
Annals of Geophysics
\textbf{54}(6)
(2012)
\doiurl{10.4401/ag-5354}
\end{botherref}
\endbibitem

\bibitem[\protect\citeauthoryear{Atkinson and Wald}{2007}]{atkinson2007}
\begin{barticle}
\bauthor{\bsnm{Atkinson}, \binits{G.M.}},
\bauthor{\bsnm{Wald}, \binits{D.J.}}:
\batitle{"did you feel it?" intensity data: A surprisingly good measure of
  earthquake ground motion}.
\bjtitle{Seismological Research Letters}
\bvolume{78}(\bissue{3}),
\bfpage{362}--\blpage{368}
(\byear{2007})
\doiurl{10.1785/gssrl.78.3.362}
\end{barticle}
\endbibitem

\bibitem[\protect\citeauthoryear{Worden et~al.}{2012}]{worden2012}
\begin{barticle}
\bauthor{\bsnm{Worden}, \binits{C.B.}},
\bauthor{\bsnm{Gerstenberger}, \binits{M.C.}},
\bauthor{\bsnm{Rhoades}, \binits{D.A.}},
\bauthor{\bsnm{Wald}, \binits{D.J.}}:
\batitle{Probabilistic relationships between ground-motion parameters and
  modified mercalli intensity in california}.
\bjtitle{Bulletin of the Seismological Society of America}
\bvolume{102},
\bfpage{204}--\blpage{221}
(\byear{2012})
\doiurl{10.1785/0120110156}
\end{barticle}
\endbibitem

\bibitem[\protect\citeauthoryear{Kim et~al.}{2005}]{kim2005landfill}
\begin{barticle}
\bauthor{\bsnm{Kim}, \binits{K.-H.}},
\bauthor{\bsnm{Choi}, \binits{Y.J.}},
\bauthor{\bsnm{Jeon}, \binits{E.C.}},
\bauthor{\bsnm{Sunwoo}, \binits{Y.}}:
\batitle{Characterization of malodorous sulfur compounds in landfill gas}.
\bjtitle{Atmospheric Environment}
\bvolume{39}(\bissue{6}),
\bfpage{1103}--\blpage{1112}
(\byear{2005})
\doiurl{10.1016/j.atmosenv.2004.09.083}
\end{barticle}
\endbibitem

\bibitem[\protect\citeauthoryear{{Agency for Toxic Substances and Disease
  Registry}}{2001}]{atsdr2001landfillgas}
\begin{botherref}
\oauthor{\bsnm{{Agency for Toxic Substances and Disease Registry}}}:
Landfill gas primer --- an overview for environmental health professionals,
  chapter 2: Landfill gas basics.
Technical report,
U.S. Department of Health and Human Services, Atlanta, GA
(2001)
\end{botherref}
\endbibitem

\bibitem[\protect\citeauthoryear{{Gov.uk}}{}]{govuk_nd_h2s}
\begin{botherref}
\oauthor{\bsnm{{Gov.uk}}}:
Hydrogen sulphide: toxicological overview.
\url{https://www.gov.uk/government/publications/hydrogen-sulphide-properties-incident-management-and-toxicology/hydrogen-sulphide-toxicological-overview}
\end{botherref}
\endbibitem

\bibitem[\protect\citeauthoryear{{National Research Council
  (US)}}{2010}]{nrc2010acute}
\begin{bbook}
\bauthor{\bsnm{{National Research Council (US)}}}:
\bbtitle{Acute Exposure Guideline Levels for Selected Airborne Chemicals:
  Volume 9}.
\bpublisher{National Academies Press},
\blocation{Washington, DC}
(\byear{2010}).
\bcomment{Chap. 4: Hydrogen Sulfide}.
\doiurl{10.17226/12978} .
\burl{https://www.ncbi.nlm.nih.gov/books/NBK208170/}
\end{bbook}
\endbibitem

\bibitem[\protect\citeauthoryear{Guidotti}{2010}]{guidotti2010}
\begin{barticle}
\bauthor{\bsnm{Guidotti}, \binits{T.L.}}:
\batitle{Hydrogen sulfide}.
\bjtitle{International Journal of Toxicology}
\bvolume{29}(\bissue{6}),
\bfpage{569}--\blpage{581}
(\byear{2010})
\doiurl{10.1177/1091581810384882}
\end{barticle}
\endbibitem

\bibitem[\protect\citeauthoryear{Nuvolone et~al.}{2019}]{nuvolone2019}
\begin{barticle}
\bauthor{\bsnm{Nuvolone}, \binits{D.}},
\bauthor{\bsnm{Petri}, \binits{D.}},
\bauthor{\bsnm{Pepe}, \binits{P.}},
\bauthor{\bsnm{Voller}, \binits{F.}}:
\batitle{Health effects associated with chronic exposure to low-level hydrogen
  sulfide from geothermoelectric power plants. a residential cohort study in
  the geothermal area of mt. amiata in tuscany}.
\bjtitle{Science of The Total Environment}
\bvolume{659},
\bfpage{973}--\blpage{982}
(\byear{2019})
\doiurl{10.1016/j.scitotenv.2018.12.363}
\end{barticle}
\endbibitem

\bibitem[\protect\citeauthoryear{Schiffman and Williams}{2005}]{schiffman2005}
\begin{barticle}
\bauthor{\bsnm{Schiffman}, \binits{S.S.}},
\bauthor{\bsnm{Williams}, \binits{C.M.}}:
\batitle{Science of odor as a potential health issue}.
\bjtitle{Journal of Environmental Quality}
\bvolume{34}(\bissue{1}),
\bfpage{129}--\blpage{138}
(\byear{2005})
\doiurl{10.2134/jeq2005.0129a}
\end{barticle}
\endbibitem

\bibitem[\protect\citeauthoryear{Hirasawa
  et~al.}{2019}]{hirasawa2019subjective}
\begin{barticle}
\bauthor{\bsnm{Hirasawa}, \binits{Y.}},
\bauthor{\bsnm{Shirasu}, \binits{M.}},
\bauthor{\bsnm{Okamoto}, \binits{M.}},
\bauthor{\bsnm{Touhara}, \binits{K.}}:
\batitle{Subjective unpleasantness of malodors induces a stress response}.
\bjtitle{Psychoneuroendocrinology}
\bvolume{106},
\bfpage{206}--\blpage{215}
(\byear{2019})
\doiurl{10.1016/j.psyneuen.2019.03.018}
\end{barticle}
\endbibitem

\bibitem[\protect\citeauthoryear{Martuzzi et~al.}{2010}]{martuzzi2010}
\begin{barticle}
\bauthor{\bsnm{Martuzzi}, \binits{M.}},
\bauthor{\bsnm{Mitis}, \binits{F.}},
\bauthor{\bsnm{Forastiere}, \binits{F.}}:
\batitle{Inequalities, inequities, environmental justice in waste management
  and health}.
\bjtitle{The European Journal of Public Health}
\bvolume{20}(\bissue{1}),
\bfpage{21}--\blpage{26}
(\byear{2010})
\doiurl{10.1093/eurpub/ckp216}
\end{barticle}
\endbibitem

\bibitem[\protect\citeauthoryear{Mohai et~al.}{2009}]{mohai2009}
\begin{barticle}
\bauthor{\bsnm{Mohai}, \binits{P.}},
\bauthor{\bsnm{Pellow}, \binits{D.}},
\bauthor{\bsnm{Roberts}, \binits{J.T.}}:
\batitle{Environmental justice}.
\bjtitle{Annual Review of Environment and Resources}
\bvolume{34},
\bfpage{405}--\blpage{430}
(\byear{2009})
\doiurl{10.1146/annurev-environ-082508-094348}
\end{barticle}
\endbibitem

\bibitem[\protect\citeauthoryear{Quist and Johnston}{2024}]{carson2024malodors}
\begin{barticle}
\bauthor{\bsnm{Quist}, \binits{A.J.L.}},
\bauthor{\bsnm{Johnston}, \binits{J.E.}}:
\batitle{Malodors as environmental injustice: health symptoms in the aftermath
  of a hydrogen sulfide emergency in carson, california, usa}.
\bjtitle{Journal of Exposure Science \& Environmental Epidemiology}
\bvolume{34}(\bissue{6}),
\bfpage{935}--\blpage{940}
(\byear{2024})
\doiurl{10.1038/s41370-023-00561-x}
\end{barticle}
\endbibitem

\bibitem[\protect\citeauthoryear{Elliott et~al.}{2001}]{elliott2001}
\begin{barticle}
\bauthor{\bsnm{Elliott}, \binits{P.}},
\bauthor{\bsnm{Briggs}, \binits{D.}},
\bauthor{\bsnm{Morris}, \binits{S.}},
\bauthor{\bsnm{Hoogh}, \binits{C.}},
\bauthor{\bsnm{Hurt}, \binits{C.}},
\bauthor{\bsnm{Jensen}, \binits{T.K.}},
\bauthor{\bsnm{Maitland}, \binits{I.}},
\bauthor{\bsnm{Richardson}, \binits{S.}},
\bauthor{\bsnm{Wakefield}, \binits{J.}},
\bauthor{\bsnm{Jarup}, \binits{L.}}:
\batitle{Risk of adverse birth outcomes in populations living near landfill
  sites}.
\bjtitle{BMJ}
\bvolume{323}(\bissue{7309}),
\bfpage{363}--\blpage{368}
(\byear{2001})
\doiurl{10.1136/bmj.323.7309.363}
\end{barticle}
\endbibitem

\bibitem[\protect\citeauthoryear{{Environment Agency}}{2011}]{EAH4_2011}
\begin{botherref}
\oauthor{\bsnm{{Environment Agency}}}:
Additional guidance for H4 Odour Management: how to comply with your
  environmental permit (GEHO0411BTQM-E-E).
grey literature (regulator or professional guidance)
(2011).
\url{https://web.archive.org/web/2020id_/https://assets.publishing.service.gov.uk/media/5a7ba9a2ed915d1311060b16/geho0411btqm-e-e.pdf}
\end{botherref}
\endbibitem

\bibitem[\protect\citeauthoryear{{Institute of Air Quality
  Management}}{2018}]{IAQM2018}
\begin{botherref}
\oauthor{\bsnm{{Institute of Air Quality Management}}}:
Guidance on the assessment of odour for planning, Version 1.1.
grey literature (regulator or professional guidance)
(2018).
\url{https://iaqm.co.uk/text/guidance/odour-guidance-2014.pdf}
\end{botherref}
\endbibitem

\bibitem[\protect\citeauthoryear{Pedde and
  Adar}{2024}]{LITPedde2024Representativeness}
\begin{barticle}
\bauthor{\bsnm{Pedde}, \binits{M.}},
\bauthor{\bsnm{Adar}, \binits{S.D.}}:
\batitle{Representativeness of the us epa pm monitoring site locations to the
  us population: implications for air pollution prediction modeling}.
\bjtitle{Journal of Exposure Science \& Environmental Epidemiology}
\bvolume{34}(\bissue{5}),
\bfpage{821}--\blpage{826}
(\byear{2024})
\doiurl{10.1038/s41370-024-00644-3}
\end{barticle}
\endbibitem

\bibitem[\protect\citeauthoryear{Goodchild}{2007}]{goodchild2007}
\begin{barticle}
\bauthor{\bsnm{Goodchild}, \binits{M.F.}}:
\batitle{Citizens as sensors: the world of volunteered geography}.
\bjtitle{GeoJournal}
\bvolume{69}(\bissue{4}),
\bfpage{211}--\blpage{221}
(\byear{2007})
\doiurl{10.1007/s10708-007-9111-y}
\end{barticle}
\endbibitem

\bibitem[\protect\citeauthoryear{Wing et~al.}{2008}]{wing2008}
\begin{barticle}
\bauthor{\bsnm{Wing}, \binits{S.}},
\bauthor{\bsnm{Horton}, \binits{R.A.}},
\bauthor{\bsnm{Marshall}, \binits{S.W.}},
\bauthor{\bsnm{Thu}, \binits{K.}},
\bauthor{\bsnm{Tajik}, \binits{M.}},
\bauthor{\bsnm{Schinasi}, \binits{L.}},
\bauthor{\bsnm{Schiffman}, \binits{S.S.}}:
\batitle{Air pollution and odor in communities near industrial swine
  operations}.
\bjtitle{Environmental Health Perspectives}
\bvolume{116}(\bissue{10}),
\bfpage{1362}--\blpage{1368}
(\byear{2008})
\doiurl{10.1289/ehp.11250}
\end{barticle}
\endbibitem

\bibitem[\protect\citeauthoryear{Heaney et~al.}{2011}]{heaney2011}
\begin{barticle}
\bauthor{\bsnm{Heaney}, \binits{C.D.}},
\bauthor{\bsnm{Wing}, \binits{S.}},
\bauthor{\bsnm{Campbell}, \binits{R.L.}},
\bauthor{\bsnm{Caldwell}, \binits{D.}},
\bauthor{\bsnm{Hopkins}, \binits{B.}},
\bauthor{\bsnm{Richardson}, \binits{D.}},
\bauthor{\bsnm{Yeatts}, \binits{K.}}:
\batitle{Relation between malodor, ambient hydrogen sulfide, and health in a
  community bordering a landfill}.
\bjtitle{Environmental Research}
\bvolume{111}(\bissue{6}),
\bfpage{847}--\blpage{852}
(\byear{2011})
\doiurl{10.1016/j.envres.2011.05.021}
\end{barticle}
\endbibitem

\bibitem[\protect\citeauthoryear{Lee et~al.}{2013}]{lee2013}
\begin{barticle}
\bauthor{\bsnm{Lee}, \binits{H.-D.}},
\bauthor{\bsnm{Jeon}, \binits{S.-B.}},
\bauthor{\bsnm{Choi}, \binits{W.-J.}},
\bauthor{\bsnm{Lee}, \binits{S.-S.}},
\bauthor{\bsnm{Lee}, \binits{M.-H.}},
\bauthor{\bsnm{Oh}, \binits{K.-J.}}:
\batitle{A novel assessment of odor sources using instrumental analysis
  combined with resident monitoring records for an industrial area in korea}.
\bjtitle{Atmospheric Environment}
\bvolume{74},
\bfpage{277}--\blpage{290}
(\byear{2013})
\doiurl{10.1016/j.atmosenv.2013.04.001}
\end{barticle}
\endbibitem

\bibitem[\protect\citeauthoryear{Blanes-Vidal et~al.}{2012}]{blanesvidal2012}
\begin{botherref}
\oauthor{\bsnm{Blanes-Vidal}, \binits{V.}},
\oauthor{\bsnm{Nadimi}, \binits{E.S.}},
\oauthor{\bsnm{Ellermann}, \binits{T.}},
\oauthor{\bsnm{Andersen}, \binits{H.V.}},
\oauthor{\bsnm{Løfstrøm}, \binits{P.}}:
Perceived annoyance from environmental odors and association with atmospheric
  ammonia levels in non-urban residential communities: a cross-sectional study.
Environmental Health
\textbf{11}(1)
(2012)
\doiurl{10.1186/1476-069x-11-27}
\end{botherref}
\endbibitem

\bibitem[\protect\citeauthoryear{Rico et~al.}{2025}]{rico2025}
\begin{barticle}
\bauthor{\bsnm{Rico}, \binits{B.}},
\bauthor{\bsnm{Barsanti}, \binits{K.C.}},
\bauthor{\bsnm{Porter}, \binits{W.C.}},
\bauthor{\bsnm{Carvalho}, \binits{K.}},
\bauthor{\bsnm{Stigler-Granados}, \binits{P.}},
\bauthor{\bsnm{Prather}, \binits{K.A.}}:
\batitle{Heavily polluted tijuana river drives regional air quality crisis}.
\bjtitle{Science}
\bvolume{389}(\bissue{6763}),
\bfpage{1343}
(\byear{2025})
\doiurl{10.1126/science.adv1343}
\end{barticle}
\endbibitem

\bibitem[\protect\citeauthoryear{Cartelle et~al.}{2016}]{Cartelle2016}
\begin{barticle}
\bauthor{\bsnm{Cartelle}, \binits{D.}},
\bauthor{\bsnm{Vellon}, \binits{J.M.}},
\bauthor{\bsnm{Rodriguez}, \binits{A.}},
\bauthor{\bsnm{Valino}, \binits{D.}},
\bauthor{\bsnm{Gonzalez}, \binits{J.A.}},
\bauthor{\bsnm{Casas}, \binits{C.}},
\bauthor{\bsnm{Carrera-Chapela}, \binits{F.}}:
\batitle{Prolor: a modelling approach for environmental odor forecast}.
\bjtitle{Chemical Engineering Transactions}
\bvolume{54},
\bfpage{229}--\blpage{234}
(\byear{2016})
\doiurl{10.3303/CET1654039}
\end{barticle}
\endbibitem

\bibitem[\protect\citeauthoryear{Diaz~Jimenez et~al.}{2016}]{Jimenez2016}
\begin{barticle}
\bauthor{\bsnm{Diaz~Jimenez}, \binits{C.N.}},
\bauthor{\bsnm{Izquierdo~Zamora}, \binits{C.}},
\bauthor{\bsnm{Cartelle}, \binits{D.}},
\bauthor{\bsnm{Vellon}, \binits{J.M.}},
\bauthor{\bsnm{Rodriguez}, \binits{A.}}:
\batitle{Comparison of predicted versus real odour impacts in a rendering plant
  with prolor}.
\bjtitle{Chemical Engineering Transactions}
\bvolume{54},
\bfpage{199}--\blpage{204}
(\byear{2016})
\doiurl{10.3303/CET1654034}
\end{barticle}
\endbibitem

\bibitem[\protect\citeauthoryear{Capelli and Sironi}{2018}]{capelli2018}
\begin{barticle}
\bauthor{\bsnm{Capelli}, \binits{L.}},
\bauthor{\bsnm{Sironi}, \binits{S.}}:
\batitle{Combination of field inspection and dispersion modelling to estimate
  odour emissions from an italian landfill}.
\bjtitle{Atmospheric Environment}
\bvolume{191},
\bfpage{273}--\blpage{290}
(\byear{2018})
\doiurl{10.1016/j.atmosenv.2018.08.007}
\end{barticle}
\endbibitem

\bibitem[\protect\citeauthoryear{Üstün and Özkal}{2026}]{Ustun2026}
\begin{barticle}
\bauthor{\bsnm{Üstün}, \binits{M.A.}},
\bauthor{\bsnm{Özkal}, \binits{C.B.}}:
\batitle{Predicting odor nuisance levels using meteorological data and citizen
  complaints records: A machine learning approach}.
\bjtitle{Atmospheric Pollution Research}
\bvolume{17},
\bfpage{102809}
(\byear{2026})
\doiurl{10.1016/j.apr.2025.102809}
\end{barticle}
\endbibitem

\bibitem[\protect\citeauthoryear{Gavros et~al.}{2024}]{Gavros2024}
\begin{barticle}
\bauthor{\bsnm{Gavros}, \binits{A.}},
\bauthor{\bsnm{Hsu}, \binits{Y.-C.}},
\bauthor{\bsnm{Karatzas}, \binits{K.}}:
\batitle{Modelling smell events in urban pittsburgh with machine and deep
  learning techniques}.
\bjtitle{Atmosphere}
\bvolume{15}(\bissue{6}),
\bfpage{731}
(\byear{2024})
\doiurl{10.3390/atmos15060731}
\end{barticle}
\endbibitem

\bibitem[\protect\citeauthoryear{Hsu et~al.}{2020}]{Hsu2020}
\begin{barticle}
\bauthor{\bsnm{Hsu}, \binits{Y.-C.}},
\bauthor{\bsnm{Cross}, \binits{J.}},
\bauthor{\bsnm{Dille}, \binits{P.}},
\bauthor{\bsnm{Tasota}, \binits{M.}},
\bauthor{\bsnm{Dias}, \binits{B.}},
\bauthor{\bsnm{Sargent}, \binits{R.}},
\bauthor{\bsnm{Huang}, \binits{T.-H.K.}},
\bauthor{\bsnm{Nourbakhsh}, \binits{I.}}:
\batitle{Smell pittsburgh}.
\bjtitle{ACM Transactions on Interactive Intelligent Systems}
\bvolume{10}(\bissue{4}),
\bfpage{1}--\blpage{49}
(\byear{2020})
\doiurl{10.1145/3369397}
\end{barticle}
\endbibitem

\bibitem[\protect\citeauthoryear{Mulrow et~al.}{2020}]{Mulrow2020}
\begin{barticle}
\bauthor{\bsnm{Mulrow}, \binits{J.}},
\bauthor{\bsnm{Kshetry}, \binits{N.}},
\bauthor{\bsnm{Brose}, \binits{D.A.}},
\bauthor{\bsnm{Kumar}, \binits{K.}},
\bauthor{\bsnm{Jain}, \binits{D.}},
\bauthor{\bsnm{Shah}, \binits{M.}},
\bauthor{\bsnm{Kunetz}, \binits{T.E.}},
\bauthor{\bsnm{Varshney}, \binits{L.R.}}:
\batitle{Prediction of odor complaints at a large composite reservoir in a
  highly urbanized area: A machine learning approach}.
\bjtitle{Water Environment Research}
\bvolume{92}(\bissue{3}),
\bfpage{418}--\blpage{429}
(\byear{2020})
\doiurl{10.1002/wer.1191}
\end{barticle}
\endbibitem

\bibitem[\protect\citeauthoryear{Veratti et~al.}{2026}]{Veratti2026}
\begin{botherref}
\oauthor{\bsnm{Veratti}, \binits{G.}},
\oauthor{\bsnm{Abita}, \binits{A.}},
\oauthor{\bsnm{Tirone}, \binits{N.}},
\oauthor{\bsnm{Resci}, \binits{G.}},
\oauthor{\bsnm{Guidi}, \binits{G.}},
\oauthor{\bsnm{Bonasoni}, \binits{P.}},
\oauthor{\bsnm{Landi}, \binits{T.C.}}:
Tracking Industrial Emissions and Odor Nuisance through Integrated Modeling and
  Citizen Reporting.
EGU General Assembly 2026, abstract EGU26-21101
(2026).
\url{https://meetingorganizer.copernicus.org/EGU26/EGU26-21101.html}
\end{botherref}
\endbibitem

\bibitem[\protect\citeauthoryear{Ginsberg et~al.}{2009}]{ginsberg2009}
\begin{barticle}
\bauthor{\bsnm{Ginsberg}, \binits{J.}},
\bauthor{\bsnm{Mohebbi}, \binits{M.H.}},
\bauthor{\bsnm{Patel}, \binits{R.S.}},
\bauthor{\bsnm{Brammer}, \binits{L.}},
\bauthor{\bsnm{Smolinski}, \binits{M.S.}},
\bauthor{\bsnm{Brilliant}, \binits{L.}}:
\batitle{Detecting influenza epidemics using search engine query data}.
\bjtitle{Nature}
\bvolume{457}(\bissue{7232}),
\bfpage{1012}--\blpage{1014}
(\byear{2009})
\doiurl{10.1038/nature07634}
\end{barticle}
\endbibitem

\bibitem[\protect\citeauthoryear{Lazer et~al.}{2014}]{lazer2014}
\begin{barticle}
\bauthor{\bsnm{Lazer}, \binits{D.}},
\bauthor{\bsnm{Kennedy}, \binits{R.}},
\bauthor{\bsnm{King}, \binits{G.}},
\bauthor{\bsnm{Vespignani}, \binits{A.}}:
\batitle{The parable of google flu: Traps in big data analysis}.
\bjtitle{Science}
\bvolume{343}(\bissue{6176}),
\bfpage{1203}--\blpage{1205}
(\byear{2014})
\doiurl{10.1126/science.1248506}
\end{barticle}
\endbibitem

\bibitem[\protect\citeauthoryear{Wichmann and Hill}{2001}]{wichmann2001}
\begin{barticle}
\bauthor{\bsnm{Wichmann}, \binits{F.A.}},
\bauthor{\bsnm{Hill}, \binits{N.J.}}:
\batitle{The psychometric function: {I}. {F}itting, sampling, and goodness of
  fit}.
\bjtitle{Perception \& Psychophysics}
\bvolume{63}(\bissue{8}),
\bfpage{1293}--\blpage{1313}
(\byear{2001})
\doiurl{10.3758/BF03194544}
\end{barticle}
\endbibitem

\bibitem[\protect\citeauthoryear{Pearce et~al.}{2026}]{Pearce2026}
\begin{barticle}
\bauthor{\bsnm{Pearce}, \binits{T.C.}},
\bauthor{\bsnm{Smith}, \binits{D.J.T.}},
\bauthor{\bsnm{Dobney}, \binits{A.}},
\bauthor{\bsnm{Freddo}, \binits{A.}}:
\batitle{Meteorology-driven causal nowcasting of fugitive landfill emissions
  from measured coupling timescales}.
\bjtitle{arXiv}
(\byear{2026})
\doiurl{10.48550/arXiv.2608.14254}
\end{barticle}
\endbibitem

\bibitem[\protect\citeauthoryear{Grange and Carslaw}{2019}]{Grange2019}
\begin{barticle}
\bauthor{\bsnm{Grange}, \binits{S.K.}},
\bauthor{\bsnm{Carslaw}, \binits{D.C.}}:
\batitle{{Using meteorological normalisation to detect interventions in air
  quality time series}}.
\bjtitle{Science of The Total Environment}
\bvolume{653},
\bfpage{578}--\blpage{588}
(\byear{2019})
\doiurl{10.1016/j.scitotenv.2018.10.344}
\end{barticle}
\endbibitem

\bibitem[\protect\citeauthoryear{Sucker et~al.}{2008}]{sucker2007b}
\begin{barticle}
\bauthor{\bsnm{Sucker}, \binits{K.}},
\bauthor{\bsnm{Both}, \binits{R.}},
\bauthor{\bsnm{Bischoff}, \binits{M.}},
\bauthor{\bsnm{Guski}, \binits{R.}},
\bauthor{\bsnm{Krämer}, \binits{U.}},
\bauthor{\bsnm{Winneke}, \binits{G.}}:
\batitle{Odor frequency and odor annoyance part ii: dose–response
  associations and their modification by hedonic tone}.
\bjtitle{International Archives of Occupational and Environmental Health}
\bvolume{81}(\bissue{6}),
\bfpage{683}--\blpage{694}
(\byear{2008})
\doiurl{10.1007/s00420-007-0262-4}
\end{barticle}
\endbibitem

\bibitem[\protect\citeauthoryear{{World Health
  Organization}}{2000}]{who2000airquality}
\begin{botherref}
\oauthor{\bsnm{{World Health Organization}}}:
Air quality guidelines for europe.
Technical report,
WHO Regional Office for Europe, Copenhagen
(2000)
\end{botherref}
\endbibitem

\bibitem[\protect\citeauthoryear{Scire et~al.}{2000a}]{Scire2000calmet}
\begin{bbook}
\bauthor{\bsnm{Scire}, \binits{J.S.}},
\bauthor{\bsnm{Robe}, \binits{F.R.}},
\bauthor{\bsnm{Fernau}, \binits{M.E.}},
\bauthor{\bsnm{Yamartino}, \binits{R.J.}}:
\bbtitle{A User's Guide for the {CALMET} Meteorological Model (Version 5)}.
\bpublisher{Earth Tech, Inc.},
\blocation{Concord, MA}
(\byear{2000}).
\bcomment{Earth Tech, Inc.}
\end{bbook}
\endbibitem

\bibitem[\protect\citeauthoryear{Scire et~al.}{2000b}]{Scire2000calpuff}
\begin{bbook}
\bauthor{\bsnm{Scire}, \binits{J.S.}},
\bauthor{\bsnm{Strimaitis}, \binits{D.G.}},
\bauthor{\bsnm{Yamartino}, \binits{R.J.}}:
\bbtitle{A User's Guide for the {CALPUFF} Dispersion Model (Version 5)}.
\bpublisher{Earth Tech, Inc.},
\blocation{Concord, MA}
(\byear{2000}).
\bcomment{Earth Tech, Inc.}
\end{bbook}
\endbibitem

\bibitem[\protect\citeauthoryear{Diggle et~al.}{2010}]{diggle2010}
\begin{barticle}
\bauthor{\bsnm{Diggle}, \binits{P.J.}},
\bauthor{\bsnm{Menezes}, \binits{R.}},
\bauthor{\bsnm{Su}, \binits{T.-l.}}:
\batitle{Geostatistical inference under preferential sampling}.
\bjtitle{Journal of the Royal Statistical Society Series C: Applied Statistics}
\bvolume{59}(\bissue{2}),
\bfpage{191}--\blpage{232}
(\byear{2010})
\doiurl{10.1111/j.1467-9876.2009.00701.x}
\end{barticle}
\endbibitem

\bibitem[\protect\citeauthoryear{Liu et~al.}{2023}]{liu2023}
\begin{barticle}
\bauthor{\bsnm{Liu}, \binits{Z.}},
\bauthor{\bsnm{Bhandaram}, \binits{U.}},
\bauthor{\bsnm{Garg}, \binits{N.}}:
\batitle{Quantifying spatial under-reporting disparities in resident
  crowdsourcing}.
\bjtitle{Nature Computational Science}
\bvolume{4}(\bissue{1}),
\bfpage{57}--\blpage{65}
(\byear{2023})
\doiurl{10.1038/s43588-023-00572-6}
\end{barticle}
\endbibitem

\bibitem[\protect\citeauthoryear{Murphy}{1988}]{murphy1988}
\begin{barticle}
\bauthor{\bsnm{Murphy}, \binits{A.H.}}:
\batitle{Skill scores based on the mean square error and their relationships to
  the correlation coefficient}.
\bjtitle{Monthly Weather Review}
\bvolume{116}(\bissue{12}),
\bfpage{2417}--\blpage{2424}
(\byear{1988})
\doiurl{10.1175/1520-0493(1988)116<2417:ssbotm>2.0.co;2}
\end{barticle}
\endbibitem

\bibitem[\protect\citeauthoryear{Jolliffe}{2007}]{jolliffe2007}
\begin{barticle}
\bauthor{\bsnm{Jolliffe}, \binits{I.T.}}:
\batitle{Uncertainty and inference for verification measures}.
\bjtitle{Weather and Forecasting}
\bvolume{22}(\bissue{3}),
\bfpage{637}--\blpage{650}
(\byear{2007})
\doiurl{10.1175/waf989.1}
\end{barticle}
\endbibitem

\bibitem[\protect\citeauthoryear{Zhu et~al.}{2002}]{zhu2002}
\begin{barticle}
\bauthor{\bsnm{Zhu}, \binits{Y.}},
\bauthor{\bsnm{Toth}, \binits{Z.}},
\bauthor{\bsnm{Wobus}, \binits{R.}},
\bauthor{\bsnm{Richardson}, \binits{D.}},
\bauthor{\bsnm{Mylne}, \binits{K.}}:
\batitle{The economic value of ensemble-based weather forecasts}.
\bjtitle{Bulletin of the American Meteorological Society}
\bvolume{83}(\bissue{1}),
\bfpage{73}--\blpage{83}
(\byear{2002})
\doiurl{10.1175/1520-0477(2002)083<0073:tevoeb>2.3.co;2}
\end{barticle}
\endbibitem

\bibitem[\protect\citeauthoryear{Garner and Thompson}{2012}]{Garner2012}
\begin{barticle}
\bauthor{\bsnm{Garner}, \binits{G.G.}},
\bauthor{\bsnm{Thompson}, \binits{A.M.}}:
\batitle{The value of air quality forecasting in the mid-atlantic region}.
\bjtitle{Weather, Climate, and Society}
\bvolume{4},
\bfpage{69}--\blpage{79}
(\byear{2012})
\doiurl{10.1175/WCAS-D-10-05010.1}
\end{barticle}
\endbibitem

\bibitem[\protect\citeauthoryear{Ross et~al.}{2022}]{ross2022beachlevel}
\begin{barticle}
\bauthor{\bsnm{Ross}, \binits{S.D.}},
\bauthor{\bsnm{Fish}, \binits{J.}},
\bauthor{\bsnm{Moeltner}, \binits{K.}},
\bauthor{\bsnm{Bollt}, \binits{E.M.}},
\bauthor{\bsnm{Bilyeu}, \binits{L.}},
\bauthor{\bsnm{Fanara}, \binits{T.}}:
\batitle{Beach-level 24-hour forecasts of florida red tide-induced respiratory
  irritation}.
\bjtitle{Harmful Algae}
\bvolume{111},
\bfpage{102149}
(\byear{2022})
\doiurl{10.1016/j.hal.2021.102149}
\end{barticle}
\endbibitem

\bibitem[\protect\citeauthoryear{Kirchner}{2006}]{Kirchner2006}
\begin{barticle}
\bauthor{\bsnm{Kirchner}, \binits{J.W.}}:
\batitle{Getting the right answers for the right reasons: linking measurements,
  analyses, and models to advance the science of hydrology}.
\bjtitle{Water Resources Research}
\bvolume{42}(\bissue{3}),
\bfpage{03}--\blpage{04}
(\byear{2006})
\doiurl{10.1029/2005WR004362}
\end{barticle}
\endbibitem

\bibitem[\protect\citeauthoryear{Dai
  et~al.}{2026}]{aldegunde2024pollutionalerts}
\begin{barticle}
\bauthor{\bsnm{Dai}, \binits{Y.}},
\bauthor{\bsnm{Qian}, \binits{J.}},
\bauthor{\bsnm{Yang}, \binits{Y.}},
\bauthor{\bsnm{Liu}, \binits{B.}},
\bauthor{\bsnm{Li}, \binits{S.}},
\bauthor{\bsnm{Zhang}, \binits{K.}},
\bauthor{\bsnm{Xie}, \binits{Q.}},
\bauthor{\bsnm{Tong}, \binits{C.}},
\bauthor{\bsnm{Chen}, \binits{Y.}},
\bauthor{\bsnm{MacKenzie}, \binits{A.R.}},
\bauthor{\bsnm{Shi}, \binits{Z.}}:
\batitle{Significant benefits of pollution alerts for cleaner air and better
  health}.
\bjtitle{PNAS Nexus}
\bvolume{5}(\bissue{3}),
\bfpage{054}
(\byear{2026})
\doiurl{10.1093/pnasnexus/pgag054}
\end{barticle}
\endbibitem

\bibitem[\protect\citeauthoryear{{Environment Agency}}{2024}]{EA2024h2scal}
\begin{botherref}
\oauthor{\bsnm{{Environment Agency}}}:
Hydrogen sulphide calibration expert group report.
Technical report,
Environment Agency
(2024).
8 August 2024
\end{botherref}
\endbibitem

\bibitem[\protect\citeauthoryear{Vosper et~al.}{2013}]{LITVosper2013High}
\begin{barticle}
\bauthor{\bsnm{Vosper}, \binits{S.}},
\bauthor{\bsnm{Carter}, \binits{E.}},
\bauthor{\bsnm{Lean}, \binits{H.}},
\bauthor{\bsnm{Lock}, \binits{A.}},
\bauthor{\bsnm{Clark}, \binits{P.}},
\bauthor{\bsnm{Webster}, \binits{S.}}:
\batitle{High resolution modelling of valley cold pools}.
\bjtitle{Atmospheric Science Letters}
\bvolume{14}(\bissue{3}),
\bfpage{193}--\blpage{199}
(\byear{2013})
\doiurl{10.1002/asl2.439}
\end{barticle}
\endbibitem

\bibitem[\protect\citeauthoryear{Harvey and Fernandes}{1989}]{harvey1989}
\begin{barticle}
\bauthor{\bsnm{Harvey}, \binits{A.C.}},
\bauthor{\bsnm{Fernandes}, \binits{C.}}:
\batitle{Time series models for count or qualitative observations}.
\bjtitle{Journal of Business \& Economic Statistics}
\bvolume{7}(\bissue{4}),
\bfpage{407}--\blpage{417}
(\byear{1989})
\doiurl{10.1080/07350015.1989.10509750}
\end{barticle}
\endbibitem

\bibitem[\protect\citeauthoryear{de~Vos and Rientjes}{2005}]{deVos2005}
\begin{barticle}
\bauthor{\bsnm{Vos}, \binits{N.J.}},
\bauthor{\bsnm{Rientjes}, \binits{T.H.M.}}:
\batitle{Constraints of artificial neural networks for rainfall-runoff
  modelling: trade-offs in hydrological state representation and model
  evaluation}.
\bjtitle{Hydrology and Earth System Sciences}
\bvolume{9}(\bissue{1/2}),
\bfpage{111}--\blpage{126}
(\byear{2005})
\doiurl{10.5194/hess-9-111-2005}
\end{barticle}
\endbibitem

\bibitem[\protect\citeauthoryear{de~Vos and Rientjes}{2008}]{deVos2008}
\begin{barticle}
\bauthor{\bsnm{Vos}, \binits{N.J.}},
\bauthor{\bsnm{Rientjes}, \binits{T.H.M.}}:
\batitle{Multiobjective training of artificial neural networks for
  rainfall-runoff modeling}.
\bjtitle{Water Resources Research}
\bvolume{44}(\bissue{8}),
\bfpage{08434}
(\byear{2008})
\doiurl{10.1029/2007WR006734}
\end{barticle}
\endbibitem

\bibitem[\protect\citeauthoryear{Akyol et~al.}{2023}]{Akyol2023Avoiding}
\begin{barticle}
\bauthor{\bsnm{Akyol}, \binits{H.B.}},
\bauthor{\bsnm{Preist}, \binits{C.}},
\bauthor{\bsnm{Schien}, \binits{D.}}:
\batitle{Avoiding overconfidence in predictions of residential energy demand
  through identification of the persistence forecast effect}.
\bjtitle{IEEE Transactions on Smart Grid}
\bvolume{14},
\bfpage{228}--\blpage{238}
(\byear{2023})
\doiurl{10.1109/tsg.2022.3198326}
\end{barticle}
\endbibitem

\bibitem[\protect\citeauthoryear{Cohen}{1968}]{Cohen1968}
\begin{barticle}
\bauthor{\bsnm{Cohen}, \binits{J.}}:
\batitle{Weighted kappa: nominal scale agreement provision for scaled
  disagreement or partial credit}.
\bjtitle{Psychological Bulletin}
\bvolume{70}(\bissue{4}),
\bfpage{213}--\blpage{220}
(\byear{1968})
\doiurl{10.1037/h0026256}
\end{barticle}
\endbibitem

\bibitem[\protect\citeauthoryear{Epstein}{1969}]{epstein1969scoring}
\begin{barticle}
\bauthor{\bsnm{Epstein}, \binits{E.S.}}:
\batitle{A scoring system for probability forecasts of ranked categories}.
\bjtitle{Journal of Applied Meteorology}
\bvolume{8}(\bissue{6}),
\bfpage{985}--\blpage{987}
(\byear{1969})
\doiurl{10.1175/1520-0450(1969)008<0985:ASSFPF>2.0.CO;2}
\end{barticle}
\endbibitem

\bibitem[\protect\citeauthoryear{Gneiting and Raftery}{2007}]{gneiting2007}
\begin{barticle}
\bauthor{\bsnm{Gneiting}, \binits{T.}},
\bauthor{\bsnm{Raftery}, \binits{A.E.}}:
\batitle{Strictly proper scoring rules, prediction, and estimation}.
\bjtitle{Journal of the American Statistical Association}
\bvolume{102}(\bissue{477}),
\bfpage{359}--\blpage{378}
(\byear{2007})
\doiurl{10.1198/016214506000001437}
\end{barticle}
\endbibitem

\bibitem[\protect\citeauthoryear{Laperre et~al.}{2020}]{Laperre2020}
\begin{barticle}
\bauthor{\bsnm{Laperre}, \binits{B.}},
\bauthor{\bsnm{Amaya}, \binits{J.}},
\bauthor{\bsnm{Lapenta}, \binits{G.}}:
\batitle{Dynamic time warping as a new evaluation for {Dst} forecast with
  machine learning}.
\bjtitle{Frontiers in Astronomy and Space Sciences}
\bvolume{7},
\bfpage{39}
(\byear{2020})
\doiurl{10.3389/fspas.2020.00039}
\end{barticle}
\endbibitem

\bibitem[\protect\citeauthoryear{Hamill}{1999}]{hamill1999}
\begin{barticle}
\bauthor{\bsnm{Hamill}, \binits{T.M.}}:
\batitle{Hypothesis tests for evaluating numerical precipitation forecasts}.
\bjtitle{Weather and Forecasting}
\bvolume{14}(\bissue{2}),
\bfpage{155}--\blpage{167}
(\byear{1999})
\doiurl{10.1175/1520-0434(1999)014<0155:htfenp>2.0.co;2}
\end{barticle}
\endbibitem

\bibitem[\protect\citeauthoryear{Diebold and Mariano}{1995}]{diebold1995}
\begin{barticle}
\bauthor{\bsnm{Diebold}, \binits{F.X.}},
\bauthor{\bsnm{Mariano}, \binits{R.S.}}:
\batitle{Comparing predictive accuracy}.
\bjtitle{Journal of Business \& Economic Statistics}
\bvolume{13}(\bissue{3}),
\bfpage{253}--\blpage{263}
(\byear{1995})
\doiurl{10.1080/07350015.1995.10524599}
\end{barticle}
\endbibitem

\bibitem[\protect\citeauthoryear{Kunsch}{1989}]{kunsch1989}
\begin{botherref}
\oauthor{\bsnm{Kunsch}, \binits{H.R.}}:
The jackknife and the bootstrap for general stationary observations.
The Annals of Statistics
\textbf{17}(3)
(1989)
\doiurl{10.1214/aos/1176347265}
\end{botherref}
\endbibitem

\bibitem[\protect\citeauthoryear{Nagata}{2003}]{Nagata2003}
\begin{bchapter}
\bauthor{\bsnm{Nagata}, \binits{Y.}}:
\bctitle{Measurement of odor threshold by triangle odor bag method}.
In: \bbtitle{Odor Measurement Review},
pp. \bfpage{118}--\blpage{127}.
\bpublisher{Office of Odor, Noise and Vibration, Environmental Management
  Bureau, Ministry of the Environment, Japan}, \blocation{???}
(\byear{2003})
\end{bchapter}
\endbibitem

\bibitem[\protect\citeauthoryear{Amoore}{1985}]{Amoore1985}
\begin{botherref}
\oauthor{\bsnm{Amoore}, \binits{J.E.}}:
The perception of hydrogen sulfide odor in relation to setting an ambient
  standard.
Final report, contract a4-046-33,
California Air Resources Board
(1985)
\end{botherref}
\endbibitem

\end{thebibliography}
